\documentclass[12pt]{article}

\pdfoutput=1

\usepackage[english]{babel}
\usepackage{amsmath,amssymb,amsbsy,amstext}
\usepackage{graphicx}
\usepackage{amsfonts}
\usepackage{amssymb}
\usepackage{subfigure}
\usepackage{color}
\usepackage{float}

\numberwithin{equation}{section}
\newcommand{\be}{\begin{eqnarray}}

\newcommand{\V}{\mathcal{V}}

\newcommand{\ee}{\end{eqnarray}}
\newcommand{\bdm}{\begin{displaymath}}
\newcommand{\edm}{\end{displaymath}}

\newcommand{\ba}{\begin{array}}
\newcommand{\ea}{\end{array}}

\newcommand{\al}{\alpha'}

\newcommand{\FF}{\mathcal{F}}

\newcommand{\U}{\mathcal{U}_{mod}}

\newcommand{\Mpl}{M_{p}}

\newcommand{\A}{f}

\newcommand{\RR}{R_{\perp}}
\newcommand{\X}{X}

\newcommand{\lo}{\lambda_1}
\newcommand{\lt}{\lambda_2}

\newcommand{\cbm}{c^{(-)}}
\newcommand{\cbp}{c^{(+)}}

\def\bea{\begin{eqnarray}}
\def\eea{\end{eqnarray}}

\DeclareRobustCommand{\SkipTocEntry}[4]{}

\begin{document}
\begin{titlepage}

\setcounter{page}{1} \baselineskip=15.5pt \thispagestyle{empty}

\begin{flushright}
UTTG-08-09\\
TCC-24-09\\
SU-ITP-09-34
\end{flushright}
\vfil

\begin{center}
{\LARGE\bf  Oscillations in the CMB\\[.3cm] from Axion Monodromy Inflation}
\vskip 15pt
\end{center}

\vspace{0.5cm}
\begin{center}
{\large Raphael Flauger,$^{1}$  Liam McAllister,$^{2}$ Enrico Pajer,$^{2}$}
\vskip 3pt
{\large Alexander Westphal,$^{3}$ and Gang Xu$^{2}$}
\end{center}

\vspace{0.3cm}

\begin{center}
\textit{${}^1$ Department of Physics, University of Texas at Austin,
Austin, TX 78712}\\

\vskip 4pt
\textit{${^2}$ Department of Physics, Cornell University,
Ithaca, NY 14853}\\

\vskip 4pt
\textit{${}^3$ Department of Physics, Stanford University,
Stanford, CA 94305}\\

\end{center}
\vspace{0.8cm}

\noindent
We study the CMB observables in axion monodromy inflation.  These well-motivated scenarios for inflation in string theory have monomial potentials over super-Planckian field ranges, with superimposed sinusoidal modulations from instanton effects.  Such periodic modulations of the potential can drive resonant enhancements of the correlation functions of cosmological perturbations, with characteristic modulations of the amplitude as a function of wavenumber.  We give an analytical result for the scalar power spectrum in this class of models, and we determine the limits that present data places on the amplitude and frequency of modulations. Then, incorporating an improved understanding of the realization of axion monodromy inflation in string theory, we perform a careful study of microphysical constraints in this scenario.  We find that detectable modulations of the scalar power spectrum are commonplace in well-controlled examples, while resonant contributions to the bispectrum are undetectable in some classes of examples and detectable in others. We conclude that resonant contributions to the spectrum and bispectrum are a characteristic signature of axion monodromy inflation that, in favorable cases, could be detected in near-future experiments.

\vfil

\end{titlepage}

\newpage
\tableofcontents
\newpage

\section{Introduction}

Inflation \cite{Guth:1980zm, Linde:1981mu, Albrecht:1982wi} is a successful paradigm for describing the early universe, but it is sensitive to the physics of the ultraviolet completion of gravity. This motivates pursuing realizations of inflation in string theory, a candidate theory of quantum gravity.  Considerable progress has been made on this problem in recent years, so much so that the most pressing task, particularly in view of upcoming CMB experiments, is to learn how to distinguish various incarnations of inflation in string theory from each other and from related models constructed directly in quantum field theory.

Fortunately, the additional constraints inherent in realizing inflation in an ultraviolet-complete framework can leave imprints in the low-energy Lagrangian, and hence ultimately in the cosmological observables.  In favorable cases, a given class of models may make distinctive predictions for a variety of correlated observables, allowing one to exclude this class of models given adequate data.

One decisive observable for probing inflation is the tensor-to-scalar ratio, $r$. A promising class of string inflation models producing a detectable tensor signature are those involving {\it monodromy} \cite{SW}, in which the potential energy is not periodic under transport around an angular direction in the configuration space.  The first examples \cite{SW} involved monodromy under transport of a wrapped D-brane in a nilmanifold, and a subsequent class of examples invoked monodromy in the direction of a closed string axion \cite{MSW}.

The axion monodromy inflation scenario of \cite{MSW} is falsifiable on the basis of its tensor signature, $r\approx 0.07$.  However, primordial tensor perturbations have not been detected at present, while the temperature anisotropies arising from scalar perturbations have been mapped in great detail \cite{Komatsu:2008hk}. One could therefore hope to constrain axion monodromy inflation more effectively by understanding the signatures that it produces in the scalar power spectrum and bispectrum. Characterizing these signatures is the subject of the present paper.

As we shall explain, the potential in axion monodromy inflation is approximately linear, but periodically modulated: each circuit of the loop in configuration space can provide a bump on top of the otherwise linear potential. Modulations of the inflaton potential with suitable frequency and amplitude can yield two striking signatures: periodic undulations in the spectrum of the scalar perturbations, and resonant enhancement \cite{Chen:2008wn} of the bispectrum.  Let us stress that the presence of some degree of modulations of the potential is automatic, and is an example of the situation described above in which traces of ultraviolet physics remain in the low-energy Lagrangian. We do not introduce modulations in order to make the scalar perturbations more interesting.  However, it is important to examine the typical amplitude and frequency of modulations in models that are under good microphysical control, in order to ascertain whether well-motivated models produce signatures that can be detected in practice.

To achieve this, we first investigate in detail the realization of axion monodromy inflation in string theory.  We compute the axion decay constants in terms of compactification data, we assess the importance of higher-derivative terms, and we estimate the amplitude of modulations for the case of Euclidean D1-brane contributions to the K\"ahler potential.  We also identify a potentially-important contribution to the inflaton potential, arising from backreaction in the compact space, and we present a model-building solution that suppresses this contribution.

We find that detectable modulations of the scalar power spectrum and bispectrum are possible in models that are consistent with all current data and that are under good microphysical control.  In fact, we find substantial parameter ranges that are excluded not by microphysics, but by observational constraints on modulations of the scalar power spectrum.

The organization of this paper is as follows.  We begin in \S\ref{s:back} by describing the classical evolution of the homogeneous background in axion monodromy inflation with a modulated linear potential. We then solve, in \S\ref{s:spectrum},  the Mukhanov-Sasaki equation governing the evolution of scalar perturbations, giving an analytical result for the spectrum in terms of the frequency and amplitude of the modulations of the potential.  Next, we briefly discuss the bispectrum and express the amplitude of the non-Gaussianity in terms of the model parameters. We then present, in \S\ref{s:numerics},  an analysis of the constraints imposed on axion monodromy inflation by the WMAP5 data (for prior work constraining similar oscillatory power spectra, see {\it e.g.}  \cite{Martin:2003sg, Martin:2004yi, Easther:2004vq, Chen:2006xjb, Hamann:2007pa, Bean:2008na, Pahud:2008ae}).
Then, in \S\ref{s:micro} and \S\ref{s:const}, we present a comprehensive analysis of the constraints imposed by the requirements of computability and of microphysical consistency, including validity of the string loop and $\alpha'$ perturbation expansions, successful moduli stabilization, and bounds on higher-derivative terms. In \S\ref{s:7} we combine the observational and theoretical constraints, with results presented in figure \ref{theory}.

\subsection{Review of axion monodromy inflation}

In this section we will briefly review the motivation for axion monodromy inflation, as well as the most salient phenomenological features. We will postpone until \S\ref{s:micro} a more comprehensive discussion of the realization of this model in string theory.

Inflation is sensitive to Planck-scale physics:  contributions to the effective action arising from integrating out degrees of freedom with masses as large as the Planck scale play a critical role in determining the background evolution, and hence the observable spectrum of perturbations (see \cite{BM} for a review of this issue). A central problem in inflationary model-building is establishing knowledge of Planck-suppressed terms in the effective action with accuracy sufficient for making predictions.  The most elegant solution to this problem is to provide a symmetry that forbids such Planck-suppressed contributions. Because invoking such a symmetry amounts to forbidding couplings of the inflaton to Planck-scale degrees of freedom, it is important to understand this issue in an ultraviolet-complete theory, such as string theory.

One promising mechanism for inflation in string theory involves the shift symmetry of an axion.
Axions are numerous in string compactifications and generally enjoy continuous shift symmetries $a \to a + constant$ that are valid to all orders in perturbation theory, but are broken by nonperturbative effects to discrete shifts $a \to a + 1$.  As noted in \cite{MSW}, the shift symmetries of axions descending from two-forms are also broken by suitable space-filling fivebranes (D5-branes or NS5-branes) wrapping two-cycles in the compact space.

In axion monodromy inflation \cite{MSW}, an NS5-brane wrapped on a two-cycle $\Sigma$ breaks the shift symmetry of the Ramond-Ramond two-form potential $C_2$, inducing a potential that is asymptotically linear in the corresponding canonically normalized field $\phi$,
\begin{equation}
V=\mu^3\phi \, ,
\end{equation}
with $\mu$ a constant mass scale.
Inflation begins with a large expectation value for the inflaton, $\phi \propto \int_{\Sigma} C_2 \gg 1$, and proceeds as this expectation value diminishes; note that the NS5-brane, like any D-branes that may be present in the compactification, remains fixed in place during inflation.  As argued in \cite{MSW}, this gives rise to a natural model of inflation, with the residual shift symmetry of the axion protecting the potential from problematic corrections that are endemic in string inflation scenarios.

In this paper we perform a careful analysis of the consequences of nonperturbative effects for the axion monodromy scenario.  Such effects are generically present: specifically, Euclidean D-branes make periodic contributions to the potential in most realizations of axion monodromy inflation.  However, the size of these contributions is model-dependent.  It was shown in \cite{MSW} that there exist classes of examples in which nonperturbative effects are practically negligible, but we expect -- as explained in detail in \S\ref{ss:toy} -- that in generic configurations, periodic terms in the potential make small, but not necessarily negligible, contributions to the slow roll parameters.

Therefore, it is of interest to understand the consequences of small periodic modulations of the inflaton potential in axion monodromy inflation. In this paper we address this question in two ways: first, in \S\ref{s:back}-\S\ref{s:numerics}, by studying a phenomenological potential that captures the essential effects; and second, in \S\ref{s:micro} and \S\ref{s:const}, by investigating the ranges of the phenomenological parameters that satisfy all known microphysical consistency requirements dictated by the structure of string compactifications in which axion monodromy inflation can be realized.


\section{Background Evolution} \label {s:back}

\par
In this section we will study the background evolution of the inflaton in the presence of small periodic modulations of
the potential. We will focus on modulations in axion monodromy inflation with a linear potential, but our derivations are easily modified to account for other models with a modulated potential. We will denote the size of the modulation by $\Lambda^4$, and write our potential as in \cite{MSW},
\be \label{V}
V(\phi)=\mu^3\phi+\Lambda^4\cos \left(\frac{\phi}{f}\right)=\mu^3\left[\phi+b f\cos \left(\frac{\phi}{f}\right)\right]\,,
\ee
where we defined the parameter $b\equiv\frac{\Lambda^4}{\mu^3 f}$.
The equation of motion for the inflaton is then
\be\label{eom}
\ddot{\phi}+3H\dot{\phi}+\mu^3-\mu^3 b \sin \left(\frac{\phi}{f}\right)=0\,.
\ee
To solve (\ref{eom}), we begin with two approximations.  Monotonicity of the potential requires\footnote{The case of non-monotonic potentials may also be interesting. On the one hand, for sufficiently large $b>1$, it may be possible to realize chain inflation~\cite{Freese:2004vs, Freese:2006fk, Chialva:2008zw} in our model. In this scenario, the inflaton would tunnel from minimum to minimum, with the universe expanding by less than one third of an e-fold per tunneling event. This requires a more careful analysis, and we will leave this for future studies. On the other hand, for $b\gtrsim 1$ the model essentially turns into a small-field model of inflation because the inflaton gets trapped at the peaks for a large number of e-folds. It seems hard to distinguish this from other models of small field inflation, but it may be interesting to take a closer look at this as well.} $b<1$, and as we will see in \S\ref{s:numerics}, for the case $b<1$ observational constraints in fact imply $b\ll1$. This suggests treating the oscillatory term in the potential as a perturbation.  Furthermore, the COBE normalization implies that $\phi\gg M_p$ during the era when the modes that are observable in the cosmic microwave background exit the horizon. This allows us to drop terms of higher order in $M_p/\phi$.

Under these conditions, it is straightforward to solve for the evolution of the homogeneous background. Expanding the field as $\phi=\phi_0+b \phi_1+{\cal O}(b^2)$, the equations of motion of zeroth and first order in $b$ become
\be\label{phi0}
\dot{\phi}_0=-\sqrt{\frac{\mu^3}{3\phi_0}}\,,
\ee
\be\label{phi1}
\ddot{\phi}_1+\sqrt{3\mu^3\phi_0}\dot{\phi}_1-\frac{\mu^3}{2\phi_0}\phi_1=\mu^3\sin \left(\frac{\phi_0}{f}\right)\,,
\ee
where we have neglected terms of higher order in $M_p/\phi$ and we have made use of the slow roll approximation for $\phi_0$.\footnote{In approximating $\sin\left( \phi/f\right)\simeq \sin\left(\phi_0/f\right)$ on the right hand side of \eqref{phi1}, we have assumed not only that $b\ll1$ but also that $b\phi_1/f\ll1$. As we will see from the solution \eqref{rfback}, $\phi_1$ is of order $f^2\phi_*$. Hence the mild assumption $bf\phi_*\ll1$ justifies this approximation.}
Using equation \eqref{phi0}, we can rewrite equation \eqref{phi1} with $\phi_0$ as an independent variable instead of $t$, yielding
\be
\phi_1''-3\phi_0\phi_1'-\frac{3}{2}\phi_1=3\phi_0\sin \left(\frac{\phi_0}{f}\right)\,.
\ee
where primes denote derivatives with respect to $\phi_0$.
For the period of interest, in which the modes now visible in the CMB exit the horizon, it is a good approximation to neglect the motion of $\phi_0$ everywhere except in the driving term. The inhomogeneous solution is then given by
\be
\phi_1(t)=f\frac{6f\phi_*}{(2+3f^2)^2+36f^2 {\phi_*}^2}\left[-(2+3f^2)\sin \left(\frac{\phi_0(t)}{f}\right)+6f\phi_*\cos\left(\frac{\phi_0(t)}{f}\right)\right]\,,
\ee
where $\phi_*$ denotes the value of the field $\phi_0$ at the time at which the pivot scale $k_*$ exits the horizon. Assuming 60 e-foldings of inflation, this happens around $\phi_*\simeq 11 \Mpl$. For decay constants $f$ obeying $f\gtrsim M_p/10$, there is less than one oscillation in the range of modes that are observable in the cosmic microwave background, leading to an uninteresting modulation with very long wavelength. We will thus make the additional assumption that $f\ll M_p$. Assuming that $\phi_0\gg M_p$ and $f\ll1$, using the slow roll approximation for $\phi_0(t)$, and working to first order in $b$, the solution thus becomes
\be\label{rfback}
\phi(t)=\phi_0(t)+b\phi_1(t)=\phi_0(t)+ b f\frac{3f\phi_*}{1+(3f \phi_*)^2}\left[-\sin \left(\frac{\phi_0(t)}{f}\right)+3f\phi_*\cos\left(\frac{\phi_0(t)}{f}\right)\right]\,,
\ee
with $\phi_0(t)$ given by
\be
\phi_0(t)=\left[\phi_*^{3/2}-\frac{\sqrt{3}}{2}\mu^{3/2}(t-t_*)\right]^{2/3}\,.
\ee

In the absence of oscillations, {\it i.e.} for $b=0$, axion monodromy provides a model of large field inflation that is easily studied using the slow roll expansion. Assuming for concreteness that the CMB scales left the horizon 60 e-foldings before the end of inflation, we are interested in the perturbations around $\phi_*\simeq11\Mpl$. After imposing the COBE normalization, one finds that CMB perturbations are produced at a scale $V^{1/4}\simeq 7\cdot 10^{-3}\Mpl\simeq 1.7 \cdot 10^{16}$ GeV with a spectral tilt $n_s\simeq 0.975$ and a tensor-to-scalar ratio $r\simeq 0.07$. For reference, the Hubble constant during inflation is then $H\simeq 2.8 \cdot 10^{-5}
\Mpl\simeq 6.8 \cdot 10^{13}$ GeV.

One can then ask what happens once the oscillations are switched on, {\it i.e.} when $b\neq0$. It turns out that the effect on the number of e-foldings is negligible as long as $b\ll 1$. Hence the inflationary scale is well-approximated by the slow roll analysis. On the other hand, the detailed properties of the perturbations are very different from the slow roll case and cannot be calculated in that expansion. We turn to this issue in the next section.


\section{Spectrum of Scalar Perturbations} \label {s:spectrum}

Having understood the background evolution, we are now in a position to calculate the power spectrum in axion monodromy inflation. One might be tempted to do this by brute-force numerical calculation, but we find it more instructive to have an analytic result.  We will show that under the same assumptions made in calculating the background evolution, {\it i.e.} slow roll for $\phi_0(t)$, $\phi_0\gg M_p$, $f\ll M_p$, and to first order in $b$, the scalar power spectrum is of the form
\be\label{Psk}
\Delta_\mathcal{R}^2(k)=\Delta_{\cal R}^2(k_*)\left(\frac{k}{k_*}\right)^{n_s - 1}\left[1+\delta n_s \cos\left(\frac{\phi_k}{f}\right)\right]\approx\Delta_{\cal R}^2\left(\frac{k}{k_*}\right)^{n_s - 1+ \frac{\delta
n_s}{\ln(k/k_*)}\cos\left(\frac{\phi_k}{f}\right)}\,,
\ee
where the quantity $\Delta_{\cal R}^2(k_*)$ parameterizes the strength of the scalar perturbations and will be introduced in detail in the next subsection. The second equality is valid as long as $\delta n_s\ll1$, and $\delta n_s$ is given by
\be
\label{dns}
\delta n_s=\frac{12b}{\sqrt{(1+(3f \phi_*)^2)}}\sqrt{\frac{\pi}{8}\coth\left(\frac{\pi}{2 f\phi_*}\right) f\phi_*}\,,
\ee
where
\be \label{phik}
\phi_k=\sqrt{\phi_*^2-2\ln k/k_*} \simeq\phi_*-\frac{\ln k/k_*}{\phi_*}
\ee
is the value of the scalar field at the time when the mode with comoving momentum $k$ exits the horizon.

In \S\ref{exact} we will give a derivation of this result that makes no further approximations. In \S\ref{saddle} and \S\ref{B} we will present two additional derivations of \eqref{Psk} that are valid only as long as $f\phi_*\ll1$ but that lead to a better understanding of the relevant physical effects behind the power spectrum \eqref{Psk}. Let us at this point briefly summarize the scales that will be relevant for our discussion in the next subsections.

Given the potential \eqref{V}, the time frequency of the oscillations of the inflaton is $\omega=\dot \phi/ f$. This is also the time frequency of the oscillations of the background. Perturbations around this background can be quantized in terms of the solutions of the Mukhanov-Sasaki equation, assuming an asymptotic Bunch-Davies vacuum. Every perturbation mode with comoving momentum $k$ oscillates with a time frequency $k/a$ that is redshifted by the expansion of the universe until the mode exits the horizon and freezes when $k=aH$.

Then, if $H<\omega<\Mpl$, every mode will at a certain time resonate with the background, as stressed by Chen, Easther, and Lim in \cite{Chen:2008wn}.
Using the slow roll equation of motion and the COBE normalization,
\be
  3H\dot{\phi}\simeq -V'(\phi)\,,\quad \dot{\phi}^2\simeq\frac23 \epsilon V\,, \quad V \simeq  5\cdot 10^{-7}\,\epsilon \, \Mpl^4\,,
\ee
the requirement $H<\omega<\Mpl$ can be re-expressed as
\be \label{omega}
\frac{\omega}{H}&\simeq&\frac{\Mpl^2}{\phi f}\simeq \sqrt{2\epsilon}\frac{\Mpl}{f}>1\,,\\
\frac{\omega}{\Mpl}&\simeq&\sqrt{\frac{2\epsilon V}{3}}\frac{1}{f M_{pl}}<1\,,
\ee
hence defining a range of values for the axion decay constant $f$ for which resonances occur. Using $\sqrt{2\epsilon}\simeq \Mpl/\phi_*\simeq .09$, we obtain $2.4\cdot 10^{-6}<\frac{f}{M_{pl}}<0.09$. We will show in \S\ref{s:micro} and \S\ref{s:const} that $f$ falls in this range in a class of microphysically well-controlled examples.

Going beyond our approximations, the model also predicts a small amount of running of the scalar spectral index, of order $10^{-4}$, from terms of higher order in the $M_p/\phi$ expansion. Furthermore, $\delta n_s$ develops a very mild momentum dependence. We will neglect these effects because these will most likely not be observable in current or near-future CMB experiments.



\subsection{Analytic solution of the Mukhanov-Sasaki equation}\label{exact}

We begin our study of the spectrum by choosing a gauge
such that the scalar field is unperturbed, $\delta\phi({\bf x},t)=0$, and the scalar perturbations in the spatial part of the metric take the form
\be
\delta g_{ij}({\bf x},t)=2a(t)^2\mathcal{R}({\bf x},t)\delta_{ij}\,.
\ee
The quantity $\mathcal{R}({\bf x},t)$ is a gauge-invariant quantity and in the case of single-field inflation is conserved outside the horizon. It is closely related to the scalar curvature of the spatial slices, but we will not need its precise geometric interpretation at this point.

\begin{figure}
\begin{center}
\includegraphics[width=4in]{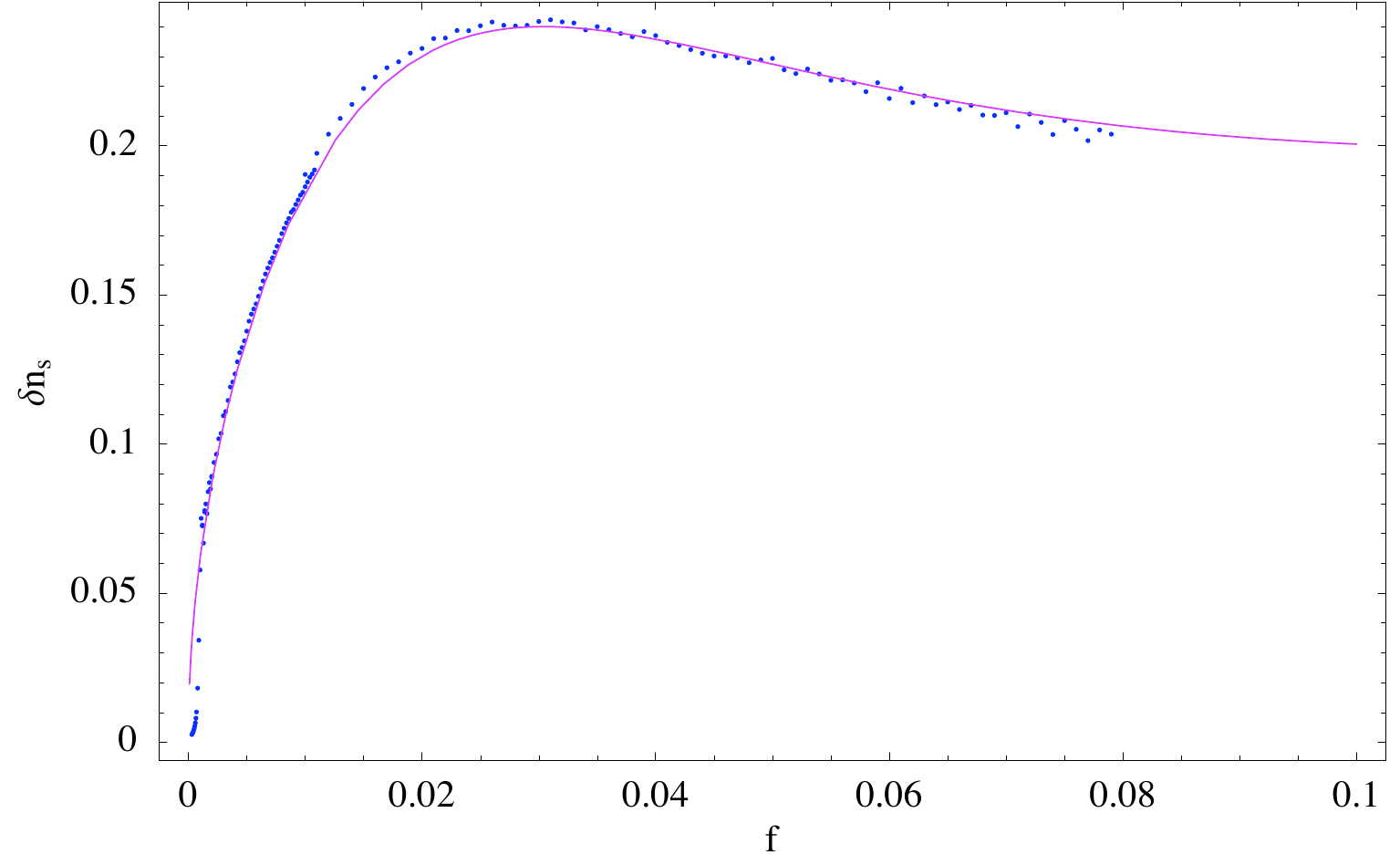}
        \caption{The solid line is the analytical result for $\delta n_s$ as a function of $f$, for $b=0.08$, while the dots are the numerical result obtained from an adaptation of the code used in \cite{inflation}.}

      \label{fig:dnsvsfp}
\end{center}

\end{figure}

The translational invariance of the background and thus the equations of motion governing the time evolution of the perturbations make it convenient to look for solutions of the linearized Einstein equations in Fourier space. One defines
\be
\mathcal{R}({\bf x},t)=\int\frac{d^3k}{(2\pi)^{3/2}}\left[\mathcal{R}_k(t)e^{i{\bf k}\cdot{\bf x}}\alpha({\bf k})+\mathcal{R}_k(t)^*e^{-i{\bf k}\cdot{\bf x}}\alpha^*({\bf k})\right]\,,
\ee
where ${\bf k}$ is the comoving momentum, and $k$ is its magnitude. The rotational invariance of the background ensures that $\mathcal{R}_k(t)$ can depend only on the magnitude of the comoving momentum but not on its direction. Directional dependence can only be contained in the stochastic parameter $\alpha({\bf k})$ that parameterizes the initial conditions and is normalized so that
\be
\left\langle\alpha({\bf k})\alpha^*({\bf k}')\right\rangle=\delta({\bf k}-{\bf k'})\,,
\ee
where the average denotes the average over all possible histories.
With this ansatz, the Einstein equations turn into an ordinary differential equation, the Mukhanov-Sasaki equation, governing the time evolution of $\mathcal{R}_k(t)$.  We will use it in the form\footnote{We use the same definitions for the slow roll parameters as in \cite{Weinberg}, {\it i.e.} $\epsilon\equiv -\frac{\dot{H}}{H^2}$, $\delta\equiv \frac{\ddot{H}}{2H\dot{H}}$. $\delta$ is related to the Hubble slow-roll parameters $\eta\equiv\dot \epsilon/\epsilon H$ by $\delta=\eta/2-\epsilon$. The other slow-roll parameters that are sometimes used are $\epsilon_V\equiv (V'/V)^2/2$ and $\eta_V\equiv V''/V$. When the slow roll expansion is valid they are related to the Hubble slow-roll parameters by $\epsilon_V=\epsilon$ and $\eta_V=4\epsilon-\eta$.}
\be \label{MSx}
\frac{d^2\mathcal{R}_k}{dx^2}-\frac{2(1+2\epsilon+\delta)}{x}\frac{d\mathcal{R}_k}{dx}+\mathcal{R}_k=0\,,
\ee
where $x\equiv-k\tau$, with the conformal time $\tau$ given as usual by $\tau\equiv \int^t\frac{dt'}{a(t')}$.
Outside the horizon, {\it i.e.} for $x\ll1$ or equivalently $k/a\ll H$, the quantity $\mathcal{R}_k(x)$ approaches a constant which we denote by $\mathcal{R}_k^{(o)}$.
In terms of $\mathcal{R}_k^{(o)}$ we define the primordial power spectrum for the scalar modes as
\be
\left|\mathcal{R}_k^{(o)}\right|^2=2\pi^2 \frac{\Delta_\mathcal{R}^2(k)}{k^3}\,.
\ee
To evaluate this quantity, it will again turn out to be sufficient to solve to first order in $b$. We therefore expand the slow roll parameters,
\begin{equation}
\epsilon=\epsilon_0 + \epsilon_1 + {\cal{O}}(b^2)\,,
\end{equation}
\begin{equation}
\delta=\delta_0 + \delta_1 + {\cal{O}}(b^2)\,.
\end{equation}
For the background solution \eqref{rfback}, the first-order terms are given by
\be
\epsilon_1=-\frac{3b f}{\phi_* [1+(3f \phi_*)^2]}\left[\cos\left(\frac{\phi_0}{f}\right)+(3f\phi_*)\sin\left(\frac{\phi_0}{f}\right)\right]\,,
\ee
\be
\delta_1=-\frac{3b}{[1+(3f \phi_*)^2]}\left[\sin\left(\frac{\phi_0}{f}\right)-(3f\phi_*)\cos\left(\frac{\phi_0}{f}\right)\right]\,.
\ee
We now consider an ansatz of the form
\be\label{Ra}
\mathcal{R}_k=\mathcal{R}_{k,0}^{(o)}\left[i\sqrt{\frac{\pi}{2}}x^{\nu_0} H_{\nu_0}^{(1)}(x)+g(x)\right]\,.
\ee
Here the index $\nu_0$ on the Hankel function, $H_{\nu_0}^{(1)}(x)$, is given by $\nu_{0}=\frac32+2\epsilon_0+\delta_0$, $g(x)$ is a perturbation of order $b$, and $\mathcal{R}_{k,0}^{(o)}$ is the value of $\mathcal{R}_k(t)$ outside the horizon in the absence of modulations, {\it i.e.} for $b=0$. To be explicit, it is given by\footnote{As mentioned earlier, we will ignore the running of the scalar spectral index, but it may be worth pointing out that the information about the running is contained in this formula.}
\be
\mathcal{R}_{k,0}^{(o)}=\mp i \sqrt{\frac{\mu^3 \phi_k^3}{6}}\frac{1}{k^{3/2}}\,,
\ee
where $\phi_k\approx\phi_*-\frac{\ln k/k_*}{\phi_*}$ once again is the value of the scalar field at the time the mode with comoving momentum $k$ exits the horizon. The quantity of interest to first order in $b$ is then
\be \label{Rq2}
\left|\mathcal{R}_k^{(o)}\right|^2=\left|\mathcal{R}_{k,0}^{(o)}\right|^2\Bigl[1+2\,\mathrm{Re}\,g(0)\Bigr]\approx\left|\mathcal{R}_{k,0}^{(o)}\right|^2e^{2\,\mathrm{Re}\,g(0)}=\left|\mathcal{R}_{k,0}^{(o)}\right|^2\left(\frac{k}{k_*}\right)^{\frac{2\,\mathrm{Re}\,g(0)}{\ln(k/k_*)}}\,.
\ee
Our ansatz automatically solves the equation of order $b^0$. To first order in $b$ and in the slow roll parameters, the Mukhanov-Sasaki equation leads to an equation for $g(x)$ of the form
\be \label{MSr}
\frac{d^2g}{dx^2}-\frac{2}{x}\frac{dg}{dx}+g=2 e^{ix}(2\epsilon_1+\delta_1)\,.
\ee
In writing this equation, we have dropped terms of order $\mathcal{O}(b \epsilon_0,b \delta_0)$, which amounts to setting $\nu_0=3/2$. Next, we notice that $\epsilon_1$ is suppressed relative to $\delta_1$ by a factor $\frac{f}{\phi_*}$. Since we are interested in the regime $\frac{f}{\phi_*}\ll 1$, we can thus drop the term proportional to $\epsilon_1$ on the right hand side of equation \eqref{MSr}. Furthermore, it turns out to be convenient to rewrite $\delta_1$ using trigonometric identities. Ignoring an unimportant phase, one finds
\be
\delta_1= -\frac{3b}{\sqrt{1+(3f \phi_*)^2}} \cos\left(\frac{\phi_{0}}{f}\right) \,.
\ee
It will be convenient to write $\phi_0(x)$ as $\phi_0(x)=\phi_*-\frac{\ln(k/k_*)}{\phi_*}+\frac{\ln x}{\phi_*}=\phi_k+\frac{\ln x}{\phi_*}$. Introducing $r(x)\equiv \mathrm{Re}\,(g(x))$, equation \eqref{MSr} becomes
\be \label{rtrig}
\frac{d^2r}{dx^2}-\frac{2}{x}\frac{dr}{dx}+r=-\frac{6b}{\sqrt{1+(3f \phi_*)^2}}\cos(x)\cos\left(\frac{\phi_k}{f}+\frac{\ln x}{f\phi_*}\right)\,.
\ee
The solution to this equation can be found {\it e.g.} using Green's functions. We are particularly interested in the inhomogeneous solution at late times, {\it i.e.} in the limit of vanishing $x$. Using more trigonometric identities, we find that the solution in this limit can be brought into the form
\be \label{r0}
r(0)=\frac{6b|\mathcal{I}(f\phi_*)|}{\sqrt{1+(3f \phi_*)^2}}\cos\left(\frac{\phi_{k}}{f}+\beta(f\phi_*)\right)\,,
\ee
where $\beta(f\phi_*)$ is an unimportant phase that we will ignore, and $\mathcal{I}$ is the integral
\be\label{eq:msint}
\mathcal{I}(f\phi_*)=\frac{\pi}{2}\int^\infty_0 dx  J_\frac32(x)J_{-\frac12}(x)\,  x^{\frac{i}{f\phi_*}}\,.
\ee
Written in this form, the integral can be recognized as a Weber-Schafheitlin integral and can be done analytically (see {\it e.g.}  \cite{bateman}). One finds
\be\label{Ics}
|\mathcal{I}|=\sqrt{\frac{\pi}{8}\coth\left(\frac{\pi}{2 f\phi_*}\right) f\phi_*}\,.
\ee
Combining equations \eqref{Rq2}, \eqref{r0} and \eqref{Ics}, we finally obtain an expression for $\delta n_s$,
\be \label{dns2}
\delta n_s=\frac{2r(0)}{\cos\left(\frac{\phi_k}{f}\right)}=\frac{12b}{\sqrt{1+(3f \phi_*)^2}}\sqrt{\frac{\pi}{8}\coth\left(\frac{\pi}{2 f\phi_*}\right) f\phi_*}\,.
\ee
Once again, this derivation is valid to first order in $b$ and assumes slow roll for $\phi_0(t)$, $\phi_0\gg M_p$, and $f\ll M_p$.
In particular, it makes no use of an $f\phi_*\ll1$ expansion, although this approximation will be needed in the derivations in \S\ref{saddle} and \S\ref{B}. A comparison between our analytical result for $\delta n_s$ as a function of $f\phi_*$ for a fixed value of $b$ and the result of a numerical calculation using a slight modification of the code described in~\cite{inflation} is shown in Figure~\ref{fig:dnsvsfp}.

\subsection{Saddle-point approximation}\label{saddle}

As we have seen in the last subsection, it is possible to calculate the power spectrum analytically to first order in $b$, assuming slow roll for $\phi_0(t)$, $\phi_0\gg M_p$, and $f\ll M_p$, but the derivation sheds little light on the physics behind the results. To get a better understanding, it is instructive to look at the integral~\eqref{eq:msint} more explicitly. For this purpose, it is convenient to separate $\mathcal{I}$ into its real and imaginary parts, $\mathcal{I}=\mathcal{I}_c+i\mathcal{I}_s$, with
\begin{eqnarray}
\mathcal{I}_c&=&\int^\infty_0 dx\,  \frac{(\sin x-x \cos x)\cos x}{x^2}\,  \cos\left({\frac{\ln x}{f\phi_*}}\right)\,,\\
\mathcal{I}_s&=&\int^\infty_0 dx\,  \frac{(\sin x-x \cos x)\cos x}{x^2}\,  \sin\left({\frac{\ln x}{f\phi_*}}\right)\,.
\end{eqnarray}
For ranges of the axion decay constant such that $f\phi_*\ll 1$, these integrals can be done in a stationary phase approximation. Using trigonometric identities to rewrite the products of trigonometric functions appearing in the integrands into sums of trigonometric functions with combined arguments, one finds that the stationary phase occurs at $\bar{x}=\frac{1}{2f\phi_*}$. Expanding around the stationary point and performing the integral as usual, one finds to leading order in $f\phi_*$
\begin{eqnarray}
\mathcal{I}_c&=&\sqrt{\frac{\pi}{8}f\phi_*}\sin\left[\frac{1+\ln (2f\phi_*)}{f\phi_*}-\frac{\pi}{4}\right]\,,\\
\mathcal{I}_s&=&\sqrt{\frac{\pi}{8}f\phi_*}\cos\left[\frac{1+\ln (2f\phi_*)}{f\phi_*}-\frac{\pi}{4}\right]\,,
\end{eqnarray}
which leads to
\be
|\mathcal{I}|=\sqrt{{\mathcal{I}_c}^2+{\mathcal{I}_s}^2}=\sqrt{\frac{\pi}{8} f\phi_*}\,.
\ee
This agrees with our previous result, equation~\eqref{Ics}, as long as $f\phi_*\ll1$. We have not only reproduced our earlier results, however: we also learn that at least for small $f\phi_*$, the integral is dominated by a period of time around $\bar{\tau}=-\frac{1}{2kf\phi_*}$. Up to the factor of two in the denominator, this corresponds to the period when the frequency of the oscillations of the scalar field background equals the frequency of the oscillations of a mode with comoving momentum $k$.\footnote{This factor of two can be understood from
momentum conservation, as will become clear in \S\ref{B}.}
The stationary phase approximation thus captures a resonance between the oscillations of the background and the oscillations of the fluctuations, and is good as long as $f\phi_*\ll1$, {\it i.e.} as long as the resonance occurs while the mode is still well inside the horizon. One might suspect that this has an interpretation in terms of particle production, and we shall make this more precise in what follows.

Recall that our ansatz for $\mathcal{R}_k$ was given in \eqref{Ra}, where $g(x)$ is the solution of the equation
\be\label{eq:g}
\frac{d^2g}{dx^2}-\frac{2}{x}\frac{dg}{dx}+g=2e^{ix}\delta_1\,,
\ee
with $\delta_1$ again given by
\be
\delta_1=-\frac{3b}{\sqrt{1+(3f \phi_*)^2}}\cos\left(\frac{\phi_k}{f}+\frac{\ln x}{f\phi_*}\right)\,,
\ee
and initial conditions given by $\lim\limits_{x\to\infty}g(x)=0$ and $\lim\limits_{x\to\infty}g'(x)=0$.
As we have just learned, the effect of the driving term can be ignored long after the resonance has occurred, {\it i.e.} for $x\ll\frac{1}{2f\phi_*}$.\footnote{One should note that this is not because the driving term goes to zero, but because its frequency becomes too high for the system to keep up with it.}
This implies that at late times, $g(x)$ must be a solution of the homogeneous equation which can be written as
\be\label{eq:ghom}
g(x)=c_k^{(+)}\left(i\sqrt{\frac{\pi}{2}}x^{\frac32}H_{3/2}^{(1)}(x)\right)+c_k^{(-)}\left(-i\sqrt{\frac{\pi}{2}}x^{\frac32}H_{3/2}^{(2)}(x)\right)\,,
\ee
where $c_k^{(\pm)}$ are momentum dependent coefficients.
The solution for equation~\eqref{eq:g} can also be written explicitly as
\begin{eqnarray}
g(x)&=&(x\cos x-\sin x)\int\limits_{x}^\infty\frac{2 e^{iy}(\cos y+y\sin y)}{y^2}\delta_1\\ \nonumber
&&\quad+(\cos x+x\sin x)\int\limits_{x}^\infty\frac{2 e^{iy}(\sin y-y\cos y)}{y^2}\delta_1\,.
\end{eqnarray}
For $x\ll\frac{1}{2f\phi_*}$ we can take the lower limit in the integrals to zero and this can be brought into the form
\be\label{eq:gsol}
g(x)=\frac12(\mathcal{I}^{2}+i\mathcal{I}^{1})\left(i\sqrt{\frac{\pi}{2}}x^{\frac32}H_{3/2}^{(1)}(x)\right)+\frac12(\mathcal{I}^{2}-i\mathcal{I}^{1})\left(-i\sqrt{\frac{\pi}{2}}x^{\frac32}H_{3/2}^{(2)}(x)\right)\,,
\ee
where the integrals $\mathcal{I}^{1}$ and $\mathcal{I}^{2}$ are given by
\begin{eqnarray}
\mathcal{I}^{1}&=&-\frac{6b}{\sqrt{1+(3f \phi_*)^2}}\int\limits_{0}^\infty\frac{ e^{iy}(\cos y+y\sin y)}{y^2}\cos\left(\frac{\phi_k}{f}+\frac{\ln x}{f\phi_*}\right)\,,\\
\mathcal{I}^{2}&=&-\frac{6b}{\sqrt{1+(3f \phi_*)^2}}\int\limits_{0}^\infty\frac{ e^{iy}(\sin y-y\cos y)}{y^2}\cos\left(\frac{\phi_k}{f}+\frac{\ln x}{f\phi_*         }\right)\,.
\end{eqnarray}
In the saddle point approximation these evaluate to
\be\label{eq:intsol}
\mathcal{I}^{1}=i\mathcal{I}^{2}=-\frac{6b}{\sqrt{(1+(3f \phi_*)^2)}}\sqrt{\frac{\pi}{8} f\phi_*}e^{-i\left(\frac{\phi_k}{f}-\frac{1+\ln 2f\phi_*}{f\phi_*}+\frac{\pi}{4}\right)}\,.
\ee
Combining equations~\eqref{Ra}, \eqref{eq:gsol}, and \eqref{eq:intsol}, we finally find that the curvature perturbation for $x\ll\frac{1}{2f\phi_*}$ takes the form
\be\label{eq:rsol}
\mathcal{R}_k=\mathcal{R}_{k,0}^{(o)}\left(i\sqrt{\frac{\pi}{2}}x^{\nu_0} H_{\nu_0}^{(1)}(x)-c_k^{(-)}i\sqrt{\frac{\pi}{2}}x^{\nu_0} H_{\nu_0}^{(2)}(x)\right)\,,
\ee
with $c_k^{(-)}$ given, up to an unimportant momentum-independent overall phase, by
\be
c_k^{(-)}=\frac{6b}{\sqrt{(1+(3f \phi_*)^2)}}\sqrt{\frac{\pi}{8} f\phi_*}e^{-i\left(\frac{\phi_k}{f}\right)}\,.
\ee
One might now interpret the coefficient $c_k^{(-)}$ of the negative frequency mode as a Bogoliubov coefficient that measures the amount of particles with comoving momentum $k$ being produced while this mode is in resonance with the background. It seems hard to make this precise as one really is comparing mode solutions of different backgrounds rather than mode solutions of different asymptotically Minkowski regions in the same background.

Equation~\eqref{eq:rsol} also shows that instead of starting in the Bunch-Davies state and then following the mode through the resonance, one may start the evolution after the resonance has occurred but use a state that is different from the Bunch-Davies state, which is similar to what is considered in~\cite{Easther:2001fi,Kaloper:2002uj,Martin:2003sg,Martin:2004yi,Easther:2004vq}. The departure from the Bunch-Davies state is of course quantified by $c_k^{(-)}$.




\subsection{Particle production and deviations from the Bunch-Davies state}\label{B}

Here we will deal with a conceptual question that generically arises in inflationary models with oscillations in the scalar potential. Driven by the background motion of the inflaton,
the oscillating contributions constitute a time-oscillating perturbation to the Hamiltonian of the system. Now, perturbations oscillating in time will generically induce transitions, in our case from the original vacuum state to some excited states. This implies that the vacuum state of the full system will deviate from the Bunch-Davies vacuum of the homogeneous background inflationary evolution.
We will now estimate the resulting quantity of particle production and relate the result to the derivation of the scalar power spectrum given in the preceding sections.

To lowest order the oscillating perturbation is given by
\be
\Delta H^{(2)}=\frac{1}{2}V''(\phi_0(t))\cdot \delta\phi^2\,,
\ee
implying that the lowest-order transitions will be from the vacuum $|0\rangle$ to two-particle states $|\vec p,-\vec p\rangle$. As the physical momentum $\vec p=\vec k/a$ corresponding to a given comoving momentum $\vec k$ is exponentially decaying in the inflationary regime, any two-particle state with given comoving momentum $|\vec k,-\vec k\rangle$ will be in resonance with the oscillating perturbation only for a short period of time which we will have to estimate in due course.

In transforming the Hamiltonian of the fluctuations into Fourier space
\be
H[\delta\phi_{\vec p}]=\frac12\delta\dot\phi_{\vec p}^2+\frac12\vec p\,^2\delta\phi_{\vec p}^2+\frac{1}{2}V''(\phi_0(t))\cdot \delta\phi_{\vec p}^2\,,
\ee
we find that the system takes the form of a perturbed harmonic oscillator with eigenfrequency $\omega_p=p\equiv|\vec p|$ for each momentum mode $\delta_{\vec p}$ separately,
\be
H[\delta\phi_p]=\frac12\delta\dot\phi_p^2+\frac12\omega_p^2\delta\phi_p^2+\underbrace{\frac{1}{2}V''(\phi_0(t))\cdot \delta\phi_p^2}_{\Delta H^{(2)}[\delta\phi_p]}\, .
\ee

Now we compare this to the perturbed harmonic oscillator in one-dimensional quantum mechanics,
\be
H=\frac12\dot x^2+\frac12\omega_0^2 x^2 +\frac12 \delta\omega(t)^2x^2\, .
\ee
In going to dimensionless variables $q,p$ we can write this as
\be
H=\frac12\omega_0\left(p^2+ q^2\right)+\underbrace{\frac{\delta\omega(t)^2}{2\omega_0}q^2}_{\Delta H^{(2)}}\,,
\ee
where for our case of a periodic perturbation periodic with frequency $\omega$ we have
\be
\delta\omega(t)^2=\delta\omega^2\cos(\omega t)\,.
\ee
We want to determine the time-dependent transition matrix element in time-dependent perturbation theory for a periodic perturbation. To do so, we first write the perturbation in standard form for time-dependent perturbation theory as
\be
\Delta H^{(2)}(t)=\frac{\delta\omega(t)^2}{2\omega_0}q^2=\frac{\delta\omega^2}{4\omega_0}q^2(e^{i\omega t}+e^{-i\omega t})\equiv F\, (e^{i\omega t}+e^{-i\omega t})\,,
\ee
in the notation of equations (40.1) through (40.9) of~\cite{Landau}.
The Hamiltonian and the transition matrix elements can be written in terms of creation and annihilation operators $a$ and $a^\dagger$ using $q=(a+a^\dag)/\sqrt 2$ and $p=-i(a-a^\dag)/\sqrt 2$. Then, canonical quantization of the unperturbed part yields a discrete spectrum $|n\rangle$ of eigenstates with energy spectrum $E_n=\omega_0(n+1/2)$.

If we compare this with our actual case above, we see that for each momentum mode $\delta\phi_{\vec p}$, $q$ and $p$ are replaced by appropriate
dimensionless fields $\delta\varphi_{\vec p}$ and $\Pi_{\delta\varphi_{\vec p}}$. In complete analogy to the simple quantum mechanical oscillator, there will be a tower of discrete states $|n\rangle_p$ with energies $E_{n,p}=\omega_p(n+1/2)=p (n+1/2)$.
In particular, $|2\rangle_p$ labels
the two-particle state $|p,-p\rangle$ which has energy difference $\Delta E_{2,p}=2\omega_p=2p$ with respect to the ground state. We thus have for the perturbation in our actual case
\be
\Delta H^{(2)}(t)=\frac{\delta\omega(t)^2}{2\omega_p}\delta\varphi_{\vec p}^2=\frac{\delta\omega^2}{4\omega_p}\delta\varphi_{\vec p}^2\,(e^{i\omega t}+e^{-i\omega t})\equiv F\, (e^{i\omega t}+e^{-i\omega t})\,.
\ee
For the transition matrix element one then finds
\be\label{matrixelement}
\langle p,-p|\Delta H^{(2)}|0\rangle=F_{20}(e^{i\omega t}+e^{-i\omega t})\quad{\rm with}\quad
F_{20}=\frac{\delta\omega^2}{4\omega_p}\langle 0|\frac{a^2}{\sqrt 2}\left(\frac{a+a^\dag}{\sqrt 2}\right)^2|0\rangle=\frac{\delta\omega^2}{4\sqrt 2\omega_p}\, .
\ee
Here we have used that
\be
|p,-p\rangle=\frac{(a^\dag)^2}{\sqrt 2}|0\rangle\,,
\ee
and
\be
\langle0|a^2(a+a^\dag)^2|0\rangle=\langle0|a^2(a^\dag)^2|0\rangle=2\,.
\ee

If the energy of the two-particle state $E_{2p}=2k/a$ were not too close to the perturbation frequency $\omega$, we could use time-dependent perturbation theory with the above matrix element and obtain the first order transition probability $P_{0\to2k}$,
\begin{eqnarray}\label{timepert1}
P_{0\to2k}&=&\left| -i \int^t dt' \langle k,-k|\Delta H^{(2)}(t')|0\rangle e^{i\omega_{20} t'}  \right|^2\nonumber \\
&=&2 |F_{20}|^2\frac{\left(\frac{2k}{a}\right)^2+\omega^2+\left[\left(\frac{2k}{a}\right)^2-\omega^2\right]\cos(2\omega t)}{\left[\left(\frac{2k}{a}\right)^2-\omega^2\right]^2}\nonumber \\
&=&\frac{\delta\omega^4}{16(k/a)^2}\frac{\left(\frac{2k}{a}\right)^2+\omega^2+\left[\left(\frac{2k}{a}\right)^2-\omega^2\right]\cos(2\omega t)}{\left[\left(\frac{2k}{a}\right)^2-\omega^2\right]^2}\quad,
\end{eqnarray}
where $\omega_{20}=E_{2,p=k/a}-E_{0,p=k/a}=2k/a$. This gives the resonance line feature characteristic of transition processes.

However, as for any given $k$ the physical momentum and frequency $k/a$ will decrease extremely rapidly with $1/a$, we can approximate the amount of transition happening in the short time interval $\Delta t_{res}$ during which the two-particle state of given $k$ is in near-resonance $\omega\approx 2k/a$. Close to resonance, time-dependent perturbation theory breaks down (visible in the singularity of the above result for $\omega=2k/a$); however, for periodic perturbations one can solve the Schr\"odinger equation of the coupled two-state system exactly \cite{Landau}. One finds that on resonance the transition probability is
\be
P_{0\to2k}=\frac{1}{2}\left[1-\cos\left(2\Omega t\right)\right]=\frac{1}{2}\left[1-\cos\left(\frac{\delta\omega^2}{2\sqrt2 k/a}t\right)\right]\,,\,{\rm where}\;\;\Omega\equiv F_{20}\,.
\ee
That is, near resonance the system effectively oscillates with frequency $2\Omega=2\frac{\delta\omega^2}{4\sqrt2\omega_p}=\frac{\delta\omega^2}{2\sqrt2 k/a}$ between the vacuum and the two-particle state.

We now have to estimate the time $\Delta t_{res}$ during which a two-particle state of comoving momentum $k$ stays in near-resonance. We will follow the analysis in~\cite{Chen:2008wn} and look at the interference terms induced between the $\cos(\omega t)$ perturbation and the $\exp(i\omega_{02} t)$ periodicity of the interaction matrix element in \eqref{timepert1}. We note that, on the one hand, the two-particle state with frequency $2k/a=\omega-\Delta\omega$ stays in resonance with the perturbation with frequency $\omega$ only for a time roughly estimated to be (for the relative phase shifting from $-\pi$ to $\pi$)
\be
\Delta t_1\sim\frac{2\pi}{\Delta\omega}\, .
\ee
On the other hand, in the inflating universe it takes very roughly a time
\be
\Delta t_2\sim\frac{2\Delta \omega}{\omega H}
\ee
to change the frequency of the two-particle state from, say, $\omega+\Delta\omega$ to $\omega-\Delta\omega$. Equating the two provides us with the effective duration of near-resonance,
\be
\Delta t_{res}\equiv\Delta t_1=\Delta t_2\sim 2\sqrt{\pi\frac{H}{\omega}}H^{-1}\, .
\ee
Plugging this into the above transition result and remembering that near resonance $k/a\approx\omega/2$, we get
\be
P_{0\to2k}\simeq\frac{1}{2}\left[1-\cos\left(\sqrt 2\frac{\delta\omega^2}{\omega}\sqrt{\pi\frac{H}{\omega}}H^{-1}\right)\right]\, .
\ee
Now, in our case above we see that $p=k/a$ in terms of comoving momenta $k$, and further
\be\label{V2prime}
\delta\omega(t)^2=V''(\phi_0(t))=\frac{\Lambda^4}{f^2}\cos\left(\frac{\phi_0(t)}{f}\right)=\delta\omega^2\cos(\omega t)\quad,\quad \omega=\frac{H}{f\phi_0}\quad.
\ee

Noting that in our scenario of interest we have $H<\omega$ and that $\delta\omega\ll H$, we can expand the argument of the cosine around zero. If we then plug in the microscopic definitions of the quantities $\delta\omega^2=\Lambda^4/f^2$ and $\omega=H/(f\phi)$, we get
\be
P_{0\to2k}\simeq\frac{\pi}{2}\frac{\Lambda^8}{f^4\omega^2H^2}\cdot\frac{H}{\omega}=\frac{\pi}{2}\frac{\Lambda^8 f\phi_*^3}{f^2H^4}=\frac{\pi}{2}\frac{9\Lambda^8 f\phi_*^3}{f^2\mu^6\phi_*^2}=\frac{36\pi}{8} b^2f\phi_*\,,
\ee
where $\phi_*\simeq 11 \Mpl$ denotes the vev of the inflaton field around 60 e-foldings before the end of inflation.

Next, because $P_{0\to2k}$ characterizes the transition probability to the two-particle states, it may be related to the negative frequency Bogoliubov coefficient $\cbm$ that relates the out-vacuum to the in-vacuum. Specifically,
the out-vacuum is specified by the modes
\be
u_k(out)=\cbm u_{-k}+\cbp u_k\,,
\ee
whereas the original Bunch-Davies in-vacuum had modes
\be
u_k(in)=u_k\quad,\quad \cbp(in)=1\, .
\ee
We therefore find that
\be
|\cbm|\simeq \sqrt{P_{0\to2k}}=6b \sqrt{\frac{\pi}{8} f\phi_*}\, .
\ee

In comparing these results with the general treatment of the Mukhanov-Sasaki equation above, we see by looking at \eqref{eq:rsol} and \eqref{Rq2} that we can identify
\be
u_k &=& i\sqrt{\frac{\pi}{2}}x^{\nu_0} H_{\nu_0}^{(1)}(x)\\
u_{-k}&=&-i\sqrt{\frac{\pi}{2}}x^{\nu_0} H_{\nu_0}^{(2)}(x)
\ee
and thus from \eqref{Rq2} we conclude that
\be
\delta n_s=\frac{2 {\rm Re}\,g(x)}{\cos\left(\frac{\phi_k}{f}\right)}\underset{x\to 0}{=}\frac{2 {\rm Re}(\cbp \cbm)}{\cos\left(\frac{\phi_k}{f}\right)}\simeq2 |\cbm|\simeq 12 b \sqrt{\frac{\pi}{8} f\phi_*}
\ee
which agrees with the general result \eqref{dns2} in the appropriate limit $\omega > H$ and $\delta\omega\ll H$, corresponding to $f\phi_* <1$, where $\coth\left(\pi/2 f\phi_*\right)\to 1$.

Note that in calculating the transition probability we lose information about the phase of the transition matrix element as given in \eqref{matrixelement}. Therefore, if we estimate the population coefficient $\cbm$ from $\sqrt{P_{0\to2k}}$, we get only an estimate for $|\cbm|$ without the phase information. A more complete derivation using the full information in the transition matrix element should also yield the information about the phase as derived in the previous subsection.

Thus, we see that in the regime of rapid oscillations, $f\phi_*<1$, the induced $\delta  n_s$ is due to a time-localized deviation from the Bunch-Davies state, which may be interpreted as being due to resonant bursts of particle production happening well before a given mode leaves the horizon during inflation.

\subsection{Bispectrum of scalar perturbations} \label{s:bispectrum}

We start by reviewing how resonance can drive the production of large non-Gaussianity during inflation, as proposed in \cite{Chen:2008wn}. We then present an estimate for the size of the non-Gaussianity for the model \eqref{V}.

The three-point function can be calculated as \cite{Maldacena:2002vr}
\be\label{3}
\langle \mathcal{R}(\tau,{\bf k}_1)\mathcal{R}(\tau,{\bf k}_2)\mathcal{R}(\tau,{\bf k}_3)\rangle=-i\int_{\tau_0}^{\tau} \langle \left[ \mathcal{R}(\tau,{\bf k}_1)\mathcal{R}(\tau,{\bf k}_2)\mathcal{R}(\tau,{\bf k}_3),H_I(\tau')\right]\rangle \,a \,d\tau'\,,
\ee
where $H_I$ is the interacting part of the Hamiltonian. $H_I$ was calculated for a generic potential (see e.g. \cite{Maldacena:2002vr,Chen:2008wn}) at cubic order in the perturbations; it takes the form
\be \label{Hbisp}
H_I&=&-\int d^3x\Big[ a\epsilon^2\mathcal{R}\mathcal{R}'^2+a\epsilon^2\mathcal{R}(\partial\mathcal{R})^2-2\epsilon\mathcal{R}'(\partial\mathcal{R})(\partial\chi)\nonumber\\
&&\qquad +\frac a 2\epsilon \eta'\mathcal{R}^2\mathcal{R}' +\frac{\epsilon}{2a}(\partial\mathcal{R})(\partial\chi)(\partial^2\chi)+\frac{\epsilon}{4a}(\partial^2\mathcal{R})(\partial\chi)^2\Big]\,,
\ee
where $\partial$ denote space derivatives,
\be
\chi \equiv a^2 \epsilon \partial^{-2}  \mathcal{ \dot R} \,,
\ee
and we used the Hubble slow-roll parameter $\eta\equiv\dot \epsilon/(\epsilon H)=2(\epsilon+\delta)$ because formulas in this subsection are simpler in terms of $\eta$ than in terms of $\delta$.

We would like to stress that \eqref{Hbisp} is exact for arbitrary values of the slow roll parameters $\epsilon$ and $\eta$. Substituting $H_I$ into \eqref{3} produces six terms, plus an additional term coming from a field redefinition. For the modulated linear potential \eqref{V}, $\epsilon$ is small, as in standard slow roll inflation. On the other hand, contrary to the standard slow-roll approximation, $\dot{\eta}$ can be much larger than $\epsilon^2$.
This suggests that the leading term comes from the $\epsilon\dot{\eta}$ term in the Hamiltonian.\footnote{In (3.9) of \cite{Maldacena:2002vr} this term was written as
\be
\frac{\dot{\phi}^2}{\dot{\rho}^2}e^{3\rho}\dot{\mathcal{R}}\mathcal{R}^2\frac{d}{dt}\left(\frac{\ddot\phi}{2\dot\phi\dot\rho}+\frac{\dot\phi^2}{4\dot\rho^2}\right)\,,
\ee
which can be reduced to the term in \eqref{Hbisp} using $H'=-\dot\phi^2/2$.}
Hence we have \cite{Chen:2006xjb,Chen:2008wn}
\be\label{44}
\langle \mathcal{R}(t,{\bf k}_1)\mathcal{R}(t,{\bf k}_2)\mathcal{R}(t,{\bf k}_3)\rangle\simeq i  \left(\prod_i u_i(\tau_{end})\right)\times \nonumber \\
\int_{-\infty}^{\tau_{end}}d\tau \epsilon \eta' a^2\left(u_1^\ast(\tau)u_2^\ast(\tau)\frac{d}{d\tau}u_3^\ast(\tau)+\mathrm{sym}\right)\delta^3({\bf K})(2\pi)^3+c.c.\,.
\ee
As in \cite{Chen:2008wn}, we parameterize the non-Gaussianity as
\be\label{5}
\langle \mathcal{R}(\tau,{\bf k}_1)\mathcal{R}(\tau,{\bf k}_2)\mathcal{R}(\tau,{\bf k}_3)\rangle\equiv\frac{G(k_1,k_2,k_3)}{(k_1k_2k_3)^3}\delta^3({\bf K})\,\Delta_{\mathcal{R}}^4(2\pi)^7\,,
\ee
where ${\bf K}={\bf k}_1+{\bf k}_2+{\bf k}_3$. We take as an ansatz for the shape of the non-Gaussianity for our modulated linear potential

\be\label{an}
\frac{G(k_1,k_2,k_3)}{k_1k_2k_3}=f_{res}\sin\left(\frac{2}{\phi \A}\ln K+\mathrm{phase}\right)
\ee
Following \cite{Chen:2008wn} and comparing \eqref{44}, \eqref{5} and \eqref{an}, we obtain the estimate
\be
f_{res}\simeq\frac{3\,\dot\eta_1}{8H\sqrt{\phi\A}}\,,
\ee
where we have again used the notation $\eta=\eta_0+b\eta_1+\dots$. Using the background solution obtained in \S\ref{s:back}, it is straightforward to find
\be\label{etadot}
\dot\eta_1\simeq 2 \dot \delta_1\simeq-\sqrt{\frac{\mu^3}{3\phi_*}}\frac{6b}{f[1+(3f \phi_*)^2]}\left[\cos\left(\frac{\phi_0}{f}\right)+(3f\phi_*)\sin\left(\frac{\phi_0}{f}\right)\right]\,.
\ee
It is not hard to convince oneself that in the region of parameter space where $f_{res}>1$ and $b\ll 1$, the second term in \eqref{etadot} is always negligible, {\it{i.e.}} $3f\phi\ll1$. Hence our estimate for the non-Gaussianity is
\be \label{nGres}
f_{res}\simeq \frac{9b}{4(f\phi)^{3/2}}=\frac{9}{4}b\left(\frac{\omega}{H}\right)^{3/2}.
\ee
where we remind the reader that $\omega=\dot{\phi}/f$.
As we will often refer to this equation, let us pause and comment on it. The resonant non-Gaussianity vanishes when the modulation is switched off, {\it{i.e.}} for $b=0$.
It is inversely proportional to some power of $f$ (depending on which quantity is held fixed). Hence the smaller the axion decay constant $f$, the larger the non-Gaussianity. On the other hand, as we will see in \S\ref{s:micro}, there are theoretical lower (as well as upper) bounds on $f$, so that the non-Gaussian signal cannot be made arbitrarily large.

No complete analysis of the observational constraints on resonant non-Gaussianity has been performed to date (however, see \cite{Meerburg:2009ys}), and such an analysis is beyond the scope of the present work.  Based on a rough comparison with known shapes of non-Gaussianity, we estimate that $f_{res}\gtrsim 200$ might be at the borderline of being excluded by the current data, while $f_{res}\lesssim 1$ would be difficult to detect in the next generation of experiments.  A comprehensive analysis of the detectability of resonant non-Gaussianity is a very interesting topic for future research.

\section{Observational Constraints} \label{s:numerics}

In the last section, we derived the theoretical predictions of axion monodromy inflation for the primordial power spectrum. We will now use these predictions to compare the model with the five-year WMAP data \cite{Komatsu:2008hk}. While the data in principle allows for a variety of statistics to be extracted, we will limit ourselves to the most fundamental one, the angular power spectrum. The reason for this is that the data is
not now adequate for the polarization data or the three-point correlations to place meaningful additional constraints on the model. This will change as soon as the Planck data becomes available, and will be an interesting problem especially given the unusual shape of the non-Gaussianities the model predicts.

For the benefit of the less cosmologically-inclined reader, we now briefly summarize the basic observables relevant to our analysis.
In the ideal scenario, in which a full-sky map is available, the temperature of the cosmic microwave background as a function of the position in the sky can be expanded in spherical harmonics as
\begin{equation}
T(\hat{n})=\sum_{\ell  m}a_{\ell m}Y_{\ell m}(\hat{n})
\end{equation}
The theoretical counterparts of these measured expansion coefficients, which we will denote $a_{\ell m}^{th}$, should be thought of as random variables satisfying a (possibly only nearly) Gaussian distribution. Each realization of these coefficients corresponds to a possible history of the universe. In the Gaussian case, all the information about the theory is contained in the two-point correlations of these, as the odd n-point functions vanish, and the even n-point functions are sums of products of the two-point functions. Assuming an isotropic background, the two-point correlations must take the form
\be
\langle a_{\ell m}^{th}{a_{\ell' m'}^{th\,*}}\rangle=C_\ell\delta_{\ell\ell'}\delta_{mm'}\,,
\ee
where the brackets denote an average over all possible histories or equivalently (by the ergodic theorem) all possible positions.
The $a_{\ell m}^{th}$'s themselves, being random variables encoding initial conditions, cannot be predicted from a given cosmological model, and only  the multipole coefficients, $C_\ell$, encoding their correlations are of interest.
These multipole coefficients $C_\ell$ can be estimated from the measured expansion coefficients $a_{\ell m}$ via
\be
C_{\ell}^\text{sky}=\frac{1}{2\ell+1}\sum_m |a_{\ell m}|^2\,.
\ee
For noiseless, full-sky CMB data, these provide an unbiased estimate of the true power spectrum in the sense that the average of the analogously defined quantity for the $a_{\ell m}^{th}$'s satisfies
\be
\langle C_\ell^\text{sky,th}\rangle\equiv\frac{1}{2\ell+1}\left\langle\sum_m |a_{\ell m}^{th}|^2\right\rangle=C_\ell\,.
\ee
Since there are only $2\ell+1$ modes per $\ell$,  even for the ideal noiseless full-sky map the estimate of the multipole coefficient has the cosmic variance uncertainty
\be
\left\langle\left(\frac{C_\ell^\text{sky,th}-C_\ell}{C_\ell}\right)^2\right\rangle=\frac{2}{2\ell+1}\,.
\ee

In a more realistic setting with noise and sky cuts, this estimator is no longer unbiased and more sophisticated estimators have to be used. The current state of the art is to use a pixel-based maximum likelihood estimator for low $\ell$ (specifically, for $\ell\leq32$), and a pseudo-$C_\ell$ estimator for higher $\ell$. For details we refer the reader to \cite{Hinshaw:2006ia} and references therein.

After this quick review of the basic relevant quantities, let us describe our analysis.
We work on a grid of model parameters.
For each point on the grid, we compute the theoretical angular power spectrum with the publicly-available CAMB code \cite{Lewis:1999bs,camb}\footnote{Of course, we modify the CAMB code to calculate all the multipole coefficients rather than calculating some and interpolating.}, using the primordial power spectrum derived in the previous section in the form
\be
\Delta_\mathcal{R}^2(k)=\Delta_\mathcal{R}^2(k_*)\left(\frac{k}{k_* }\right)^{n_s - 1+ \frac{\delta n_s}{\ln(k/k_*)}\cos\left(\frac{\phi_k}{f}+\Delta\varphi\right)} \,.
\ee
The likelihood for a given theoretical power spectrum is calculated with a modified version of the WMAP five-year likelihood code that is now available on the LAMBDA webpage~\cite{lambda}.
The power spectrum in our model contains additional parameters beyond those of the WMAP five-year $\Lambda$CDM fit (namely, $\{\Omega_Bh^2,\Omega_ch^2,\Omega_\Lambda,\tau,n_s,\Delta_\mathcal{R}^2\}$ and the marginalization parameter $\{A_{SZ}\}$). The additional parameters are $\delta n_s$, $f$ and a phase $\Delta \varphi$. This phase parameterizes both our uncertainty in the number of e-folds needed, which originates in our poor understanding of reheating, and a microscopically determined phase offset in the sinusoidal modulation of the scalar potential arising in the string theory construction.

\begin{figure}[H]
\begin{center}
\includegraphics[width=6.1in]{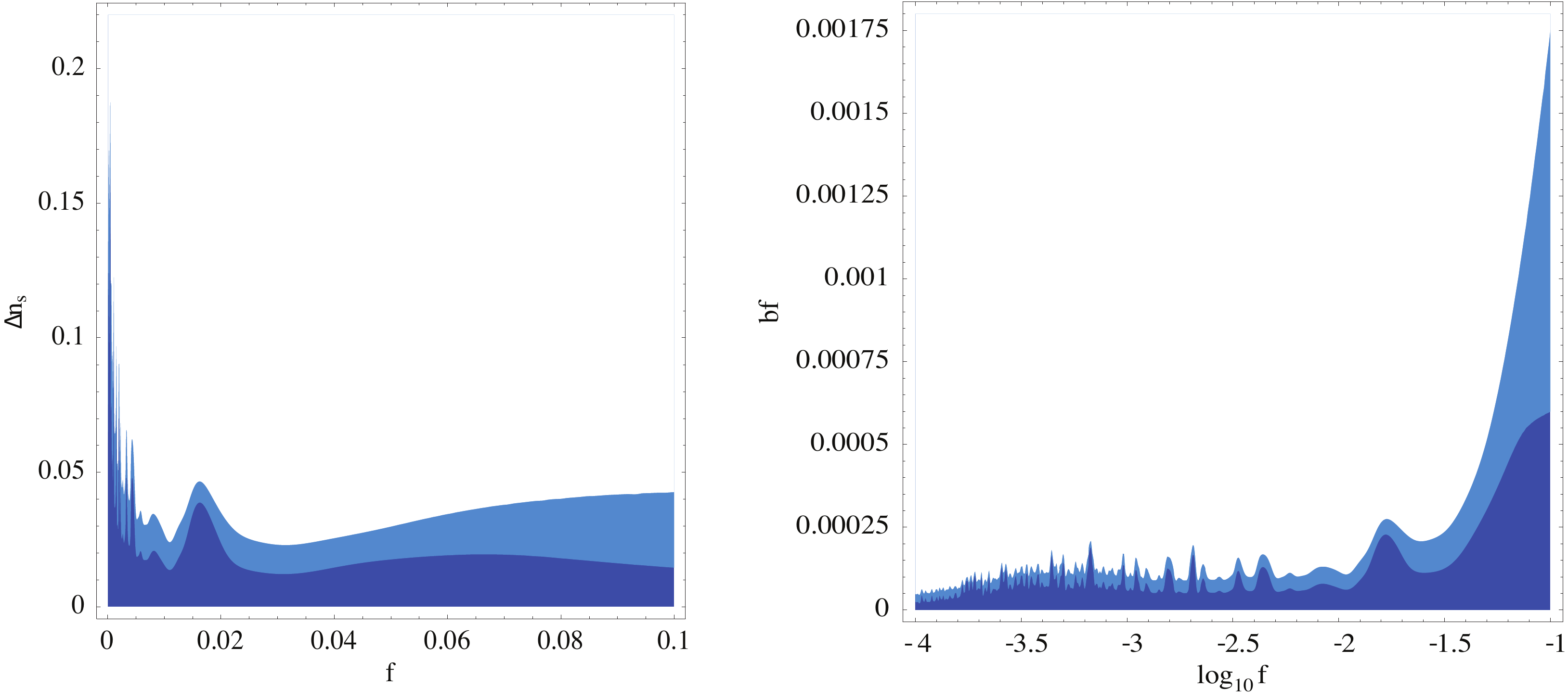}
\caption{This plot shows the 68\% and 95\% likelihood contours in the $\delta n_s$-$f$, and $bf$-$\log_{10}f$ plane, respectively, from the five-year WMAP data on the temperature angular power spectrum.}
\label{gridresult}
\end{center}
\end{figure}

We fix the value of the scalar spectral index $n_s=0.975$. As in any model of large-field inflation, the spectral index is a prediction of the model that depends only on the physics of reheating and, correspondingly, on the total amount of inflation since the observable  modes exited the horizon.  The value we choose corresponds to the situation in which the pivot scale exits the horizon 60 e-folds before the end of inflation.
The results turn out to be fairly independent of the precise value chosen for the scalar spectral index and we could have chosen the value corresponding to any number of e-folds between 50 and 60.
We fix $\{\Omega_ch^2,\Omega_\Lambda,\tau,A_{SZ}\}$ to the WMAP five-year best-fit values for the $\Lambda$CDM fit.
We allow ${f, \delta n_s,\Omega_B h^2,\Delta\varphi}$ to vary on the grid, and we also marginalize over the scalar amplitude $\{\Delta_\mathcal{R}^2\}$ in the likelihood code. To obtain Figure~\ref{gridresult}, we thus marginalize over $\{\Omega_Bh^2,\Delta_\mathcal{R}^2\}$ and over the unknown phase $\Delta\varphi$, while we fix $\{\Omega_ch^2,\Omega_\Lambda,\tau,A_{SZ}\}$, as we expect at most mild degeneracies between these parameters and the primordial ones.

The grid consists of 16 equidistantly spaced points in $\Omega_Bh^2$ between $\Omega_Bh^2\approx0.0212$ and $\Omega_Bh^2\approx0.0266$, 128 equidistantly spaced points in $\delta n_s$ between $\delta n_s=0$ and $\delta n_s=0.44$, 512 logarithmically spaced points in the axion decay constant $f$ between $f=9\times 10^{-5}$ and $f=10^{-1}$, as well as 32 points for the phase $\Delta \varphi$ between $\Delta\varphi=-\pi$ and $\Delta\varphi=\pi$.
This leads to a grid with a total of 33,554,432 points. The analysis was run on 64 of the compute nodes of the Ranger supercomputer at the Texas Advanced Computing Center. The compute nodes are SunBlade x6420 blades, and each of the nodes provides four AMD Opteron Quad-Core 64-bit processors with a core frequency of $2.3\, \text{GHz}$.

\begin{figure}[h]
\begin{center}
\includegraphics[width=6.7in]{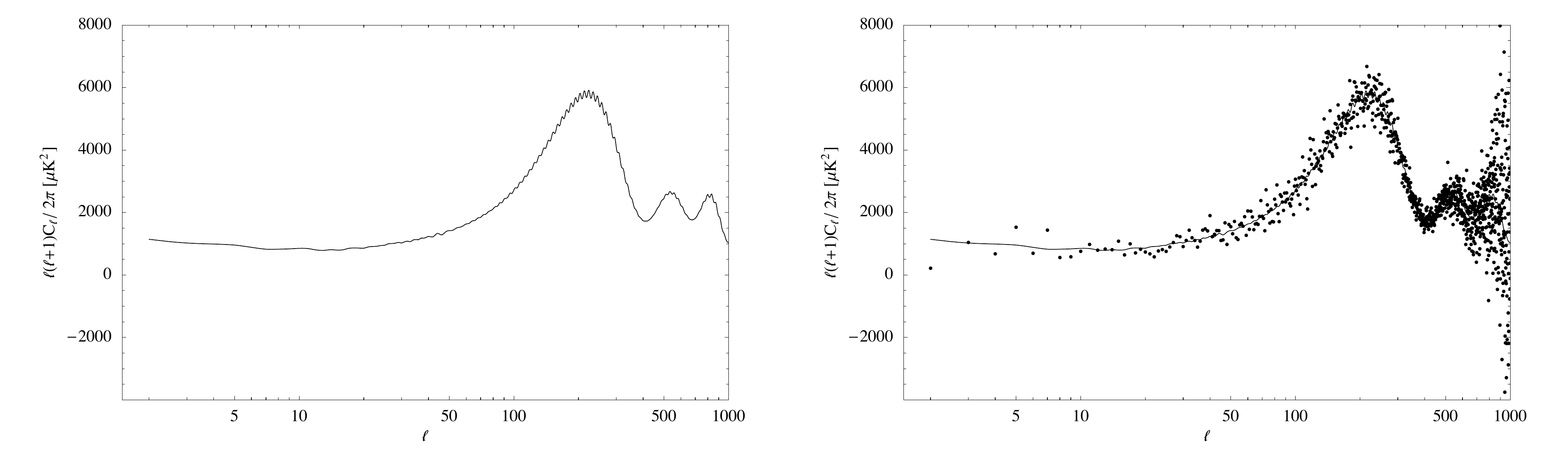}
\caption{The left plot shows the angular power spectrum for the best fit point $f=6.67\times 10^{-4}$, and $\delta n_s=0. 17$. The right plot shows the angular power spectrum for the best fit point together with the unbinned WMAP five-year data.}
\label{clbest}
\end{center}
\end{figure}

The resulting 68\% and 95\% contours in the $\delta n_s-f$ plane are shown in the left plot of Figure~\ref{gridresult}.
To convert the resulting observational constraints on $\delta n_s$ as a function of $f$ into constraints on the microscopic parameter $bf$ as a function of $f$, we make use of equation~\eqref{dns2}.
The resulting 68\% and 95\% contours in the $bf$-$f$ plane are shown in the right plot of Figure~\ref{gridresult}.
Roughly, the results can be summarized as $bf\lesssim10^{-4}$ for $f\lesssim0.01$ at 95\% confidence level.
Our best fit point is at a rather small value of the axion decay constant, $f=6.67\times 10^{-4}$, and a rather large amplitude for the oscillations, $\delta n_s=0.17$. The fit improves by $\Delta\chi^2\simeq11$ over the fit in the absence of oscillations. The corresponding angular power spectrum is shown in Figure~\ref{clbest}.

The improvement can be traced to a better fit to the data around the first peak. We would like to stress, however, that we do not take this as an indication of oscillations in the observed angular power spectrum.  Similar spikes in the likelihood function occur quite generally when fitting an oscillatory model to toy data generated with the conventional power spectrum without any oscillations, because the oscillations fit some features in the noise. The polarization data could provide a cross check, but we find that it is presently not good enough to do so in a meaningful way.

Let us say a few words motivating the necessity of marginalizing over $\Omega_Bh^2$ and $\Delta\varphi$.  There is a known degeneracy in the angular power spectrum between $\Omega_Bh^2$ and $n_s$, as changing $\Omega_Bh^2$ changes the ratio of the power in the first and second acoustic peaks, which to some extent can be undone by changing the spectral tilt $n_s$. In our case we do not vary $n_s$, but we add a sinusoidal contribution to the standard power spectrum. It is intuitively clear that by doing so we can change the ratio of power in the first and second acoustic peak by choosing the right oscillation frequency (controlled by $f$) and phase $\Delta\varphi$, leading to a degeneracy between $\Omega_Bh^2$ and $\delta n_s$ at least for a certain range of $f$.

\begin{figure}[h!]
\begin{center}
\includegraphics[width=5.5in]{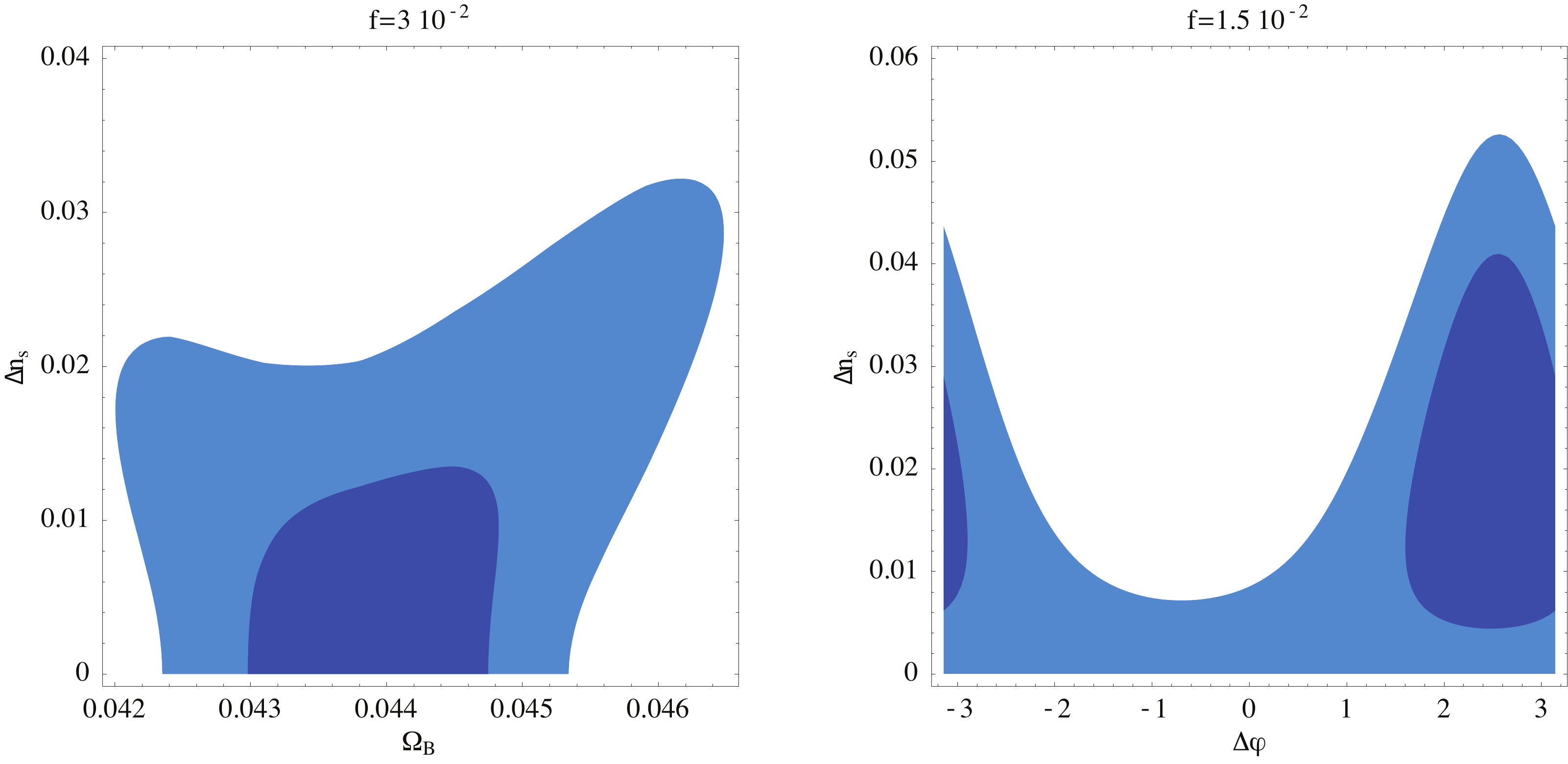}
        \caption{These plots show the 68\% and 95\% likelihood contours for the five-year WMAP data on the temperature angular power spectrum in the $\delta n_s$ - $\Omega_Bh^2$ plane and $\delta n_s$ - $\Delta\phi$ plane for an axion decay constant of $f=3\times 10^{-2}$ and $f=1.5\times 10^{- 2}$, respectively.  }
     \label{degenfig}
\end{center}

\end{figure}

The most straightforward way to demonstrate this degeneracy between $\Omega_Bh^2$ and $\delta n_s$ arising for certain `resonant' values of $f$ is to   present a likelihood plot in the $\Omega_B$-$\delta n_s$ plane for a value of $f$ for which the degeneracy is clearly visible. An example is shown in the plot on the left side of Figure~\ref{degenfig}. It shows that marginalizing over $\Omega_Bh^2$ is necessary to obtain correct exclusion contours on $\delta n_s$ and $f$.
That marginalization over the phase is necessary can easily be seen from a likelihood plot in the $\delta n_s$-$\Delta\varphi$ plane. This is shown in the plot on the right side of Figure~\ref{degenfig}.

We have also performed a Markov chain Monte Carlo analysis for the model using the publicly available CosmoMC code~\cite{Lewis:2002ah}, \cite{cosmomc}. While the      Monte Carlo has the advantage that it is less computationally intensive than a grid when varying all cosmological parameters, the likelihood function for oscillatory  models turns out to be rather spiky, making the Monte Carlo hard to set up, because the chains tend to get trapped in the spikes.

To some extent this can be overcome by taking out the problematic regions or increasing the temperature of the Monte Carlo. When run on parts of the parameter space where the Monte Carlo runs reliably, we found agreement with the grid-based results shown above.

\begin{figure}[H]
\begin{center}
\includegraphics[width=6.5in]{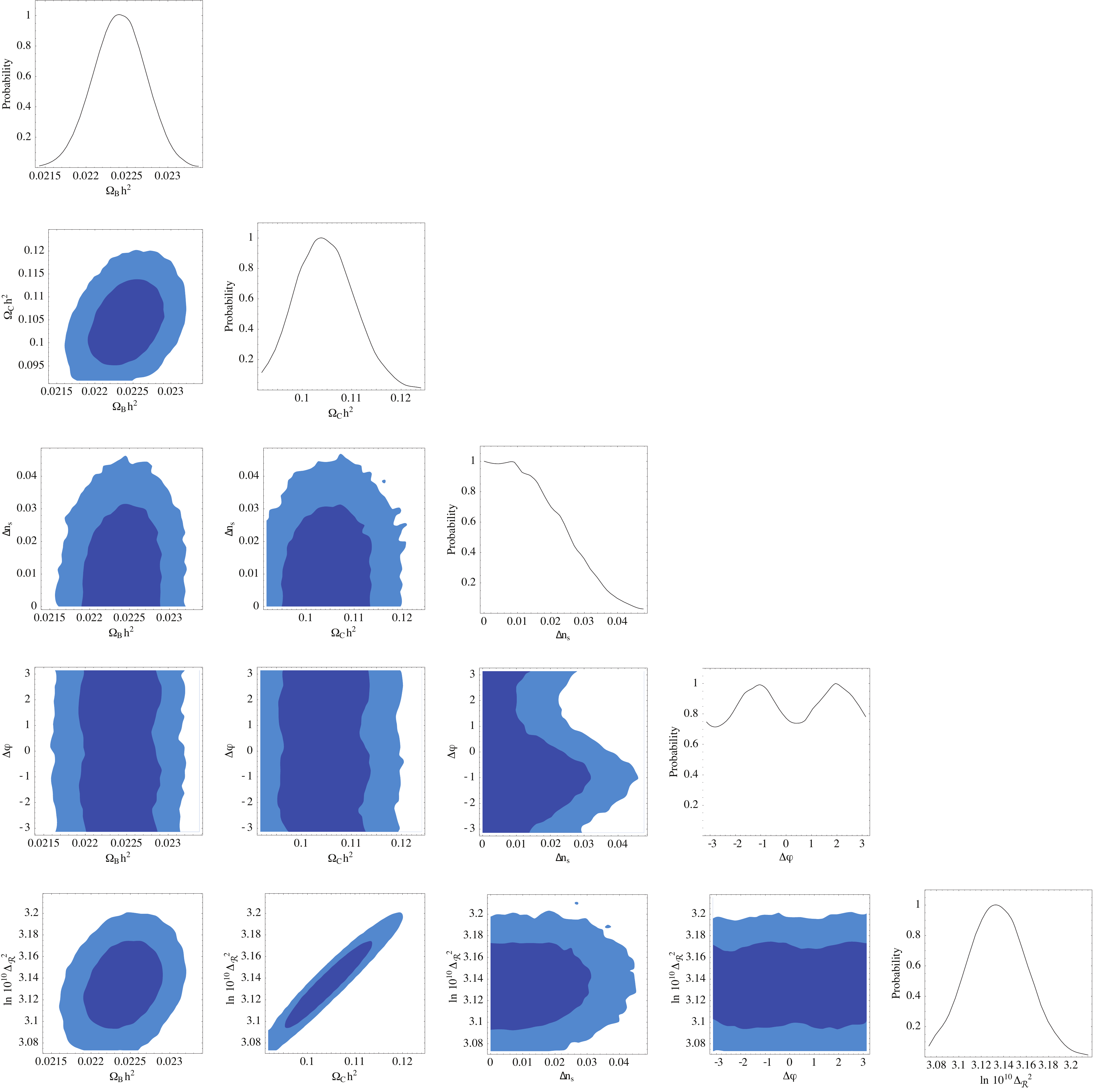}
\caption{This figure shows a triangle plot for some of the parameters that were sampled in a Markov chain Monte Carlo for an axion decay constant of $f=10^{-2}$. The contours again represent 68\% and 95\% confidence levels.}
\label{markovfp01}
\end{center}

\end{figure}

The most problematic direction to sample is that of the axion decay constant, $f$. We show the result of one of our chains for $f=0.01$ in Figure~\ref{markovfp01}. The plot shows marginalized one-dimensional distributions and two-dimensional 68\% and 95\% confidence level limits for the most important $\Lambda$CDM parameters as well as $\delta n_s$ and $\Delta\phi$.
In the Monte Carlo, we sampled the parameters $\delta n_s$, $\Delta\phi$, all parameters of the $\Lambda$CDM except the scalar spectral index, as well as the Sunyaev-Zel'dovich amplitude.

\section{Microphysics of Axion Monodromy Inflation} \label{s:micro}

In \S1.1 we briefly reviewed the properties of axion monodromy inflation, focusing on the description in effective field theory. For a general characterization of the signatures of the scenario, the phenomenological model of \S1.1 was sufficient. However, the phenomenological parameters $f,\mu,\Lambda$ are in principle
derivable from the data of a string compactification, and as such they obey nontrivial microscopic constraints: the ranges and correlations of these parameters are restricted by microphysics.

We should therefore determine the values of the phenomenological parameters allowed  in consistent, computable string compactifications.
We will begin by reviewing the string theory origin of axion monodromy inflation, both to set notation and to highlight the properties most relevant in constraining the parameters $f,\mu,\Lambda$.
For concreteness we will restrict our attention to a specific realization of the scenario, in O3-O7 orientifolds of type IIB string theory, with the K\"ahler moduli stabilized by nonperturbative effects.  Our considerations could be generalized to other compactifications, but the numerical results would differ.


\subsection{Axions in string theory}\label{ss:MSW}

Let us first review the origin of the relevant axions. Our conventions and notation are summarized in appendix \ref{s:not}. Consider type IIB string theory compactified on an orientifold of a Calabi-Yau threefold $X$.
Let the forms $\omega^I$ be a basis of the cohomology $H^2(X,\mathbb{Z})$, normalized such that $\int_{\Sigma_I}\omega^J=\delta_I^{~J} (2\pi)^2\alpha'$, where $\Sigma_I$ are a basis of the dual homology $H_2(X,\mathbb{Z})$. The RR two-form $C_2$ gives rise to a four-dimensional axion via the ansatz\footnote{The factor of $2\pi$ is introduced so that the four-dimensional axions $c_I$ have periodicity $2\pi$, as can be seen via S-duality from the world-sheet coupling $i\int B_2/(2\pi \al)$. Notice that in our conventions $C_2$ and $\omega^I$ have the dimensions of length-squared, while $c_I$ are dimensionless.}
\be
C_2=\frac{1}{2\pi} c_I(x) \omega^I\,,
\ee
where $x$ is a four-dimensional spacetime coordinate.
The ten-dimensional Einstein-frame action \cite{Polchinski:1998rr} that follows is
\be
\int d^{10}x\frac{g_s \sqrt{-g_{E}} }{2(2\pi)^7\al^4} |d C_2|^2=\int d^{10}x\frac{g_s \sqrt{-g_{E}} }{12(2\pi)^9\al^4} g_{E}^{\mu\nu}\partial_\mu c_I \partial_\nu c_J \omega^I_{ij} \omega^J_{i'j'} g_{E}^{ii'}g_{E}^{jj'}\,.
\ee
Notice that the axions only have derivative couplings, and hence enjoy a continuous shift symmetry at the level of the classical action. In \S\ref{ss:shift} we will recall the origin of this symmetry and explain how it persists to all orders in perturbation theory and is broken by nonperturbative effects.

Upon dimensional reduction, one finds a relation between the four-dimensional reduced Planck mass $\Mpl$ and $\al$,
\be
\al \Mpl^2=\frac{\V_E}{\pi}\,,
\ee
where $\V_E$ is the Einstein-frame (dimensionless) volume of the Calabi-Yau $X$ measured in units of $l_s\equiv 2\pi\sqrt{\al}$. The decay constant of the canonically normalized axion is then
\be
\frac{f^2}{\Mpl^2}&=& \frac{g_s}{48\pi^2 \V_E} \left[\frac{ \int \omega \wedge \ast \omega}{(2\pi)^{6}\al^3}\right] \,.
\ee
The present definition of the axion decay constant differs by a factor of $2 \pi$ from that in \cite{MSW}, {\it i.e.} $f^{here}=2 \pi f^{there}$. As a consequence our canonically normalized axion has periodicity $2 \pi f$, consistent with \eqref{V}.


\subsection{Dimensional reduction and moduli stabilization}

\subsubsection{Four-dimensional data of O3-O7 orientifolds} \label{sss:03}

Now we consider how to stabilize the compactification in a setup that will allow inflation. We focus on the KKLT scenario for moduli stabilization \cite{Kachru:2003aw}.
We assume that the complex structure moduli, the dilaton, and any open string moduli have been stabilized at a higher scale, and we concentrate on the remaining closed string moduli (specifically, the remaining moduli are those descending from hypermultiplets).  The ${\cal N}= 1$ supersymmetric four-dimensional theory resulting from dimensional reduction of type IIB orientifolds was worked out in detail in \cite{Grimm:2004uq}. We are
interested in orientifold actions under which the holomorphic three-form $\Omega$ of the Calabi-Yau manifold is odd, so that the fixed-point loci are O3-planes and O7-planes. The cohomology decomposes into eigenspaces of the orientifold action,
\be
H^{(r,s)}= H^{r,s}_+\oplus H^{r,s}_- \,.
\ee
We therefore divide the basis $\omega_A, A=1,\ldots,h^{1,1}$ into $\omega_\alpha, \alpha=1,\ldots h^{1,1}_+$ and $\omega_a, a=1,\ldots h^{1,1}_-$.  Working out the sign of the orientifold action on the physical fields, one finds that the two-forms $C_2$ and $B_2$ are odd, and should be expanded in terms of the $\omega_a$.  Grimm and Louis \cite{Grimm:2004uq} have derived the K\"ahler coordinates on the corresponding moduli space, {\it i.e.} the proper complex combinations of fields that appear as the lowest components of chiral multiplets:\footnote{We use the same notation as \cite{Grimm:2004uq} with two exceptions: we rescale $T_\alpha$ as $T_\alpha^{here}=(2/3)T_\alpha^{there}$,
and we add a factor of $(2\pi)^{-1}$ in the definition of $G^a$ such that the fields $c^a$ and $b^a$ have periodicity $2\pi$. See appendix \ref{s:not} for more details
on our conventions.}
\be \label{GL}
G^a&\equiv& \frac{1}{2\pi} \left(c^a-i\frac{b^a}{g_s}\right) \label{ta0} \\
T_\alpha&\equiv&i\rho_\alpha+\frac{1}{2}c_{\alpha \beta \gamma}v^\beta v^\gamma+\frac{g_s}{4}c_{\alpha b c}G^b(G-\bar G)^c \label{ta}
\ee
where $\rho_\alpha$ comes from the RR four-form $C_4$ integrated over some orientifold-even four-cycle $\Sigma_\alpha$, with $\alpha=1,\dots,h^{1,1}_+$; and $c^a$ and $b^a$ come from the RR and NS-NS two-forms $C_2$ and $B_2$ integrated over some orientifold odd two-cycle $\Sigma_a$ with $a=1,\dots,h^{1,1}_-$. The tree-level K\"ahler potential is given by\footnote{We assume that the axio-dilaton $\tau=C_0+i e^{-\phi}$ is already stabilized by fluxes at $\tau=i/g_s$ and we write down the dilaton-dependent part of the K\"ahler potential only to keep track of factors of $g_s$.}
\be
K=\log\left(\frac{g_s}{2}\right)-2\log \V_E
\ee
where the (dimensionless) Einstein-frame volume $\V_E$ of the Calabi-Yau manifold is defined in \eqref{vdef}. The dependence of this K\"ahler potential on the multiplets \eqref{ta0} and \eqref{ta}
cannot be written
down explicitly for a generic choice of the intersection numbers $c_{IJK}$. The implicit
dependence is given by writing the (Einstein-frame) volume in terms of two-cycle volumes
$v^\alpha$
\be \label{K}
K=\log\left(\frac{g_s}{2}\right)-2\log \left[\frac16 c_{\alpha \beta \gamma}v^\alpha(T,G) v^\beta(T,G) v^\gamma(T,G)\right] \,.
\ee
Then one has to solve \eqref{ta} for $v^\alpha$ and substitute the result into the above K\"ahler potential.
The K\"ahler potential is a function of $\V_E$, and hence is a function of $v^\alpha$, and in turn of $\tau_\alpha\equiv \mathrm{Re}\, T_\alpha$ and $\mathrm{Im}\, G$, but does not depend on $\mathrm{Re}\, G$ and $\mathrm{Im}\, T_\alpha$ (as can be seen by taking the real part of \eqref{ta}).  One might be tempted to conclude that $c$ enjoys a shift symmetry but that $b$ does not, but, as we will explain in \S\ref{ss:shift}, both fields have shift symmetries.

The tree-level superpotential $W_0$ does not depend on the multiplets \eqref{ta0} and \eqref{ta}. In fact it depends on the complex structure moduli and the dilaton, which we assume have already been stabilized by fluxes. Therefore we will take $W_0$ to be a discretely tunable constant.

\subsubsection{Nonperturbative stabilization of the K\"ahler moduli}

Let us now proceed to consider nonperturbative effects. We follow the KKLT strategy \cite{Kachru:2003aw} for the construction of a de Sitter vacuum. We assume that each four-cycle $T_\alpha$ is wrapped either by a Euclidean D3-brane or by a stack of D7-branes giving rise to a four-dimensional gauge theory that undergoes gaugino condensation.\footnote{In general, Euclidean D3-branes or D7-branes will wrap some linear combinations $\widetilde{T}_\alpha$ of the cycles appearing in (\ref{ta}), rather than the basis cycles $T_\alpha$ themselves, but for simplicity we will suppress this issue.}
This results in the following four-dimensional superpotential:
\be \label{W}
W=W_0+\sum_{\alpha=1}^{h^{1,1}_+} A_\alpha e^{-a_\alpha T_\alpha}\,,
\ee
where $A_\alpha$ will be treated as constants, as they depend on the complex structure moduli, which we have assumed to be stabilized;
$a_\alpha\equiv 2\pi/N_\alpha$, with $N_\alpha$ the number of D7-branes in the stack; and $N_\alpha=1$ for the case of a Euclidean D3-brane.
We can find a supersymmetric minimum by solving for the vanishing of all the F-terms: for the $h^{1,1}_+$ even K\"ahler moduli via
\be
0&=&D_\alpha W\equiv \partial_{T_\alpha} W + W \partial_{T_\alpha}K=-A_\alpha a_\alpha e^{-a_\alpha T_\alpha}-2 W \frac{\partial_{T_\alpha}\V_E}{\V_E} \nonumber \\
&=&-A_\alpha a_\alpha e^{-a_\alpha T_\alpha}- W \frac{v^\alpha}{2\V_E}\,, \label{dw}
\ee
and for the $h^{1,1}_-$ odd moduli via
\be\label{dw2}
0&=&D_a W\equiv \partial_{G^a} W + W \partial_{G^a}K= -i W  \frac{c_{\alpha a c } v^\alpha b^c}{4 \pi \V_E}
\ee
where in both cases in the last step we used the chain rule and the definitions of $G^a$ and $T_\alpha$ in terms of two-cycle volumes $v^\alpha$. The condition \eqref{dw} is simplified if we first solve for $\mathrm{Im}\,T_\alpha$, which gives
\be \label{ImT}
a_\alpha \mathrm{Im}\,T_\alpha=\theta_{A_\alpha}-\theta_{W_0}+k_\alpha \pi\,, \qquad k_\alpha \in\mathbb{Z}\,.
\ee
Then we are left with the set of real equations for each $\alpha$,
\be \label{vi}
(\pm1)_\alpha |A_\alpha| a_\alpha e^{-a_\alpha \tau_\alpha}=\partial_{T_\alpha}K \left(|W_0|+\sum_{\beta} (\pm1)_\alpha |A_\beta| e^{-a_\beta \tau_\beta}\right) \,,
\ee
where $(\pm1)$ depends on the value of $k$ in \eqref{ImT}. As long as the orientifold-even four-cycle K\"ahler moduli are defined as in \eqref{ta}, then $\partial_{T_\alpha}K=-v^\alpha/(2\V_E)<0$ for every $\alpha$. Now we prove
that in \eqref{vi} the minus sign has to be chosen for every $\alpha$ in order to have a supersymmetric solution. First we notice that the sign of the right hand side does not depend on $\alpha$, so $k_\alpha$ and hence $(\pm 1)_\alpha$ have to be the same for every $\alpha$. If we choose the positive sign in \eqref{vi}, the quantity in brackets in the right hand side is manifestly positive. Then the two sides of the equation have opposite signs
and no (compact) solution exists. To summarize, the minimization of $\mathrm{Im}\,T_\alpha$ boils down to taking all $A_\alpha$ real and negative and $W_0$ real and positive or the other way around.\footnote{We notice that if one chooses as K\"ahler variable a linear combination of the $T_\alpha$ defined in \eqref{ta}, as is done {\it e.g.}  in the large volume scenario with Swiss-cheese Calabi-Yau manifolds, then the sign of $\partial_{T_\alpha}K$ can depend on $\alpha$. In this case, the minimization of $\mathrm{Im}\,T_\alpha$ boils down to taking $W_0$ real and positive and $A_\alpha$ real with $\mathrm{Sign}(A_\alpha)=\mathrm{Sign}(\partial_{T_\alpha}K)$, up to multiplying $W$ by an overall phase.}

Concerning \eqref{dw2}, an obvious solution is given by $b^a=0$ for every $a$.
As argued in \cite{MSW}, an inflationary model with a $b$-type axion as the inflaton will generically suffer from an eta problem, and we will therefore focus on a $c$-type axion.


\subsubsection{Nonperturbative breaking of axionic shift symmetries} \label{ss:shift}

Axionic shift symmetries are central to this paper, so we will now explain how they originate and how they are ultimately broken by nonperturbative effects.  First, let us recall the classic result \cite{WW, DS} establishing the shift symmetry to all orders in perturbation theory.
Consider the axion $b =\int_{\Sigma}B/(2\pi\al)$, where $B$ is the NS-NS two-form potential and $\Sigma$ is a two-cycle in the Calabi-Yau manifold. The vertex operator representing the coupling of $b$ to the string worldsheet is \cite{WW}
\be
V(k) =       \frac{1}{2\pi \alpha'}   \int_{\Sigma} d^2\xi\, {\rm exp}\Bigl(i k\cdot X(\xi)\Bigr)\epsilon^{\alpha\beta}\partial_{\alpha}X^{\mu}\partial_{\beta}X^{\nu}B_{\mu\nu}(X)   \, .
\ee
At zero momentum, this coupling is seen to be a total derivative in the worldsheet theory. Therefore, the axion $b$ can only have derivative couplings (which vanish at zero momentum), to any order in sigma-model perturbation theory. Notice that the genus of the worldsheet did not enter in this argument, so the axion shift symmetry is also valid to all orders in string perturbation theory.

This argument fails in the presence of worldsheet boundaries ({\it i.e.}, D-branes), and also fails once worldsheet instantons, or D-brane instantons, are included.  In axion monodromy inflation, both sorts of breaking play an important role, as we shall now explain.

First, the introduction of an NS5-brane wrapping a curve $\Sigma_a$ creates a monodromy for the axion $c_a$, spoiling its shift symmetry and inducing an asymptotically linear potential \cite{MSW}.  Specifically, the potential induced by the Born-Infeld action of the NS5-brane (obtained by S-dualizing the Born-Infeld action of a D5-brane) is
\begin{equation}\label{linear}
V(c_a)=\frac{\epsilon}{g_s(2\pi)^{5}\al^2}\sqrt{\ell^{4}+(2\pi g_s c_a )^2} \,,
\end{equation}
where $\ell\sqrt{\al}$ is the size of $\Sigma_a$ and $\epsilon$ captures the possibility of suppression due to warping.  For $c_a\gg 1$, this potential is linear in $c_a$, or in the corresponding canonically normalized field, which we denoted by $\phi$ in the preceding sections.  Let us remark that the square root form of the potential can be important at the end of inflation and also makes a small change in the number of e-foldings produced for given parameter values, so that in a model that includes a specific scenario for reheating, the square root structure should be incorporated as well.  As we have not invoked a concrete reheating scenario, for our purposes the linear potential suffices, but one must still bear in mind that this form is not valid for small $\phi$.

As we will explain in detail, the D-brane instantons involved in moduli stabilization introduce sinusoidal modulations to the linear potential.  We will work exclusively in a regime in which the breaking by wrapped branes dominates over the nonperturbative breaking, although we remark in passing that the complementary regime might be interesting for realizing models involving repeated tunneling.

The breaking of the $b$ shift symmetry by Euclidean D-branes (or by gaugino condensation on D7-branes) is slightly subtle, so we will address it briefly.  As we remarked above, $b$ appears quadratically in the  classical K\"ahler potential, which seems to contradict the statement that it enjoys a shift symmetry at the perturbative level in the absence of boundaries.  However, there is no contradiction: the shift symmetry of $b$ is true at constant two-cycle volumes $v$ and \textit{not} at constant four-cycle volumes $T$.  To see this, suppose that there is a single K\"ahler modulus $T$, so that the superpotential is of the form (\ref{W}) with $h_{+}^{1,1}=1$. The K\"ahler potential is then \cite{Grimm:2004uq}
\begin{equation}
K=-3 M_{p}^2 \,{\rm log} \,(T+\bar T- d\, b^2)\,,
\end{equation}
with  $d$ a constant.  In the absence of a nonperturbative superpotential, a suitable simultaneous shift of $T +\bar T$ and  $b$ is a symmetry of the scalar potential of this system; under such a shift, the two-cycle volumes $v$ are invariant.  However, this symmetry is spoiled by the nonperturbative term in $W$, because the superpotential and the scalar potential are no longer invariant.  Therefore, in a scenario in which the four-cycle volumes are stabilized nonperturbatively, the $b$ axion receives a mass in a stabilized vacuum.

At this stage the mass-squared $m^2_b$ of $b$ is proportional to the vacuum energy and hence is negative in the supersymmetric AdS minimum. The minimum of the potential will be the final point of the inflationary dynamics, and hence we would like it to have a very small \textit{positive} cosmological constant to be consistent with the current accelerated expansion of the universe. Thus, we need to include an uplifting term. In the uplifted minimum, $m_b^2\propto V_{dS}>0$. This relation is the origin of the eta problem that was found in \cite{MSW} choosing $b$ as an axion: for a generic uplifting,\footnote{Notice that the
proportionality constant in $m_b^2\propto V_{dS}$ depends on the volume-dependence of the uplifting term and could be made small for particular choices of the latter
as proposed in \cite{Pajer:2008uy}. }
$V^{\prime\prime}(b)\sim V(b)$, so that $\eta\sim\mathcal{O}(1)$ and slow roll inflation does not take place. This is completely analogous to the eta problem of D-brane inflation
found in \cite{Kachru:2003sx} and can be intuitively understood in the same way.  Here we will take $b=0$ as the stabilized value \footnote{It is easy to check that $b=0$ is still the stabilized value after the inclusion of nonperturbative corrections to the K\"ahler potential, {\it cf.}  \S\ref{ss:toy}.} of $b$ and concentrate on $c$ as a candidate inflaton.

Let us now turn to consider $c$, which does not appear in the K\"ahler potential or superpotential at any order in perturbation theory.  To assess $c$ as an inflaton, one should determine the leading nonperturbative effects, either in the superpotential or in the K\"ahler potential, that do introduce a potential for $c$, {\it i.e.} one should identify the leading breaking of the shift symmetry.
Euclidean D3-branes carrying vanishing D1-brane charge do not induce a potential for $c$, but Euclidean D3-branes supporting worldvolume fluxes (and hence nonvanishing D1-brane charge) give rise to a dependence on $c$, via the Chern-Simons coupling $\int F_2\wedge C_2$.  As observed in \cite{MSW}, it follows that when the K\"ahler moduli are stabilized by Euclidean D3-branes, $c$ receives a mass in the stabilized vacuum: one must sum over Euclidean D-brane contributions to the superpotential, including summing over the amount $n=\int F_2$ of magnetization, and this generically introduces an eta problem for $c$.  The solution, as explained in \cite{MSW}, is to stabilize the K\"ahler moduli via gaugino condensation on D7-branes, which leads to  an exponentially smaller (and hence negligible) mass for $c$.


\subsection{Axion decay constants in string theory}

We now turn to the important task of expressing the axion decay constant, $f$, in terms of the data of a compactification.  As we reviewed in \S \ref{ss:MSW}, the decay constant of an axion $C_2= c(x) \omega /(2\pi)$ is given by
\be \label{eq:fis}
\frac{f^2}{\Mpl^2}&=&\frac{g_s}{48\pi^2 \V_E}\left[\frac{\int \omega\wedge \ast \omega}{(2\pi)^{6} \al^3} \right]\, ,
\ee
so that the primary task is to compute the norm $\int \omega\wedge \ast \omega$.  (This problem has been studied in a wide range of examples in \cite{Svrcek:2006yi}.)  We will first recall, in \S\ref{sss:dcn}, how to express the axion kinetic term, and hence also the axion decay constant, in terms of ${\cal{N}}=1$ data. This will lead us to a simple expression for the decay constant in terms of intersection numbers of the Calabi-Yau.  We will then propose a class of models in which the decay constant is rather small, motivated by the fact that with other parameters held fixed, decreasing $f$ increases the amplitude of the resonant non-Gaussianity.
Next, in \S\ref{sss:dcg}, we will present a concrete example that illustrates the geometry of a configuration that leads to small $f$.


\subsubsection{Decay constants in terms of ${\cal{N}}=1$ data} \label{sss:dcn}

In \S\ref{sss:03} we have reviewed, following \cite{Grimm:2004uq}, the four-dimensional ${\cal{N}}=1$ description of Type IIB O3-O7 orientifolds.
The multiplets relevant for us are the orientifold-odd chiral multiplets $G^a$ and the orientifold-even chiral multiplets $T_\alpha$.
The tree-level K\"ahler potential
given in \eqref{K} determines the kinetic terms for $G^a$ and hence the decay constants of the axions $b^a$ and $c^a$. First let us notice that the K\"ahler
metric in the space of the chiral multiplets $T_\alpha$ and $G^a$ factorizes in two blocks, $K_{T_\alpha \bar T_\beta}$ and $K_{G^a \bar G^b}$. The reason is that off-diagonal
terms such as $K_{T_\alpha \bar G^a}$ are proportional to intersection numbers $c_{\alpha \beta a}$ with one odd index and two even indices, which are forbidden by the orientifold action \cite{Grimm:2004uq}.
We are interested in one particular mode from among the $G^a$, which we will denote by $G^-$; $\Sigma_-$ is then the orientifold-odd two-cycle that supports our candidate inflaton $c^-$.
We now choose a basis for $G^a$ such that $K_{G^a \bar G^b}$ is block diagonal with a $1\times 1$ block $K_{G^- \bar G^-}$.  The kinetic term for $c^-$ is then given by
\be
-\frac12 f^2 \left( \partial c^-\right)^2=\Mpl^2 K_{G^- \bar G^-} \frac{1}{(2\pi)^2} \left(\partial c^- \right)^2  \subset \Mpl^2 K_{G^- \bar G^-} \left| \partial G^- \right|^2 \,,
\ee
where
\be
K_{G^- \bar G^-}=\frac{\partial^2 K(G,T)}{\partial G^- \partial \bar G^- } =-g_s   \frac{c_{\alpha --}v^\alpha}{4\V_E}\,,
\ee
and we used
\be \label{vImG}
c_{\alpha \beta \gamma}  v^\beta v^\gamma= 2 \tau_\alpha+g_s \,c_{\alpha b c}\, \mathrm{Im}\, G^b \, \mathrm{Im}\, G^c\,.
\ee
Hence we can express the decay constant of the axion $c^-$ as
\be \label{ff}
\frac{f^2}{\Mpl^2}= \frac{g_s}{8\pi^2}  \frac{c_{\alpha --}v^\alpha}{\V_E}\,.
\ee
As promised, we have expressed the norm $\int \omega\wedge \ast \omega$ in terms of the intersection numbers
\be
\frac{\int \omega\wedge \ast \omega}{(2\pi)^6 \al^3}=\frac23 c_{\alpha --}v^\alpha\,.
\ee
In \S \ref{ss:6decay} we will discuss the constraints that follow from the result \eqref{ff}. First, in the following subsection we provide some geometrical intuition for \eqref{ff}.


\subsubsection{An example: a complex plane of fixed points} \label{sss:dcg}

An instructive example arises from considering an orbifold that is locally $\mathbb{C}^2/\mathbb{Z}_2 \times \mathbb{C}$, {\it i.e.} an Eguchi-Hanson space fibered over a base $\Sigma$ of complex dimension one.  Let $\omega$ be the two-form dual to the blowup cycle of the orbifold, and let $\Sigma$ be the two-manifold of fixed points, {\it i.e.} the base over
which the Eguchi-Hanson space is fibered.\footnote {Concretely, we are imagining that $\Sigma$ extends into a warped throat region, and that an NS5-brane wraps the blowup
cycle at a particular location in the throat. The warping is invoked in order to suppress the energy density of the wrapped NS5-brane.  See \cite{MSW} for further details,
and for an example of a suitable orbifold action in a Klebanov-Strassler throat.}  We are interested in the decay constant of the axion $C_2=\frac{1}{2\pi} c(x) \omega$, so we must compute $\int \omega\wedge \ast \omega$.  In the local approximation, this is straightforward, as we shall see.  However, far from the fixed-point locus, the fiber may deviate substantially from the Eguchi-Hanson geometry, in a complicated and model-dependent way, and moreover the fixed-point locus $\Sigma$ may be embedded in the compact space in a nontrivial manner.  Happily, the integral $\int \omega\wedge \ast \omega$ has its primary support near the fixed-point locus, where the local approximation is excellent.

We recall, following the useful summary in Appendix B of \cite{Bertolini}, that the Eguchi-Hanson space has a unique homology two-cycle of radius $a/2$, where $r=a$ defines the location of the coordinate singularity; here $r$ is the standard radial coordinate.
The two-form $\omega$ corresponding to this cycle may be written
\begin{equation}
\omega = 2\pi \al \frac{a^2}{ r^2}\Bigl(\frac{dr}{r}\wedge d\psi + {\rm cos}\, \theta\frac{dr}{r}\wedge d\phi+ \frac{1}{2}{\rm sin}\, \theta\, d\theta\wedge d\phi  \Bigr)
\end{equation}
in terms of $r$ and the angular coordinates $\psi,\theta,\phi$.  By observing that $\ast_4\omega=-\omega$ and that $\int \omega\wedge\omega =-(2\pi)^4 \al^2/2$, one finds
\begin{equation}
\frac{\int_{EH} \omega\wedge\ast_{4}\omega}{(2\pi)^4 \al^2}=\frac{1}{2}\, .
\end{equation}
Clearly, given the form of $\omega$, this integral has its support in a region $a\le r\lesssim few \times a$.
This justifies the local approximation as long as the compact space has a radius that is large compared to $a$.
Next, we observe that
\begin{equation}
\frac{\int \omega\wedge\ast_{6}\omega}{(2\pi)^6\al^3}=\frac{\int_{EH}d^4 x\, \omega\wedge\ast_{4}\omega}{(2\pi)^4\al^2} \times \frac{\int_{\Sigma} \sqrt{g}}{(2\pi)^2\al} = \frac{1}{2}{\rm Vol}(\Sigma)
\end{equation}
Substituting this in (\ref{eq:fis}), we recover the parametric scaling of \S\ref{sss:dcn}.


\section{Microscopic Constraints}\label{s:const}

We now turn to determining the ranges of our phenomenological parameters that are allowed in a consistent and computable microphysical model.

Let us first remark that, as usual in string theory model building, computability imposes stringent constraints on the compactification parameters.  Because large-field inflation involves substantial energy densities and requires correspondingly steep moduli barriers, the compact space needs to be reasonably small, so that the Kaluza-Klein scale and the (necessarily lower) scale of moduli masses can be large enough to prevent runaway moduli evolution.  Clearly, one must then carefully check that the compactification is still large enough for the supergravity approximation to be valid; furthermore, backreaction of the inflationary energy on the compact space is a serious issue, particularly when this space is not large in string units.  Incorporating these requirements then leads to severe restrictions on the allowed values of the decay constant $f$.

We will begin in \S\ref{ss:constraint} by considering the constraints from computability, then give, in \S\ref{ss:backreaction}, a qualitative description of the constraints from backreaction, deferring details to appendix B.
Next, in \S\ref{ss:higher derivative}, we will verify that a two-derivative action suffices to describe this system.  This is not obvious, as rapid oscillations in the potential could enhance the importance of generic higher-derivative terms; however, we will show that the specific terms emerging from string theory are negligible in our solution.  We then apply these constraints in \S\ref{ss:6decay} to determine the range of the decay constant $f$. Finally, we estimate the size $bf$ of the modulations; as this is rather model-dependent, in \S\ref{ss:toy}, we will restrict our attention to a specific example in which a periodic contribution is generated by Euclidean D1-brane corrections to the K\"ahler potential.

\subsection{Constraints from computability}\label{ss:constraint}

In this subsection we will list several constraints coming from the consistency of the string theory setup. We will first require the validity of the string and $\al$ perturbation expansions, and the validity of neglecting higher-order corrections to the nonperturbative superpotential, and then we will require that the inflaton potential does not destabilize
the compactification.

First of all we require the validity of string perturbation theory, {\it i.e.} we require $g_s\ll1$.  We must also ensure the validity of the $\al$ expansion.  To do this including a reasonable {\it{estimate}} of numerical factors such as $2\pi$, it is convenient to use worldsheet instantons as a proxy for perturbative $\al$ corrections, because the normalization is easily determined. To get the correct coefficient, we start from the string-frame ten-dimensional metric
$g_{string}$
and impose that the worldsheet instanton action obeys $e^{-S_{WS}}\lesssim e^{-2}$, or
\be
\frac{1}{2\pi \alpha'}\int \sqrt{g_{string}}\gtrsim 2\,,
\ee
which using $g_{string}=g_{Einstein}\sqrt{g_s}$ is converted to Einstein frame
\be \label{ws}
2<\frac{\sqrt{g_s}}{2 \pi \al}\int_{\Sigma^\alpha} J=\sqrt{g_s} v^\alpha 2\pi \quad \Rightarrow  \quad v^\alpha>\frac1{\pi \sqrt{g_s}}\,,
\ee
and we used that $\int_{\Sigma^\alpha} \omega^\beta=(2\pi)^2\al \delta^{~\beta}_\alpha$.

As we invoked nonperturbative corrections to the superpotential, we must also require that any further superpotential corrections, {\it e.g.}  from multi-instantons, are negligible. For this purpose it suffices to impose
\be
e^{-a_\alpha T_\alpha}<e^{-2}\ll 1 \quad \Rightarrow \quad \tau_\alpha>\frac{N_\alpha}{\pi}\,.
\ee
Additional constraints come from the moduli stabilization process. To use the single-field inflationary analysis we have developed in \S\ref{s:back} and
\S\ref{s:spectrum}, we need to require that the
uplifted minimum is only slightly perturbed by the inflationary dynamics.
In particular, the linear potential that we have represented as $\mu^3\phi$ actually depends on the compactification volume, and hence shifts the minimized value of the volume.  In four-dimensional Einstein frame, the leading term in the inflaton potential is
\begin{equation}
V(\phi,{\cal{V}}_E)\approx \left(\frac{\langle{\cal{V}}_E\rangle}{{\cal{V}}_E}\right)^{2}\mu^{3}\phi
\end{equation}
where $\langle{\cal{V}}_E\rangle$ is the expectation value of the volume.
To ensure that the resulting contribution to the potential for the volume is unimportant, we will insist that the inflaton potential induced by the NS5-brane,  $V(\phi)$, is smaller than the moduli potential $\U$.

At the supersymmetric minimum we have
\be
V_{AdS}=-\frac{g_s}{2}\frac{3|W|^2}{\V_E^2}\,.
\ee
Without specifying the details of the uplifting mechanism, we assume that an uplifting to a small and positive cosmological constant is possible, and that the height of the potential barrier $\U$ that separates the uplifted minimum from decompactification is of the same order as $\U\sim |V_{AdS}| $.  Now, the COBE normalization tells us that
\be
V(\phi_{CMB})=\epsilon\left(\,0.027 \Mpl\right)^4\simeq 2.4 \cdot 10^{-9} \Mpl^4\,.
\ee
Hence we obtain the constraint
\be
\frac{g_s}{2}\frac{3|W|^2}{\V_E^2}=|V_{AdS}|\simeq\U\gg 2.4 \cdot 10^{-9} \Mpl^4\,.
\ee

To extract a useful form of the above constraints, let us substitute for $W$ the solution of any of the equations \eqref{dw}
\be
W=+|A_\alpha|a_\alpha \, e^{-a_\alpha \tau_\alpha} \frac{2 \V_E}{v^\alpha}\,,
\ee
with no sum over $\alpha$. We will also assume $|A_\alpha|\sim 1$ (see \cite{MSW} for a discussion of this point).
After some manipulations we find
\be
\tau_\alpha \ll -\frac{N_\alpha}{2 \pi} \log \left( N_\alpha 10^{-5} \frac{v^\alpha}{\pi \sqrt{g_s}}   \right)\,,
\ee
again with no summation over $\alpha$.  Finally, we should limit the number of D7-branes in each stack; although there plausibly exist examples with $N_\alpha$ quite large, we will impose $N_\alpha\le 50$.  This gives us
\be \label{73}
\tau_\alpha \ll 73- 8\log \left( \frac{v^\alpha \pi \sqrt{g_s}}{2 g_s}\right)\,.
\ee
We notice that $v^\alpha(\pi \sqrt{g_s}) >1$ was the condition in \eqref{ws} that enabled us to neglect $\al$ corrections,  so that as long as $g_s\leqslant0.5$ the second term on the right hand side of \eqref{73} is negative.


\subsection{Constraints from backreaction on the geometry}\label{ss:backreaction}

Another important constraint comes from the requirement that the backreaction of the inflationary energy density on the compact space is small.  In this section we will give a qualitative description of the problem and will briefly sketch a model-building solution; the interested reader is referred to appendix B for a more complete treatment.

At the time that the CMB perturbations are produced, the inflaton has a large vev in Planck units, $\phi\sim 11 \Mpl$, corresponding to a configuration of the two-form potential threading the two-cycle $\Sigma_-$ of the form
\be
\frac{1}{(2 \pi)^2 \alpha'} \int_{\Sigma_-} C_2  \equiv N_w=\frac{\phi}{2 \pi f} \gg 1
\ee
In the absence of an NS5-brane wrapping $\Sigma$, there would be no energy stored in this configuration, as $C_2$ enjoys a shift symmetry.  However, inflation is driven by the substantial energy stored in this system by the Born-Infeld action of the wrapped NS5-brane.  Moreover, there is a corresponding D3-brane charge induced by the Chern-Simons coupling $\int C_2\wedge C_4$.   Note that the net induced D3-brane charge in the total compactification is zero, as required by Gauss's law, because we have arranged for an additional, tadpole-canceling NS5-brane that wraps a distant cycle $\Sigma'_-$ homologous to $\Sigma_-$, but does so with opposite orientation.  Therefore, the Chern-Simons coupling induces a dipole configuration of D3-brane charge, with $F_5$ flux lines stretching from $\Sigma_-$ to $\Sigma'_-$.

It is essential to ensure that the inflationary energy, which is effectively localized in the compact space in the vicinity of the wrapped NS5-brane, does not substantially correct the remainder of the compact geometry.  Heuristically, one can imagine that the increased tension of the NS5-brane, as well as the induced charge, is represented by $N_w$ D3-branes dissolved in the NS5-brane.  We must therefore estimate the effect of  $N_w$ D3-branes in a warped throat (recall that we have situated each wrapped NS5-brane in a warped region in order to suppress its energy density below the string scale, as required {\it e.g.}  by the COBE normalization).  Clearly, this backreaction will be reasonably small if $N_w \ll N$, with $N$ the D3-brane charge of the background throat.

However, we must be careful about the effect of even a modest distortion of the geometry on the moduli stabilization and therefore on the four-dimensional potential.  Let us first recall that in scenarios of D3-brane inflation in nonperturbatively-stabilized vacua, even a single D3-brane moving slowly in a throat can affect the warp factor, and correspondingly the warped volumes of four-cycles bearing nonperturbative effects, to such a degree that this interaction is the leading contribution to the inflaton potential \cite{BDKMMM,explicit}.

This sensitivity originates in two facts: first, D3-branes perturb the warped metric in a manner that is not suppressed by the background warp factor at the location of the D3-branes, because D3-branes are BPS with respect to a  throat generated by D3-brane charge, and hence their contributions to the metric may simply be superposed on the background.  Second, nonperturbative effects on a four-cycle are exponentially sensitive to changes in the four-cycle volume.  Both these facts appear threatening for a situation such as ours in which the moduli are stabilized nonperturbatively and substantial D3-brane charge is induced in a throat: one can anticipate that as inflation proceeds and the D3-brane charge diminishes, the four-cycle volume changes, leading to an unanticipated, and possibly steep, contribution to the inflaton potential.

To understand this concretely, we will first consider a simpler system: an anti-D3-brane in a warped throat generated by $N$ D3-branes, or equivalently a warped throat generated by $N-1$ D3-branes, together with a brane-antibrane pair.  Furthermore, from the result of \cite{DeWolfe:2008zy} one learns that at long distances, the effect of the brane-antibrane pair on the supergravity solution is strongly suppressed by the warp factor at the location of the pair, {\it i.e.} at the tip of the throat. In contrast, the effects of D3-branes are not suppressed in this manner.  Therefore, for the purpose of computing perturbations to the bulk compact space, we may replace an anti-D3-brane in a warped throat generated by $N$ D3-branes with a warped throat generated by $N-1$ D3-branes, up to exponentially small corrections.

Equipped with this approximation, we may represent the configuration of interest as follows: two warped throats, carrying the charge of $N_1, N_2$ D3-branes respectively, are perturbed to $N_1+N_w, N_2-N_w$ by the inclusion of the NS5-brane in (say) the first throat, and the anti-NS5-brane in the second throat.  Here we are ignoring the warping-suppressed correction indicated above, and we are approximating the NS5-branes by the D3-brane charge and tension that they carry, which is an excellent approximation for $N_w\gg1$. Other effects due to the NS5-brane that do not depend on its induced D3-brane charge, {\it i.e.} on its world-volume flux, are independent of the inflaton and hence do not correct its potential. One can now easily see that the volume of a four-cycle at a generic location in the compact space will be corrected by the inclusion of the NS5-branes.  If the four-cycle happens to enter one or both throats, the change in the volume is easily computed, and is seen to be substantial ({\it cf.}  appendix B).

To control this problem, we situate the NS5-brane and the anti-NS5-brane, together with the family of homologous cycles connecting them, in a single warped region. The idea is that from the bulk of the compact space, the NS5-brane configuration will appear to be a distant dipole whose net effect, integrated over a four-cycle, averages out to be small. This setup allows us to parametrically suppress the backreaction by a small factor given by the ratio of the dipole length, {\it i.e.} the distance between two NS5-branes, to the distance between the NS5-branes and the four-cycle in question. This small factor comes in addition to the suppression by the small ratio $N_w/N$.\footnote{A further suppression can be achieved with a carefully-chosen embedding of the four-cycle, {\it e.g.}  one that is symmetric with respect to the two NS5-branes.  However, this requires fine-tuning, whereas the dipole suppression on which we have focused is parametric.}

In appendix \ref{4} we give more details about the above setup. We show, through two explicit models of increasing complexity, the robustness of the above suppression mechanism. 

\subsection{Constraints from higher-derivative terms} \label{ss:higher derivative}

The analysis presented thus far has used the two-derivative action, which is an approximation with a limited range of validity. In general, one expects an infinite series of higher-derivative terms, possibly including multiple derivatives as well as powers of the first derivative. Our background solution involves rapid oscillations, so it is reasonable to ask whether these high frequencies enhance the role of higher-derivative terms and render the two-derivative approximation invalid.  To check this, one should evaluate the higher-derivative terms on the solution and compare to the two-derivative action.  We will now show that the two-derivative approximation is valid in the scalar sector; analogous considerations apply to the gravitational action.

Rather than write down the most general higher-derivative corrections to the scalar sector, we give here the terms that end up being present in the string theory examples.  In string theory, we can directly compute the leading higher-derivative terms in the action for $b$, extending the result to $c$ using S-duality. To get the leading terms,
one considers the $\alpha'^3$ corrections to the effective action due to Gross and Sloan \cite{Gross:1986mw} (at the four-point level) and Kehagias and Partouche \cite{Kehagias:1997cq} (up to the eight-point level). These corrections are of the same lineage as the famous $\rm{Riemann}^4$ term, but involve NS-NS three-form flux. This yields corrections to the axion kinetic terms.
Following \cite{Kehagias:1997cq}, the ten-dimensional Einstein-frame action including the leading ($\alpha'^3$) corrections is
\be \label{equ:derivative}
    S_{10D,E}= \frac{1}{(2\pi)^7\alpha'^4}\int d^{10}x\sqrt{g_E}\left(R_E-\frac{1}{12g_s}H_{KMN}H^{KMN}+\frac{\zeta(3)}{3\times 2^6}g_s^{-3/2}\alpha'^3\bar{R}^4+\ldots\right)                    \, .
\ee
where
\be
\bar{R}_{MN}^{\,~~~~  PQ}=R_{MN}^{\,~~~~  PQ}+\frac{1}{2}g_s^{-1/2}\nabla_{[M}H_{N]}^{\,~~ PQ}-\frac{1}{4}g_s^{-1}H_{[M}^{\,~~ C[P}H_{N]C}^{\,~~~~ Q]} + \ldots       \,
\ee
and the square brackets are defined without the combinatorial factor $1/2$ in front.
Hence, the terms that are relevant for our axion at order $\alpha'^3$ are proportional to $H^8$ and $(\nabla  H)^4$. To estimate the importance of these terms, we will consider a special case in which the internal space is a $T^2\times T^4$, with the NS-NS two-form field only along the $T^2$ directions $8$ and $9$, {\it i.e.} $B_{89}=-B_{98}=b$. Furthermore, since the background dynamics involves large frequencies but not large spatial gradients, we are primarily interested in terms containing only time derivatives, and can therefore take $b$ to be homogeneous in the noncompact spatial directions. In this special case, making use of (2.13) in \cite{Gross:1986mw}, and using S-duality to determine the action for $c$ from that for $b$, we find that after dimensional reduction the corrected action for $c$ is
\begin{eqnarray}
S_{4D}=\int d^4x\left[ -M_p^2\frac{g_s}{2}\dot{c}^2 g^{88}g^{99}+\frac{\zeta(3)}{2^6 g_s^{3/2}}\frac{\V_E^3}{\pi^3 M_p^4}\left( \frac{1}{2}\dot{c}^8 g_s^{4} (g^{88}g^{99})^4+\frac{1}{2^4}\ddot{c}^4g_s^2(g^{88}g^{99})^2\right)\right]  \,
\end{eqnarray}
Now we use $\phi=c f$ to make the kinetic term canonical, yielding the action in terms of $\phi$,
\begin{eqnarray}
S_{4D}&=&\int d^4x\left[ -\frac{1}{2}\dot{\phi}^2 +\frac{\zeta(3)}{2^6 g_s^{3/2}}\frac{\V_E^3}{\pi^3 }\left( \frac{1}{2}\frac{\dot{\phi}^8}{M_p^{12}}  +\frac{1}{2^4}\frac{\ddot{\phi}^4}{M_p^8}\right)\right] \nonumber \\
&\equiv & \int d^4 x\left[-\frac{1}{2}\dot{\phi}^2+\frac{\dot{\phi}^8}{M_{I}^{12}}+\frac{\ddot{\phi}^4}{M_{II}^8}\right] \,,
\end{eqnarray}
where we can now calculate the scale of the higher derivative terms $M_{I}$ and $M_{II}$ to be
\be
M_I=M_p\frac{g_s^{1/8}}{\V_E^{1/4}}\left(\frac{\pi^3 2^7}{\zeta(3)} \right)^{1/12}
\ee
and
\be
M_{II}=M_p\frac{g_s^{3/16}}{\V_E^{3/8}}\left(\frac{\pi^3 2^{10}}{\zeta(3)} \right)^{1/8}
\ee
To determine whether these higher-derivative terms will become important, we compute the dimensionless quantity $\frac{\omega}{M_{I,II}}$, where $\omega=\frac{\dot{\phi}}{f}$ \eqref{omega} is the physical frequency of oscillations; we obtain

\be
\frac{\omega}{M_I}\simeq 5\cdot 10^{-3}\left(\frac{f}{10^{-3}}\right)^{-1} \left(\frac{g_s}{0.2} \right)^{-1/8} \left( \frac{\V_E}{120}\right)^{1/4}\,,
\ee
\be
\frac{\omega}{M_{II}}\simeq 6\cdot 10^{-3}\left(\frac{f}{10^{-3}}\right)^{-1} \left(\frac{g_s}{0.2} \right)^{-3/16} \left( \frac{\V_E}{120}\right)^{3/8}\,.
\ee
For the ranges of $f$ and $\V_E$ that will be of interest to us ({\it{cf.}} \S\ref{s:7}), the higher-derivative terms are not important and our two-derivative approximation is justified.

\subsection{Constraints on the axion decay constant}\label{ss:6decay}

In this section, we discuss direct constraints on the axion decay constant $f$.
We first recall a rather general (conjectured) upper bound $f<\Mpl$ \cite{Banks:2003sx}, and we then describe and incorporate a novel lower bound, specific to our setup, that arises from combining the requirements that $\al$ perturbation theory should be valid and that the inflationary energy should not drive decompactification.

Despite many attempts, at the time of writing there is no known, controllable string theory construction that provides $\A>\Mpl$.  In particular, the authors of \cite{Banks:2003sx} have scanned several classes of string theory models and found sub-Planckian axion
decay constants in every case.
However, this upper bound on $f$ is of relatively little importance for the phenomenological signatures we are considering in this paper.

On the other hand, a potential lower bound on $f$ is of considerable importance for our analysis.  Considering oscillations in the CMB spectrum, in the regime $f\ll\Mpl$ one can easily find models that range from being observationally excluded to giving undetectably small modifications,  depending on the amplitude of the ripples in the inflationary potential. Furthermore, the resonant non-Gaussianity becomes large only for small $f$ ({\it e.g.}  we will find that $f<3 \cdot 10^{-3}$ is a necessary condition to give a reasonable prospect of detectability). Hence we will move on to consider possible lower bounds on $f$.

As discussed in \cite{MSW} and in the preceding section, a direct lower bound on $f$ comes from the requirement of small backreaction.
In particular, the radius of curvature induced by the energy localized on the wrapped NS5-brane should be smaller than the smallest radius of curvature $\RR$
in a direction transverse to the NS5-brane in the compactification. This requires
\be \label{bk}
N_w \ll \frac{\RR^4 \X}{4 \pi g_s} \quad\Rightarrow \quad \frac{f}{\Mpl}\gg\frac{2 \phi g_s}{\RR^4 X}\,,
\ee
where we have defined $\X\equiv {\rm{Vol}}(X_5)/\pi^3$, with $X_5$ the base of the cone forming the warped throat.  We remark that $X\le 1$, as $S^5$ is the Sasaki-Einstein manifold with the largest volume, in the sense defined above.
We can estimate $\RR$ as being comparable to the AdS radius $R$ of the throat containing the NS5-brane.
Given that the volume\footnote{We always refer to the warped volume, calculated with the whole warped metric.} $\V$ of the Calabi-Yau has to be larger than the volume of any throat it includes, one finds that
\be \label{th}
\V > V_{throat}=\frac{\pi^3}{2}\X R^6\,,
\ee
where for simplicity we have assumed that the UV cutoff of the throat is at $r\sim R$ where the warp factor becomes of order unity. Putting together \eqref{bk} and \eqref{th}, we find
\be \label{fv}
\frac f  \Mpl>\frac{\pi^2 2^{1/3} \phi g_s}{\X^{1/3} \V^{2/3}}\simeq\frac{137 g_s}{\X^{1/3} \V^{2/3}}=\frac{0.09}{X^{1/3}\V_E^{2/3}}\,.
\ee

Although the above constraint substantially restricts our parameter space, an even stronger constraint comes from demanding the validity of $\al$ perturbation theory: using \eqref{ff} for $f$ and combining this with the lower bound on two-cycle volumes given in \eqref{ws}, we obtain
\be\label{fff}
\frac{f^2}{\Mpl^2}=\frac{\sqrt{g_s}}{(2 \pi)^3 \V_E} \left(c_{\alpha --}v^\alpha\sqrt{g_s}\pi \right)>\frac{\sqrt{g_s}}{(2\pi)^3 \V_E}\,,
\ee
where we have assumed that $c_{\alpha--}\ge 1$. \eqref{fff} turns out to give the strongest microphysical lower bound on $f$. An upper bound is harder to determine from this formula.  Assuming again that $c_{\alpha--}\ge 1$, assuming that no precise cancellations occur, and using \eqref{ws}, we find
\be\label{upper}
\frac{f}{\Mpl}<g_s\frac{\sqrt{3}}{2}\,.
\ee


\subsection{Constraints on the amplitude of the modulations}\label{ss:toy}

So far we have seen that with the K\"ahler potential and superpotential given in \eqref{K} and \eqref{W}, the axion $c$ persists as a flat direction after moduli stabilization.\footnote{As we have remarked, the axion $b$ has its flat direction lifted by nonperturbative stabilization of the K\"ahler moduli.}  As explained in \cite{MSW}, the presence of an NS5-brane wrapping the two-cycle that defines $c$ introduces a monodromy and results, for large $c$, in the linear potential in \eqref{linear}. In this section we will consider further nonperturbative corrections that will in general induce small modulations of this linear potential. These are precisely the modulations whose phenomenology we have studied in the first part of this paper.

Nonperturbative corrections could appear both in the K\"ahler potential and in the superpotential. We focus on the first possibility and comment at the end of this section on
the second. Consider the type IIB orientifolds with O3-planes and O7-planes. As we have remarked, the RR two-form $C_2$ is odd under the orientifold projection and therefore a
four-dimensional axion that survives projection comes from integrating $C_2$ over an odd two-cycle $v^-$. Such an odd cycle can be thought of as $v^-=v^1-v^2$, where $v^1$ and $v^2$
represent two two-cycles in the parent Calabi-Yau manifold that are mapped into each other by the orientifold action. Now consider a Euclidean D1-brane wrapping the even cycle $v^+=v^1+v^2$. Such an instanton feels the local ${\cal N}=1$ supersymmetry of the orientifolded theory, and it breaks this supersymmetry completely.\footnote{To see this, note that ({\it cf.} \cite{Witten}) the instanton action depends on a two-cycle volume, but the proper K\"ahler coordinates are four-cycle volumes.  Therefore, the instanton action cannot be holomorphic, so the instanton cannot contribute to a superpotential, and must instead be non-BPS.}
Hence this is a non-BPS instanton with four universal fermionic zero modes, namely the goldstini of the
broken ${\cal N}=1$ supersymmetry. If the Euclidean D1-brane wraps a minimum-volume cycle in the homology class $v^+$ then it has the right total number of fermionic zero modes (four) to contribute to a
D-term and in particular to the K\"ahler potential.

More specifically, in \cite{grimmmodularity} it was argued that nonperturbative contributions from worldsheet instantons and their $SL(2,\mathbb{Z})$ images, Euclidean $(p,q)$ strings, give rise to corrections to the prepotential of the ${\cal N}=2$ theory of the parent Calabi-Yau compactification. Such corrections are most naturally expressed inside the logarithm of the K\"ahler potential,
\be
K=-2\log\left[\V_E+g(G,\bar G) \right] \, ,
\ee
where $g$ is an appropriate function.  Invariance under $SL(2,\mathbb{Z})$, or more generally under a subgroup $\Gamma \subset SL(2,\mathbb{Z})$, is naturally achieved if $g$ is the sum of some individual correction $\tilde g$ over an orbit of $\Gamma$.

At the time of writing, the nonperturbative correction $g$ is not known explicitly, but a modular-invariant result has been conjectured in \cite{Frank}.  Inspired by the structure of this result (which we will not reproduce here), we will make a simple educated guess based on the following criteria: the non-perturbative correction should go to zero exponentially for large two-cycle
volume $v^+$; it should break the continuous shift-symmetry of $c$ to a discrete shift-symmetry $c\rightarrow c + 2\pi$; and it should be invariant under whatever discrete subgroup $\Gamma \subset SL(2,\mathbb{Z})$ of the ten-dimensional $SL(2,\mathbb{Z})$ symmetry is preserved by the compactification. The subgroup $\Gamma$ may well be trivial, and we will assume this for simplicity; note, however, that one can plausibly obtain a more constrained result when some or all of the symmetry is preserved, as in \cite{Frank}.  Moreover, notice that along the orbits of $\Gamma$, the instanton action generally increases compared to that of a single worldsheet instanton or Euclidean D1-brane; thus, when the volume $v^+$ is not too small, only a few terms make an important contribution, with the remainder enjoying further exponential suppression.

A reasonable guess satisfying these criteria, for $\Gamma$ trivial, is
\be\label{dK}
K=-2\log\left[\V_E+e^{-S_{ED1}} \cos(c) \right]=-2\log\left[\V_E+e^{-\frac{2\pi v^+}{\sqrt{g_s}}} \cos(c) \right]\,.
\ee

In light of this corrected K\"ahler potential, we should revisit the moduli stabilization before proceeding to calculate the size $b$ of the periodic contribution to the scalar potential.

We begin by noticing the following implication
\be \label{F-flat}
\left\{\begin{array}{l}
   D_{T_\alpha}W=\mathcal{O}\left(e^{-2S_{ED1}}\right)\\
  D_{G^a}W=\mathcal{O}\left(e^{-S_{ED1}}\right)\\
  W_{G^a}=0\\
\end{array}\right. \quad
\Rightarrow \quad
\left\{
\begin{array}{l}
  \partial_{T_\alpha} V=0+\mathcal{O}\left(e^{-2S_{ED1}}\right)\\

\partial_{G^a}V=-2\,e^K\,|W|^2K_{G^a}+\mathcal{O}\left(e^{-2S_{ED1}}\right)
\end{array} \right.
\ee
which can be verified by direct computation. This allows us to use the F-flatness condition to find the minimum in the $T_\alpha$-directions even when one of the F-terms, namely $D_{G_-}W$, does not vanish. Equipped with this knowledge we repeat, \textit{mutatis mutandis}, the steps of \S\ref{s:micro}.

First, the phases of the $T_\alpha$ are stabilized as in \eqref{ImT}, with $k$ being odd as explained below \eqref{vi}. The reason is that the sign of $\partial_{T_\alpha} K$ is not changed by the small nonperturbative correction $e^{-S_{ED1}}$. Second, $\mathrm{Im}\, G$ is again stabilized at $0$. Given \eqref{F-flat}, the equation one needs to solve is $D_{G^a}W=0$, which reduces to
\be\label{dgw}
0=W\partial_{G^a} K\propto\partial_{G^a} \V_E-e^{-S_{ED1} } \left[ \pi \sin(c) + \cos (c) \frac{2\pi} {\sqrt{g_s}} \partial_{G^a} v^+ \right]=0\,,
\ee
where we made use of \eqref{GL} to perform the derivative on $c$.  Since $\V_E$ and $v^+$ only depend on $\mathrm{Im}\, G$ implicitly as in \eqref{vImG}, we can take the imaginary part of \eqref{dgw},
\be\label{Imdgw}
\frac12 (\partial_{\mathrm{Im}\, G^-} v^\alpha) c_{\alpha \beta \gamma}  v^\beta v^\gamma  -e^{-S_{ED1} } \cos (c) \frac{2\pi} {\sqrt{g_s}} \partial_{\mathrm{Im}\, G^-} v^+ =0\,.
\ee
But from \eqref{vImG} we know that
\be\label{bnewK}
c_{\alpha\beta\gamma}(\partial_{\mathrm{Im}\, G^-} v^\beta) v^\gamma=g_s c_{\alpha a -}\mathrm{Im}\, G^a\,,
\ee
which means that $\mathrm{Im}\, G^a=0$ (for every $a$) is a solution to \eqref{Imdgw}. The real part of \eqref{dgw} is nonvanishing and of order $e^{-S_{ED1}}$. Again because of \eqref{F-flat}, the minimization in the $\tau_\alpha$ is obtained by imposing $D_{T_\alpha}W=0$. These equations depend on the inflaton $c$, appearing explicitly in \eqref{dK}, and hence the minimum in the $T_\alpha$ directions will be a function of $c$. Integrating out the $T_\alpha$ leads to a contribution in the effective potential $V[T(c),c]$ for $c$ which is of the same order as the contribution coming from the explicit $c$-dependence in the K\"ahler potential. Therefore this effect cannot be neglected. To take it into account, we solve the $D_{T_\alpha}W=0$ equations perturbatively in $e^{-S_{ED1}}$.

We define the coefficients of the minimum in the $\tau_\alpha$ directions in a perturbative expansion in $e^{-S_{ED1}}$ by
\be
\tau_{\alpha,min}\equiv \tau_{\alpha,(0)}+\cos(c)\, e^{-S}\tau_{\alpha,(1)}+ \dots\,,
\ee
and so on for all other variables. The zeroth-order equations are
\be\label{0th}
(D_{T_\alpha}W)_{(0)}=(\partial_{T_\alpha}W)_{(0)}+W_{(0)}(\partial_{T_\alpha}K)_{(0)}=0\,,
\ee
which can be solved numerically once the model is specified. The first-order equations are
\be \label{1st}
(D_{T_\alpha}W)_{(1)}=(\partial_{T_\alpha}W)_{(1)}+W_{(1)}(\partial_{T_\alpha}K)_{(0)}+W_{(0)}(\partial_{T_\alpha}K)_{(1)}=0\,,
\ee
which again can be solved numerically using the solutions of \eqref{0th}. We turn now to estimate the parameter $b$ defined in \eqref{V}. One finds
\be
bf&\equiv& V_{(1)} \mu^{-3} e^{-S_{ED1}}=\frac{\U\, \phi}{\mu^3 \phi} e^{-S_{ED1}}\left(K_{(1)}+2 \mathrm{Re}\,  \frac{ W_{(1)}}{W_{(0)}}\right)\,,
\ee
where we have defined $\U$ as the moduli stabilization barrier at zeroth order in $e^{-S_{ED1}}$, {{\it i.e.}}
\be
\U=\frac{g_s}{2}\left(\frac{3|W|^2}{\V_E^2}\right)_{(0)}\,.
\ee
More explicitly, using \eqref{dK} and \eqref{W},
\be\label{bf<}
bf&=&\frac{\U\, \phi}{2.4\cdot10^{-9}\Mpl^4} e^{-S_{ED1}}\left[\frac{8\pi}{\sqrt{g_s}}\frac{ (\partial_{\tau_\alpha}v^+)_{(0)}} {v^\alpha_{(0)}} -2a_\alpha \tau_{\alpha,(1)}-\frac{2 v^\alpha_{(1)}} {v^{\alpha}_{(0)}} \right] \nonumber \\
&=&\frac {\U \,\phi} {2.4\cdot10^{-9}\Mpl^4} 2 e^{-S_{ED1}} \left[ \frac{\sum_{\beta} (\partial_{T_{\beta }}W )_{(0)} \tau_{\beta,(1)} } {W_{(0)}}-\frac{\V_{E,(1)}+1}{\V_{E,(0)}} \right]\,,
\ee
where the first line is valid for any $\alpha$ and the second line (obtained using \eqref{1st}) shows that the expression for $b$ is independent of $\alpha$. Notice that $\partial_{T_\alpha} v^+=\frac{1}{2}\partial_{\tau_\alpha}v^+$ is given implicitly by
\be
c_{\alpha\beta\gamma}(\partial_{\tau_\rho}v^\gamma) v^\beta=\partial_{\tau_\rho}\tau_\alpha=\delta_\alpha^{~\rho}\,.
\ee
Some comments are in order. The size of the ripples in the potential is proportional to the ratio of the moduli stabilization barrier to the scale of inflation, which has to be large for the self-consistency of the estimate. We have used the value of the potential at the would-be AdS minimum to estimate the moduli stabilization barrier once an uplifting term is included. Due to the exponential suppression $e^{-S_{ED1}}$, the size of $bf$ is extremely sensitive to $g_s$ and $v^+$.

An upper bound can be derived from \eqref{bf<} using
the following considerations. In the KKLT construction, perturbative corrections to the K\"ahler potential can be neglected as long as $W_0\ll1$, and generically $W\sim W_0$. For larger values of $W_0$, perturbative corrections have to be included, as in the large volume scenario \cite{lvs}. In the present work, we focused on the former setup and we leave an investigation of the latter for the future. The exponential suppression in \eqref{bf<} can be bounded by \eqref{ws}. Finally, we denote the model-dependent term in square brackets in \eqref{bf<} by $c_0$. Putting things together leads to the bound
\be \label{bfbound}
bf <2 c_0\cdot 10^7 \frac{g_s}{\V_E^2} e^{-2/g_s} \left(\frac{W}{0.1}\right)^2 \,.
\ee

Even imposing all the model-independent constraints
we have described in the previous sections, one can still have $bf > 10^{-4}$, which, as shown in \S\ref{s:numerics}, is roughly the upper bound imposed by measurements of the scalar power spectrum.
Therefore,
in certain parameter ranges the primary constraint on modulations of the potential comes from the data, not from microphysics.




\section{Combined Theoretical and Observational Constraints} \label{s:7}


We now summarize our results, combining the observational constraints from \S\ref{s:numerics} with the theoretical constraints from \S\ref{s:const}.  As an aid to the reader, we will now briefly recall the qualitative properties of those results.

Axion monodromy can produce characteristic signatures in the CMB: the oscillations in the axion potential generated by nonperturbative effects source resonant contributions to the scalar power spectrum and bispectrum.
The amplitude and frequency of the oscillations in the potential can therefore be bounded by comparison to observations.  We recall from \S\ref{s:numerics} that the observational constraints take the form of exclusion contours in the space of the phenomenological parameters, after marginalization over additional model parameters that have important degeneracies with those displayed.  For convenience, we have chosen to display constraints in terms of the parameters $f$ and $bf$ defined in (\ref{V}), marginalizing over the phase $\Delta\phi$ and over $\Omega_B h^2$.

\begin{figure}
\begin{center}
      \includegraphics[width=3.2in]{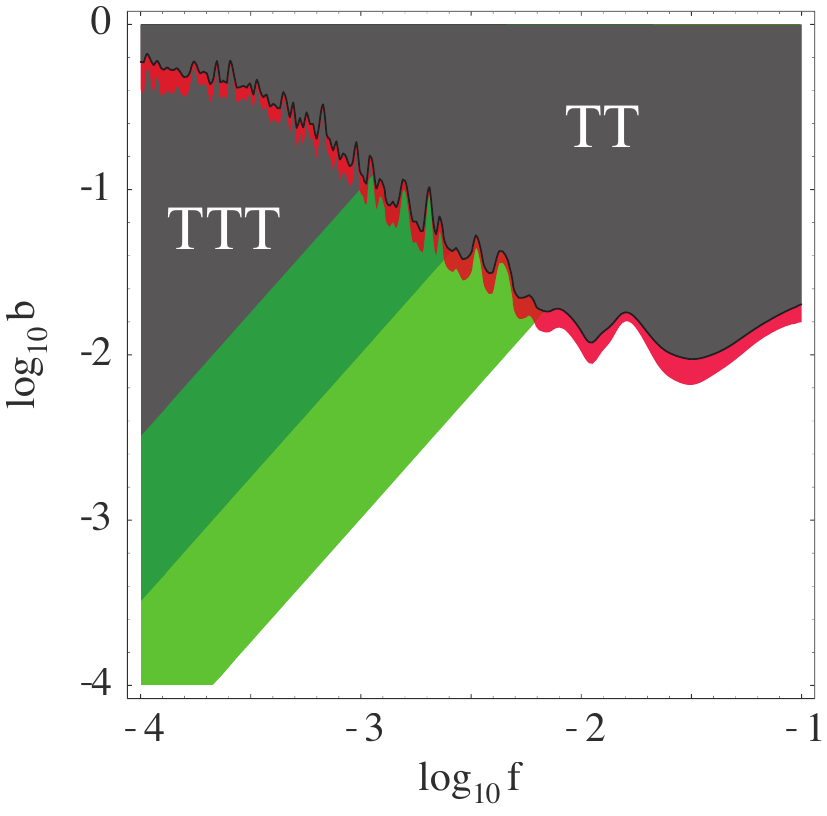}
      \caption{We show the (one- and two-sigma) likelihood contours for the temperature two-point function together with three contours that characterize the amplitude of the three point function, for $f_{res}=200,20,2$.}\label{exp}
\end{center}
\end{figure}

\begin{figure}
\begin{center}
      \includegraphics[width=\textwidth]{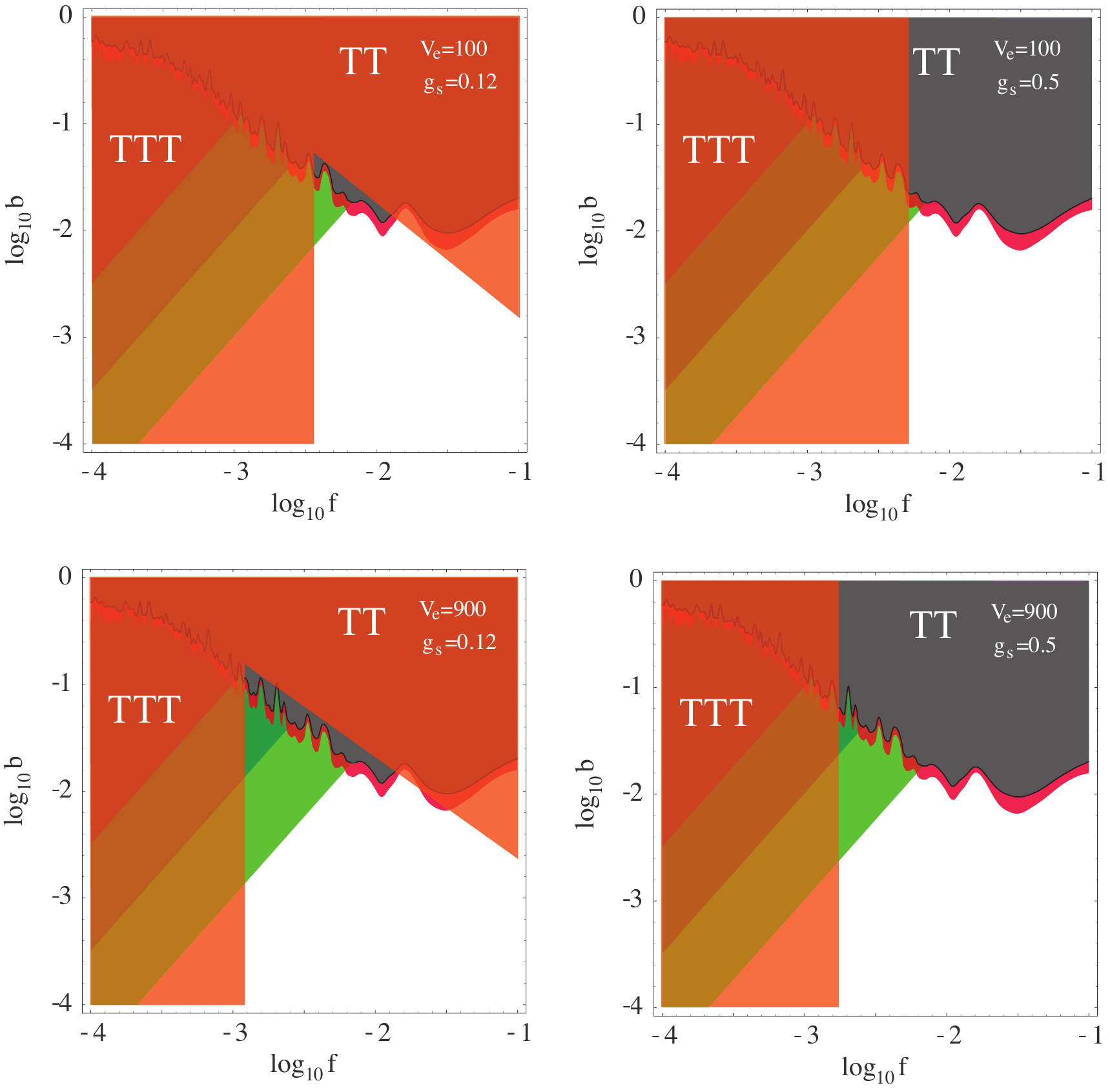}
      \caption{We superimpose the theoretical constraints, summarized in \eqref{b<} and \eqref{f<}, on the constraints imposed by observations, which are shown in figure \ref{exp}. The orange overlay indicates regions of the parameter space that are difficult to reach in the class of models considered in the present work. The theoretical constraints are shown for $g_s=0.12,0.5$ and $\V_E=100,900$.}\label{theory}
\end{center}
\end{figure}

The first new step is to combine the exclusion contours based on the temperature two-point function with estimates of constraints from the three-point function. Based on the rough estimates described in \S\ref{s:bispectrum}, we present, in figure \ref{exp}, three contours at $f_{res}=200,20,2$, with the expectation that the gray region ($f_{res}>200$) might plausibly be excluded, while the colored, lighter regions ($20<f_{res}<200$ and $2<f_{res}<20$, respectively) are possibly within detectability.  A careful study of the constraints on resonant non-Gaussianity would be a worthwhile topic for future research.

Next, we recall that in \S\ref{s:const}, we found that the requirements of consistency and computability in the string compactifications giving rise to axion monodromy models led to constraints on the parameters $f$ and $bf$.  Let us remark that as these constraints are not rooted in deep principles of string theory or of quantum field theory, but rather originate in practical limitations in our present ability to construct computable models, they may well be loosened in further work.  As such, the theoretical constraints we present here should be understood as designating {\it{included}} rather than excluded regions: in contrast to experimental contours, theoretical contours of this sort may expand rather than contract given improved understanding.

Because the parameter $b$ measures the amplitude of a nonperturbative effect, it is exponential in the natural input parameters, and can therefore be made small without substantial fine-tuning.  We therefore do not present a lower bound on $b$. However, we found the theoretical upper bound (\ref{bfbound})\footnote{We stress that this `bound' is {\it{not}} universal and depends on the assumptions enumerated in \S\ref{s:const}. We include it here as a representative example of the constraints that arise in particular scenarios.}
\be\label{b<}
bf <2 c_0\cdot 10^9 \Mpl^4 \frac{g_s}{\V_E^2} e^{-2/g_s}  \,,
\ee
with a model-dependent constant $c_0$ that can be estimated in explicit examples, and which we find to be typically of order $10^{-2}$.

Next, we obtained a lower bound for $f$ in \eqref{fff}.\footnote{The constraint from the backreaction described in \S\ref{ss:backreaction} is weaker than \eqref{fff}.}  A precise upper bound, however, is highly model-dependent. We estimate an upper limit by assuming that the intersection numbers are of order one\footnote{Larger intersection numbers are an interesting possibility that we will not investigate here.} and that no precise cancellations occur.
From \eqref{fff} and \eqref{upper}, the complement of the theoretically excluded range for $f$ is then
\be \label{f<}
 \frac{g_s^{1/4}}{(2\pi)^{3/2} \sqrt{\V_E}}<f<g_s\frac{\sqrt{3}}{2}\,.
\ee

Notice that the theoretical constraints depend mainly on two quantities: the string coupling $g_s$ and the volume $\V_E$ of the compactification. The former appears in the exponential suppression of the nonperturbative effect generating the modulations of the potential. Hence, $g_s\lesssim 0.1$ suppresses any possible signature of the modulations. For $g_s\gtrsim 0.1$, there is always a theoretically allowed region in which the oscillations in the inflaton potential lead to observable ripples in the two-point function of the CMB. On the other hand, the size of the non-Gaussianity depends critically on $\V_E$ as well. Assuming $g_s\gtrsim 0.1$, larger $\V_E$ allows for a larger range of $f$ and therefore larger non-Gaussianity (see \eqref{f<}). A way to quantify this is to use the estimate obtained in \S\ref{s:bispectrum},
\be
f_{res}\simeq \frac 94 \frac{b}{(\phi f )^{3/2}}\,,
\ee
and the lower bound in \eqref{f<}. The result is
\be
\V_E>170 \, \left(\frac{g_s}{0.2}\right)^{1/2} \left(\frac{f_{res}}{10}\right)^{4/5} \left(\frac{10^{-4}}{bf}\right)^{4/5}\,.
\ee

We now combine the theoretical and observational constraints, presenting them in the plane \mbox{$\{ \log(f),\log(b)\} $}. We choose as boundaries $10^{-4}<f\ll\Mpl$ and $10^{-4}<b\ll 1$ based on the following considerations. The number of oscillations per e-folding is roughly $10^{-2} \Mpl/f$. Hence for $f\gtrsim 0.1 \Mpl$ there is less than one oscillation in the whole range of scales probed by the CMB, and the signal from modulations becomes degenerate with the overall amplitude. Furthermore, in \S\ref{s:spectrum}, we systematically used the expansion $b\ll 1$, where $b=1$ divides monotonic from non-monotonic potentials.  Finally, the lower boundaries $10^{-4}<f$ and $10^{-4}<b$ exclude regions that are relatively uninteresting in the present context: smaller values of $b$ lead to an unobservably small signal, while smaller values of $f$ are rather difficult to obtain in the class of string theory constructions we considered.  In the $\{ \log(f),\log(b)\}$ plane, the theoretically allowed region looks like an interval in $f$, whose size is determined by $\V_E$, with an upper cut effectively determined by $g_s$ as in \eqref{b<}.

\begin{figure}
\begin{center}
      \includegraphics[width=.5\textwidth]{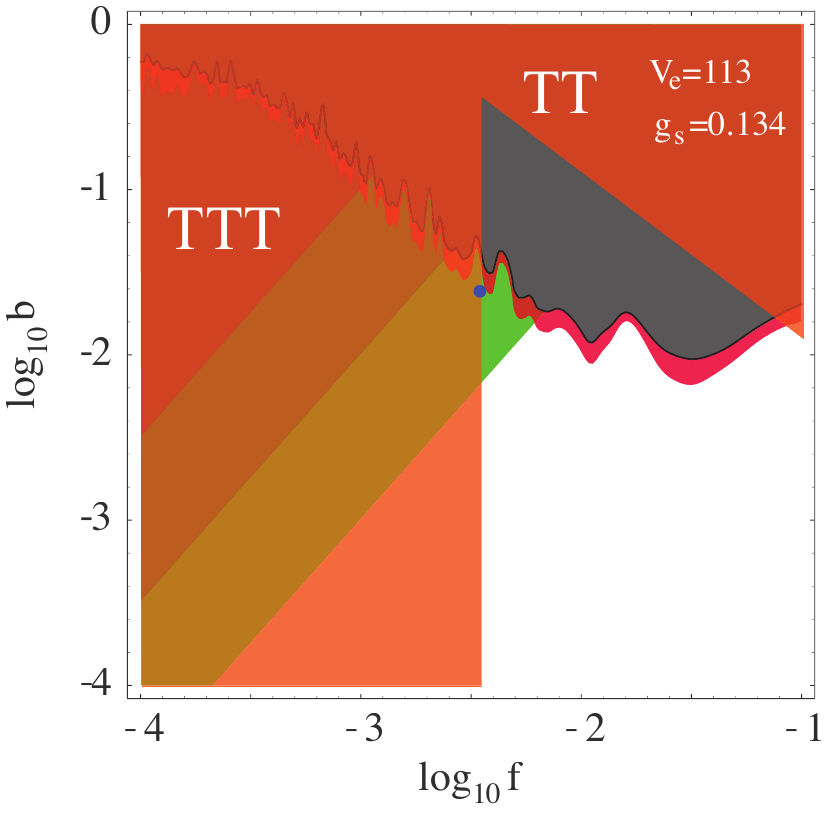}
      \caption{The blue dot represents the explicit numerical example presented in appendix \ref{a:num}. It represents a case in which upcoming experiments could detect the signatures of modulations in both the two-point function and the three-point function.\label{point}}
\end{center}
\end{figure}

Finally, in figure \ref{point} we show where a particular numerical toy example, with specific choices of the intersection numbers, lies in the $\{ \log(f),\log(b)\} $ plane.




\section{Conclusions}

The goal of this investigation was to characterize the predictions of axion monodromy inflation for the CMB temperature anisotropies.  Nonperturbative effects in these models generically introduce sinusoidal modulations of the inflaton potential, which in turn lead to resonantly-enhanced modulations of the scalar spectrum and bispectrum.

We have provided a simple analytic result for the modulated scalar power spectrum in this class of models.  We also presented an alternative derivation in terms of episodes of particle production driven by resonance between a mode inside the horizon and the driving force of the oscillatory background evolution.  We then determined in detail the constraints that the five-year WMAP data places on models with modulations of this sort.

Next, after reviewing the realization of axion monodromy inflation in string theory, we performed a comprehensive study of the parameter constraints implied by the requirements of microphysical consistency and computability.  The resulting allowed parameter regions are very plausibly realizable in sensible string theory constructions.

We also identified a new contribution to the inflaton potential in axion monodromy inflation: the backreaction of the inflationary energy on the compact space can source an important correction to the potential by correcting the volumes of four-cycles and hence affecting the scale of nonperturbative moduli-stabilizing effects.  We then presented a model-building solution to this problem, in which the NS5-brane and anti-NS5-brane driving inflation are in the same warped region, or more generally are distant from the four-cycles of interest.

Finally, we combined the observational and theoretical constraints, in order to ascertain whether detectable modulations of the scalar spectrum and/or bispectrum are possible, consistent with current observational bounds and known theoretical restrictions.  Our conclusion is that both sorts of modulations are possible, and in fact in many cases the strongest bound on the amplitude of the modulations comes from data, not from microphysics.  Moreover, even though observational limits on the amplitude and frequency of modulations in the scalar power spectrum provide a strong constraint on the parameter space of axion monodromy models, and even though microphysical constraints sharply restrict the allowed frequencies, detectably-large non-Gaussianity can indeed be produced in a class of controllable models.  Such models enjoy three nontrivial signatures: detectable tensors with $r\approx 0.07$, a modulated scalar power spectrum, and resonant non-Gaussianity.

Let us remark that even in the absence of non-Gaussianity, this class of models is eminently testable: axion monodromy inflation unambiguously predicts a large tensor signal, and the parameters of the models are already strongly constrained by limits on modulations in the scalar power spectrum.

There are several interesting directions for future work. First, we have not analyzed the constraints on the model from the three-point function; more generally, understanding the prospects for constraining or detecting resonant non-Gaussianity is an important task.  Moreover, it would be instructive to construct an explicit model in which the many theoretical constraints we have checked can be combined in a coordinated way.  In addition, it would be interesting to determine whether chain inflation can be realized in this context.

It is intriguing that the modulated power spectrum we have found is very similar in form to that proposed in the context of modifications of the initial state, as in {\it e.g.}  \cite{Easther:2001fi,Kaloper:2002uj,Martin:2003sg,Martin:2004yi,Easther:2004vq}. In light of our calculation of the power spectrum in terms of particle production (\S\ref{B}), this coincidence is not entirely surprising: the driving force of the oscillating background  eventually generates an excited state, even if one begins in the Bunch-Davies vacuum.  We leave for the future a more systematic exploration of this connection.

Finally, it would be most valuable to develop a broader understanding of the connection, if any, between symmetries and signatures in models of large-field inflation.

\section*{Acknowledgments}
We are particularly indebted to E.~Silverstein for initial collaboration and for providing several crucial insights at various stages of this project. We thank R.~Easther, E.~Silverstein and especially D.~Baumann for comments on  a draft of this paper.  We thank D.~Baumann, B.~Bellazzini, M.~Berg, C.~Burgess, R.~Easther, B.~Freivogel, S.~Kachru, E.~Lim, D.~Marsh, H.~Peiris, M.~Trodden, H.~Tye, and H.~Verlinde for helpful discussions.  R.F. thanks E.~Komatsu and P.~Steinhardt for insightful comments and helpful discussions.  L.M. thanks  M.~Cveti\v{c}, I.~Garc\'ia-Etxebarria, R.~Richter, A.~Uranga, and T. Weigand for interesting discussions on related topics.  G.X. thanks P.~Tanedo and W.~Valkenburg for assistance with the code used in \cite{inflation}.  The research of L.M., E.P., and G.X. was supported in part by the National Science Foundation through grant NSF-PHY-0757868.  The work of R.F. has been partially supported by
the National Science Foundation under Grant No. PHY-0455649. The research of A.W. is supported in part by the Alexander-von-Humboldt foundation, as well as by NSF grant PHY-0244728.  We thank the Aspen Center for Physics, the KITP, and the Perimeter Institute for hospitality while several of the results described herein were obtained.
R.F. would also like to thank the Stanford Institute for Theoretical Physics as well as Cornell University for their hospitality.

\appendix

\addtocontents{toc}{\setcounter{tocdepth}{1}}

\section{Notation and Conventions}\label{s:not}

In this appendix we review our conventions, emphasizing differences with the existing literature.

A good starting point is the ten-dimensional string-frame action\footnote{Remember that $2\kappa_{10}^2=(2\pi)^7\al^4$.} \cite{Polchinski:1998rr}
\be
S_{10}=\frac{1}{(2\pi)^7 \al^4}\int d^{10}x \sqrt{g_{string}}\left( e^{-2\Phi} R_{string}-\frac12 |dC_2|^2\right)
\ee
which after the rescaling to the ten-dimensional Einstein-frame metric $e^{-\Phi/2}g_{string, MN}=g_{E,MN}$ becomes
\be
S_{10}=\frac{1}{(2\pi)^7 \al^4}\int d^{10}x \sqrt{g_{E}}\left( R_{E}-\frac12 g_s |dC_2|^2\right) \,,
\ee
where we assumed that the axio-dilaton is $\tau=i/g_s$. Upon compactifying on a six-dimensional manifold $Y$, the resulting four-dimensional reduced Planck mass is
\be
\Mpl^2=\frac{\int_{Y} \sqrt{g_E}}{(2\pi)^6 \al^3} \frac{1}{\al \pi}\equiv \frac{\V_E}{\al \pi}\,,
\ee
where $\V_E$ is the (dimensionless) Einstein volume of $Y$ measured in units of $2\pi \sqrt{\al}$. When $Y$ is (conformally equal to) a Calabi-Yau space, we have
\be\label{vdef}
\V_E=\frac16\frac{\int J\wedge J\wedge J}{(2\pi)^6 \al^3}=\frac16 v^I v^J v^K \frac{\int \omega_I\wedge \omega_J\wedge \omega_K}{(2\pi)^6 \al^3}\equiv \frac16 v^I v^J v^K c_{IJK}\,,
\ee
where $\omega_I$ for $I=1,\dots,h^{1,1}$ are a basis of the cohomology $H^2(Y,\mathbb{Z})$ normalized such that
\be
\int_{\Sigma_I} \omega_J=(2 \pi)^2 \al \delta^{~I}_J
\ee
for a basis $\Sigma_I$ of the dual homology $H_2(Y,\mathbb{Z})$. With the ansatz for the ten-dimensional RR two-form
\be
C_2=\frac{1}{2\pi}c(x) \omega\,,
\ee
for some base two-cycle $\omega$, we get a four-dimensional axion $c(x)$ with periodicity\footnote{Note that this choice differs from that in \cite{MSW}, where the axion periodicity was $(2\pi)^2$.} $2\pi$, as can be seen {\it e.g.}  via S-duality starting from the world-sheet coupling $\int B /(2\pi \al)$. The axion decay constant of $c$ is
\be
\frac{f^2}{\Mpl^2}=\frac{g_s}{12 \V_E (2\pi)^2} \left[\frac{\int \omega\wedge \ast \omega}{(2\pi)^6 \al^3}\right]\,.
\ee
The four-dimensional ${\cal N}=1$ K\"ahler potential for the K\"ahler moduli is
\be
K= -2 \log \V_E\,.
\ee


\section{Induced Shift of the Four-Cycle Volume}\label{4}

In this appendix we address the issue raised in \S\ref{ss:backreaction}: the inflationary energy can correct the warped volumes of four-cycles in the compact space, leading to corrections to the moduli potential, and hence inducing new terms in the inflaton potential itself.

More specifically, if an NS5-brane wraps some cycle $\Sigma$, then a nonvanishing integral $\int_{\Sigma}C_2 \neq 0$ leads to the presence of energy that is localized near $\Sigma$ in the compact space; this energy corresponds to the increased tension of the NS5-brane.  Moreover, there is a corresponding induced D3-brane charge via the coupling $\int C_2\wedge C_4$.  The increased tension creates a backreaction on the metric (and in particular, on the warp factor) of the compact space, while the induced charge sources five-form flux.  We must determine whether these effects substantially correct the nonperturbative effects that are responsible, in our KKLT-like scenario, for stabilization of the K\"ahler moduli.

Whether the nonperturbative superpotential arises from gaugino condensation on D7-branes or from Euclidean D3-branes, it is exponentially sensitive to the warped volume of the four-cycle wrapped by these D-branes.  Therefore, we will carefully consider the possibility of an inflaton-dependent shift of the warped volume of various four-cycles.

Concretely, we will consider a fivebrane/anti-fivebrane pair wrapping two homologous cycles, and will compute the leading correction to the volume of a particular four-cycle in the same throat region as the fivebranes. This will serve as a conservative upper bound on the effect of the worldvolume flux, as more distant four-cycles would be more weakly affected.

It would be very interesting to perform a systematic study of this backreaction in the four-dimensional effective theory and in ten-dimensional supergravity/string theory. We leave this task for future investigation. In what follows, we simply show that the effect described above can be ameliorated by choosing an appropriate configuration.

The are two mechanisms to suppress the backreaction on a given four-cycle volume. A first improvement comes from choosing a setup in which the leading backreaction is due to a dipole as opposed to a monopole potential. This allows for a parametric suppression. The second improvement can be achieved by a carefully chosen geometry of the four-cycle under consideration. In general, this latter mechanism requires fine tuning.

The problem of estimating the backreaction from two-form flux on an NS5-brane pair may be simplified by a series of approximations. First, the inflaton-dependent backreaction is generated by the increased tension and the induced D3-brane charge of the NS5-branes, which may be understood as corresponding to some number of D3-branes (or anti-D3-branes) dissolved in the NS5-branes.  In practice, it is much simpler to study the effect of the D3-branes themselves; this captures the leading inflaton-dependent contributions.

The configuration of interest involves an NS5-brane wrapping $\Sigma$, with
\be
\frac{1}{(2 \pi)^2 \alpha'} \int_{\Sigma} C_2  \equiv N_w \,,
\ee
as well as a distant NS5-brane wrapping a homologous cycle $\Sigma'$, but with opposite orientation.
(We will refer to the latter object as the anti-NS5-brane.)
Next, we recall that the COBE normalization requires each fivebrane to be in a warped region. Let us denote by $N/2$ the amount of D3-brane charge that creates the background warping for each of the fivebranes.\footnote{More generally, one could consider different degrees of warping for each fivebrane; extending our considerations to this case is straightforward.}

In light of the above discussion, we may approximate the fivebrane by a stack of $N_w$ D3-branes and the anti-fivebrane by a stack of $N_w$ anti-D3-branes.  Combining this with the background D3-brane charge, we conclude that a convenient proxy for our system consists of two stacks of D3-branes, which we call A and B respectively. The first consists of $N/2+N_w$ D3-branes and the second of $N/2$ D3-branes and $N_w$ anti-D3-branes, which we may more conveniently represent as $N/2-N_w$ D3-branes and $N_w$ brane-antibrane pairs.

Next, using the results of \cite{DeWolfe:2008zy}, we recognize that the leading backreaction effect comes from the total D3-brane charge on each stack, while the brane-antibrane pairs lead to subleading effects that are suppressed by powers of the warp factor. Thus, we can simplify even further, so that at last we are considering a supersymmetric system involving two stacks that contain $N/2+ N_w$ and $N/2-N_w$ D3-branes, respectively.

Equipped with this much simpler system, we may now estimate the inflaton-dependent backreaction, by computing how the presence of the stacks A and B leads to a $N_w$-dependent change in the warped volume of some four-cycle.

We choose the usual D3-brane ansatz
\bea\label{ans}
ds_6^2&=&\sqrt{H^{-1}(y)}ds_4^2 +\sqrt{H(y)} ds_6^2\,\\
\tilde F_5 &=&(1+\ast)dH^{-1}\wedge {{\rm Vol}}_4\,,\quad \Phi=\textrm{const}\,.
\eea
The resulting equation of motion is linear in $H(y)$. Therefore we may simply add the solutions obtained in the presence of either of the two individual stacks. Once the resulting warp factor is used to compute the volume of a four-cycle, the $N_w$ dependence of the result gives us an estimate of the inflaton-dependence of the nonperturbative superpotential.

We tackle the problem in two steps of increasing complexity. First, in \S\ref{flat} we give a very simple, (conformally) flat toy example in which the calculations are easy. This already shows the relevant features of the more complicated solution. The inflaton-dependent shift of the volume can be suppressed by having the distance between A and B much smaller than the distance between the four-cycle and either of A and B. This corresponds to a configuration in which the leading interaction is via a dipole. In addition, one can fine-tune the four-cycle embedding so that the $N_w$-dependent correction to its volume actually cancels.\footnote{Although suppression from symmetry of the embedding appears unappealing because of the fine-tuning required, one should keep in mind that it could conceivably be enforced by a discrete symmetry of the compactification.}

Then, in \S\ref{a:res}, \S\ref{abbc} and \S\ref{fin}, we describe the case of a resolved conifold using the solution of \cite{Klebanov:2007us,Pando Zayas:2000sq}. We consider a particular holomorphic embedding of a four-cycle and compute numerically the inflaton-dependence of its warped volume.


\subsection{A simple illustration of the suppression mechanism}\label{flat}
\begin{figure} \begin{center} \includegraphics[width=2.4in]{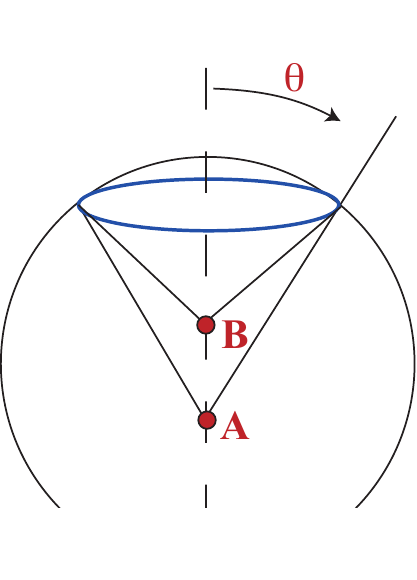} \label{spherical} \caption{This diagram illustrates the positions of the A and B stacks of D3-branes in $\mathbb{R}^6$, the choice of the angular coordinate $\theta$, and, in blue, a (topologically trivial) four-cycle.}
\end{center} \end{figure}

Consider two stacks of $N/2\pm N_w$ D3-branes, called A and B, respectively, in conformally flat space $M_4\times \mathbb{R}^6$. The A stack is located at the origin of $\mathbb{R}^6$ and the B stack is located at some position $(u,0,0,0,0,0)$ for\footnote{We choose the letter $u$ in analogy with the setup of the next subsections, where the distance between the A and the B stack is given by the resolution parameter of the resolved conifold.} $u\in \mathbb{R}^+$, where we have chosen spherical coordinates (see figure \ref{spherical}) with the metric
\be
ds_6^2=dr^2+r^2\left(d\theta^2+\sin^2 \theta d\Omega_4^2 \right)\,,
\ee
where $d\Omega_4$ is the volume form of $S^4$. With the usual D3-brane ansatz \eqref{ans} one finds the solution
\be
H&=&1+\frac{R_A^4}{r^4}+\frac{R_B^4}{(r^2+u^2-2r u \cos \theta)^2}\,,\\
R_{A,B}&=&4\pi g_s \alpha'^2 \left(\frac N2 \pm N_w\right)\,.
\ee
Let us consider a (topologically trivial) four-cycle $\Sigma_4$ defined by\footnote{We choose the letter $\mu$ in analogy with the (usually complex) parameter appearing in other known embeddings of four-cycles in the conifold \cite{Kuperstein:2004hy,Ouyang:2003df}.} $r=\mu$ and $\theta=\bar \theta$, whose unwarped volume is $V_4=\frac{8\pi^2}{3}\bar r^4  \sin^4 \bar\theta $. The warped volume is
\be
\int_{\Sigma_4} H(r,\theta) \sin^4\theta d\Omega_4&=&V_4 H(\mu,\bar \theta)\\
&=&V_4 \left[ 1+ \frac {R^4}{\mu^4} \left(1 + \mathcal{O}\left(\frac {u}{\mu}\right)\right) - N_w \frac{u}{\mu}\left (4\cos \bar \theta -2 \frac{u}{\mu}+\mathcal{O}\left(\frac {u^2}{\mu^2}\right)\right) \right]\,. \nonumber
\ee
From this result, one can see that a fine-tuning of the embedding can suppress the backreaction, {\it i.e.} if $\cos\bar \theta\simeq u/(2\mu)$. On the other hand, a parametric suppression is also clearly visible. Both the factors $N_w/N$ and $u/\mu$ can be made small by construction. The former must be small in order for the background geometry to be at all trustworthy. The latter can be made small by arranging for all the four-cycles bearing nonperturbative effects to be far away in units of the separation of the fivebranes.  Physically, this means that the four-cycle is sensitive only to the dipole field generated by the A and B stacks.


\subsection{The conifold and its resolution}\label{a:res}

In the following we review some relevant definitions and conventions regarding the resolved conifold. The treatment is based on \cite{Candelas:1989js,Pando Zayas:2000sq}. The (singular) conifold is a cone over $T^{1,1}$ (the coset space $SU(2)\times SU(2)/U(1)$, which is topologically $S^2\times S^3$). It is defined as the hyperspace in $\mathbb{C}^4$ that is a solution of the complex constraint
\be\label{sing}
\mathrm{det}W\equiv\mathrm{det}\left(
	\begin{array}{cc}
	 X & U \\
	 V & Y
	\end{array}
 \right) = XY-VU=0
\ee
where $(X,U,V,Y)$ are coordinates on $\mathbb{C}^4$. The resolved conifold can be defined as the zero locus in $\mathbb{C}^4\times \mathbb{CP}_1 $ of the two linear complex equations
\be\label{res}
\left(
	\begin{array}{cc}
	 X & U \\
	 V & Y
	\end{array}
 \right) \left(\begin{array}{c}
	 \lambda_1 \\
	 \lambda_2
	\end{array}\right)=0
\ee
where $(\lo,\lt)$ are complex coordinates on $\mathbb{CP}_1$, {\it i.e.} they are identified by $(\lo,\lt)\simeq (\alpha \lo, \alpha \lt)$, for every $\alpha\in \mathbb{C}_*$. For every $W\neq0$ \eqref{sing} and \eqref{res} are equivalent, but when $W=0$, {\it i.e.} at the tip, $(\lo,\lt)$ are arbitrary and \eqref{res} defines a $\mathbb{CP}_1 \simeq S^{2}$. The radial direction is defined by
\be
\mathrm{Tr}W^\dagger W=r^2\,.
\ee

One can check that the resolved conifold is an $\mathcal{O}(-1)\oplus \mathcal{O}(-1)$ bundle over $\mathbb{CP}_1$ with fiber $\mathbb{C}_2$. If we define $\lambda\equiv \lt/\lo$, then we can choose coordinates on a patch $H_+ \equiv\{\lambda\neq0 \}$ of the resolved conifold using the following solution of \eqref{res}:
\be
W= \left(
	\begin{array}{cc}
	 -\lambda U & U \\
	 -\lambda Y & Y
	\end{array}
	\right)\,.
\ee
Defining $\tilde \lambda\equiv \lo/\lt$, one can find coordinates on a complementary patch $H_-\equiv\{\tilde \lambda\neq0\}$ using the following solution:
\be
W= \left(
	\begin{array}{cc}
	 X & -\tilde\lambda X  \\
	 V & -\tilde\lambda V
	\end{array}
	\right)\,.
\ee
The complex structure is given by
\be\label{vic}
\Omega=dU\wedge dY\wedge d\lambda=dV \wedge dX \wedge d\tilde\lambda\,.
\ee
For later use, we introduce a parametrization of the resolved conifold in terms of real coordinates and give the explicit K\"ahler metric. We start by noting that a particular solution of \eqref{res} is given by
\be
W_0= \left(
	\begin{array}{cc}
	 0 & r  \\
	 0 & 0
	\end{array}
	\right)\,,
	\quad
	\left(
	\begin{array}{cc}
	 \lambda_0  \\
	 0
	\end{array}
	\right)\,
	\quad \Rightarrow \quad
	\lambda=0\,.
\ee
The base of the resolved conifold with respect to $r$ can be obtained by acting on this solution with two $SU(2)$ transformations, $L_1$ and $L_2$,
\be
L_i=\left(
	\begin{array}{cc}
	 \cos \frac{\theta_i} 2 e^{\frac i2 (\psi_i+\phi_i)} & -\sin \frac{\theta_i} 2 e^{-\frac i2 (\psi_i-\phi_i)}   \\
	 \sin \frac{\theta_i} 2 e^{\frac i2 (\psi_i-\phi_i)}  & \cos \frac{\theta_i} 2 e^{-\frac i2 (\psi_i+\phi_i)}
	\end{array}
	\right)\,, \quad i=1,2
\ee
written in terms of Euler angles. This gives
\be
W=L_1 W_0 L^\dagger_2 \,,\quad
	\left(
	\begin{array}{cc}
	 \lo  \\
	 \lt
	\end{array}
	\right)=
	L_2
	\left(
	\begin{array}{cc}
	 \lambda_0  \\
	 0
	\end{array}
	\right)
\ee
which depends only on the combination $\psi\equiv\psi_1+\psi_2$. A K\"ahler metric on the resolved conifold with resolution parameter $u$ is given by \cite{Pando Zayas:2000sq}
\be\label{s6b}
ds_6^2=\kappa^{-1}(\rho) d\rho^2+\frac19 \kappa(\rho)\rho^2 e_\psi^2+\frac16 \rho^2(e_{\theta_1}^2+ e_{\phi_1}^2) +\frac16 \left(\rho^2+6 u^2\right) ( e_{\theta_2}^2+ e_{\phi_2}^2)\,,
\ee
where
\be
\kappa(\rho)=\frac{\rho^2+9u^2}{\rho^2+6 u^2}\, .
\ee
Here, following \cite{Pando Zayas:2000sq}, we have defined a new radial coordinate $\rho$ by
\be
r^4=\frac49 \rho^4\left(\frac23 \rho^2+6 u^2 \right)\,,
\ee
The explicit expression for the $e$'s is
\be
e_\psi=d\psi+\sum_{i=1}^2\cos\theta_id\phi_i\quad,\quad e_{\theta_i}=d\theta_i\quad,\quad e_{\phi_i}=\sin\theta_id\phi_i\,.
\ee

\subsubsection{The $\lambda UY$ embedding}\label{abbc}

In this subsection, we consider a particular holomorphic embedding of a four-cycle. A simple embedding would be $\lambda=\mu$ because this is trivial to solve for in real coordinates, $\tan \theta_2=\mu$ and $\phi_2=0$ for $\mu\in\mathbb{R}$. The trouble is that this embedding reaches the tip, and in fact $r$ is unconstrained. This can also be seen from
\be
r^2=\left(1+|\lambda|^2\right)\left(|U|^2+|Y|^2\right)\,.
\ee
As a result, this embedding does not give us the dipole suppression factor analogous to the $(u/\mu)$ of appendix \ref{flat}. The next-simplest embedding (whose defining equation depends on $r$) is
\be
\lambda U Y=  \mu^3\,,\quad \mu\in \mathcal{R}\,,
\ee
which in real coordinates gives
\be
\psi=0\,,\quad \sin(\theta_2)\sin(\theta_1)=4 \frac{\mu^3}{r^2}\sim\frac{\mu^3}{\rho^3}\quad{\rm for\;large}\; r\,.
\ee

After some algebra (in particular, expressing $d\theta_2$ as a function of $d\theta_1$ and $dr$) one finds the metric in terms of $d\rho,\,d\theta_2,\,d\phi_1$ and $d\phi_2$. Its determinant $g_4^{ind}$ is independent of $\phi_{1,2}$ and reads
\bea
g_4^{ind}&=&\frac{\rho ^2 \csc ^4(\theta _2)}{20736\left(9 u^2+\rho ^2\right)^3 \left(\rho ^4 (9 u^2 +\rho ^2) \sin ^2(\theta _2)-54 \mu ^6\right)}\times\nonumber\\
&&\Big(\left(6 u^2+\rho ^2\right) \left(9 u^2 \rho +\rho ^3\right)^2 \cos \left(4 \theta _2\right)\nonumber\\
&&-4 \cos \left(2 \theta _2\right) \left(486 u^6 \rho ^2+189 u^4 \rho ^4+u^2 \left(24 \rho ^6-324 \mu ^6\right)-27 \mu ^6 \rho ^2+\rho ^8\right)\nonumber\\
&&+3 \left(54 \rho ^2 \left(9 u^6+2 \mu ^6\right)+189 u^4 \rho ^4+864 u^2 \mu ^6+24 u^2 \rho ^6+\rho ^8\right)\Big)^2\quad.
\eea

We see that there is a boundary beyond which the sign of the determinant becomes negative, which thus defines the integration boundary in $\rho,\theta_2$-space:
\bea
\rho_{min}(\theta_2)&=&\sqrt{3}\,\mu\; \sqrt{A-\frac{u^2}{\mu^2}\cdot\left(1-\frac{u^2}{\mu^2}\frac{1}{A}\right)}\to\sqrt{3\cdot 2^{1/3}}\,\mu\; \csc ^{1/3}(\theta_2 )\quad{\rm for}\;\frac{u}{\mu}\ll1\nonumber\\
&&{\rm with:}\quad A=\sqrt[3]{\sqrt{-\csc ^2(\theta_2 ) \left(2 \frac{u^6}{\mu^6}- \csc ^2(\theta_2 )\right)}-\frac{u^6}{\mu^6}+ \csc ^2(\theta_2 )}\,\,.
\eea



\subsection{The shift of the four-cycle volume}\label{fin}

The solution with the branes smeared over the $S^2$ was obtained in \cite{Pando Zayas:2000sq}. Later, the solutions with pointlike sources were given in \cite{Klebanov:2007us}. If the D3-brane stacks are at the north and south
pole of the resolution $S^2$, respectively, {\it i.e.} $\theta_2^A=\pi-\theta_2^B=0$, then one finds
\bea
H&=&\sum_{l}(2l+1) H_l(\rho) \left[L_A^4 P_l(\cos(\theta_2)) +L_B^4 P_l(\cos(\theta_2)) (-1)^l  \right]\,,\label{H}\\
H_l&=&\frac{2}{9 u^2}\frac{C_\beta}{\rho^{2+2\beta}} \phantom{x}_2 F_1\left(\beta,1+
\beta,1+2\beta,-\frac{9 u^2}{\rho^2}\right)\,,\label{Hl}\\
C_\beta&=&\frac{(3 u)^{2\beta} \Gamma(1+\beta)^2}{\Gamma(1+2\beta)}\,\quad \beta=\sqrt{1+(3/2)l(l+1)}\,,\\
L_{A,B}&=&\frac{27}{16}4\pi g_s (\alpha')^2 (N\mp N_w)\,,
\eea
where $\phantom{}_2F_1$ is a hypergeometric function. We want to integrate this warp factor on some supersymmetric four-cycle $\Sigma_4$. This gives us an estimate of the inflaton dependence of the gauge kinetic function of a stack of D7-branes wrapping $\Sigma_4$.

Using this information and \eqref{H} and \eqref{Hl} we can now calculate the integral
\be
{\cal V}_{warped}=\int_{\Sigma_4}d\rho \,d\theta_2\,d\phi_1\,d\phi_2\sqrt{-g_4^{ind}}H(\rho,\theta_2)=4\pi^2\int_{\Sigma_4}d\rho\, d\theta_2\sqrt{-g_4^{ind}}H(\rho,\theta_2)
\ee
numerically, as a function of $\mu$. To facilitate this we will expand \eqref{H} up to $\ell=1$, the dipole term, and take the large $\rho$ limit
\be
H(\rho,\theta_2)=\frac{L^4}{2\rho^4}\left[1+\left.3(2\ell+1) \frac{N_w}{N} \frac{u^2}{\rho^2}P_\ell(\cos\theta_2)\right|_{\ell=1}\right]=H_{\ell=0}(\rho)+\delta H_{\ell=1}(\rho,\theta_2)\quad.
\ee
\begin{figure}[t]
\begin{center}\vspace{-1.cm}
      \includegraphics[width=3.2in]{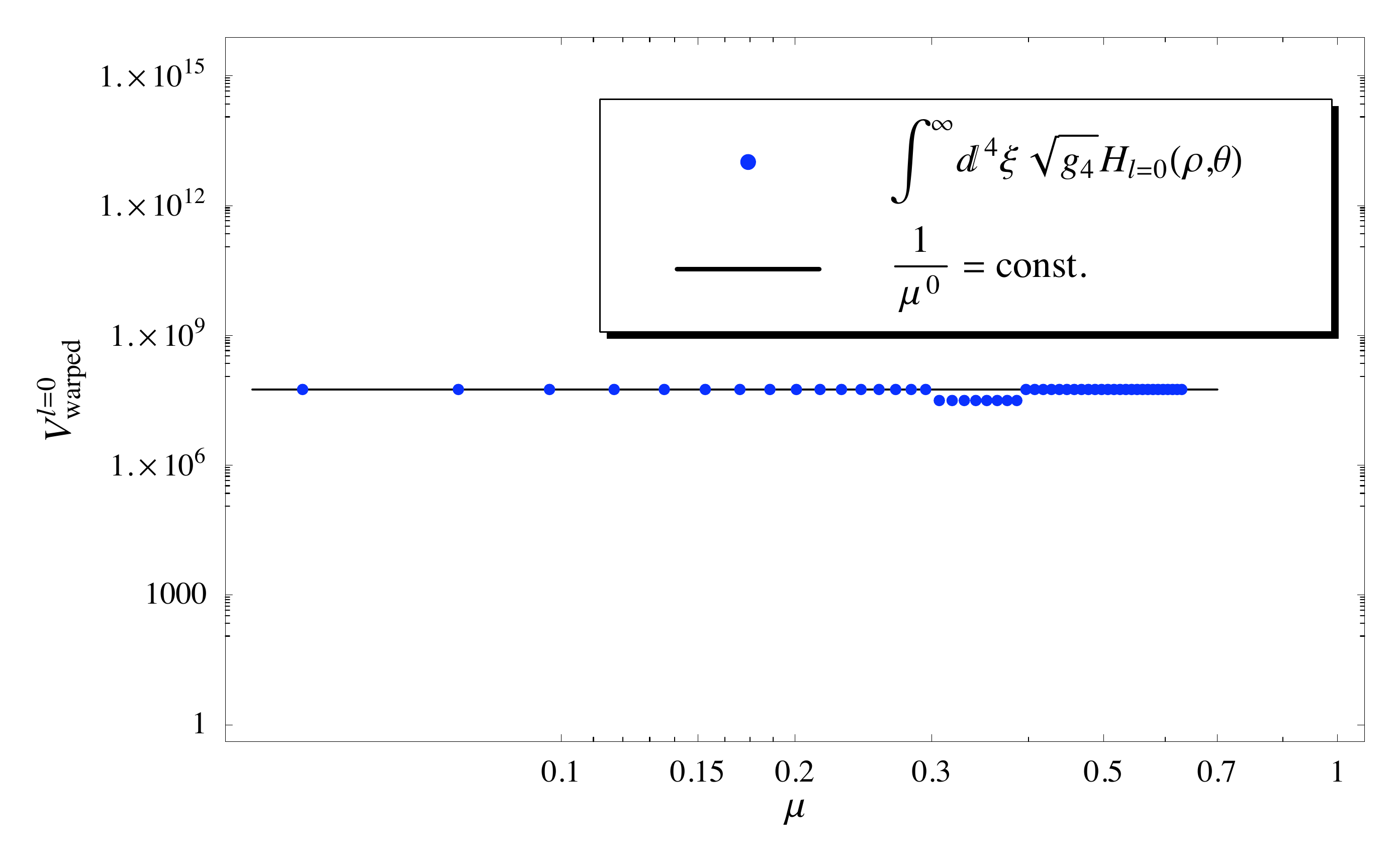}
       \includegraphics[width=3.2in]{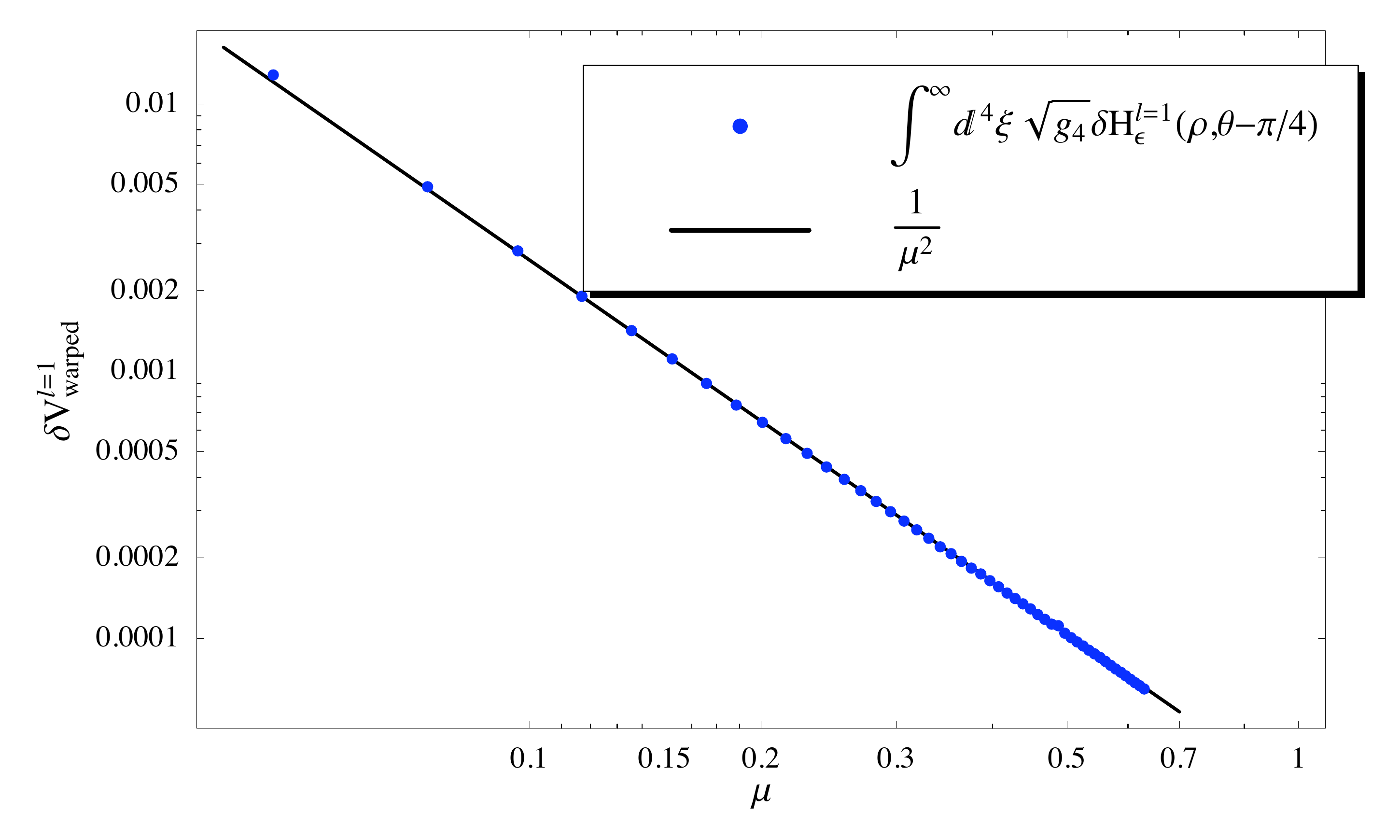}\\
       \includegraphics[width=3.2in]{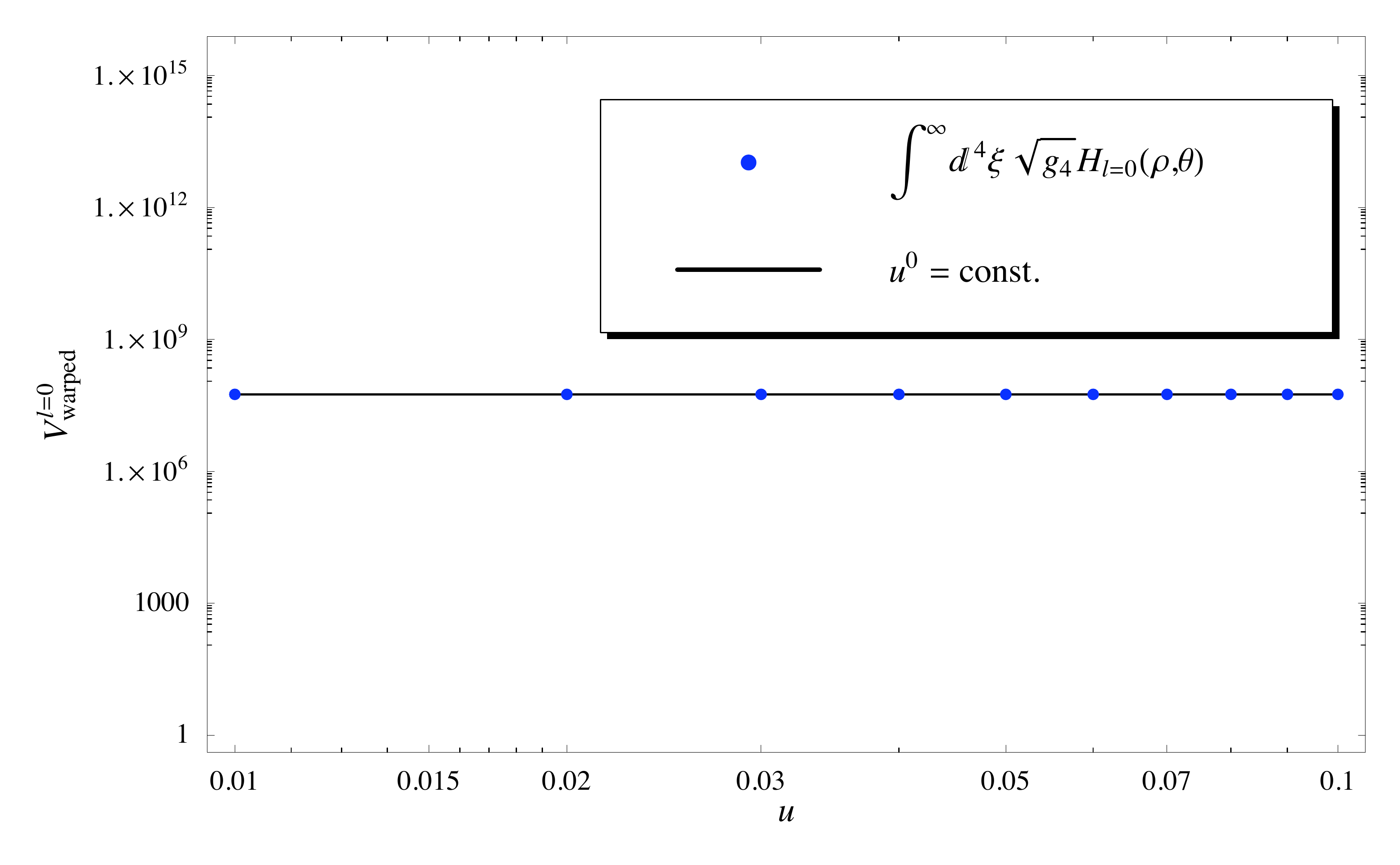}
       \includegraphics[width=3.2in]{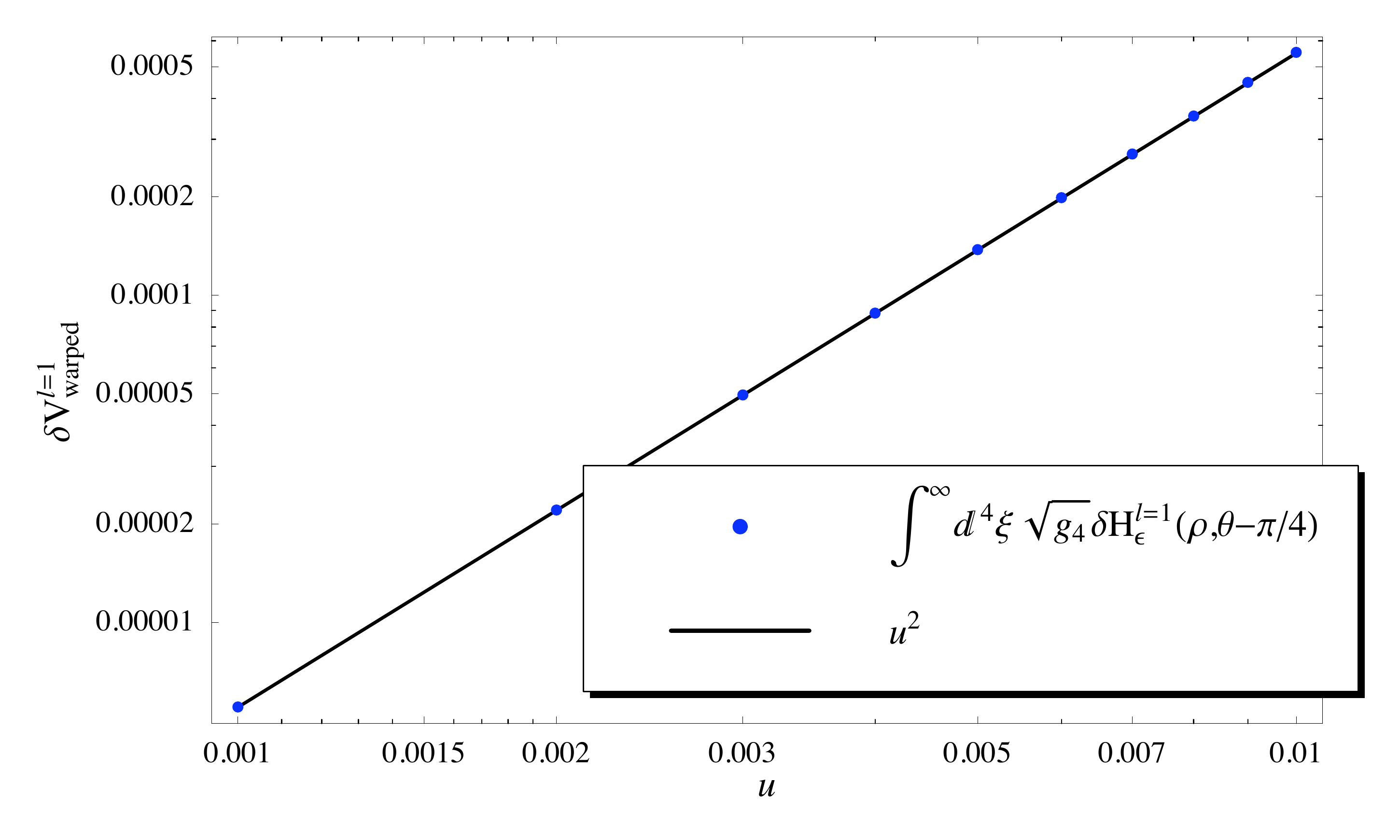}
       \caption{1st row: Plot of ${\cal V}_{warped}^{(0)}$ and $\delta{\cal V}_{warped}(\delta\theta_2)$ as functions of $\mu$ at constant $u=0.01$. 2nd row: Plot of ${\cal V}_{warped}^{(0)}$ and $\delta{\cal V}_{warped}(\delta\theta_2)$ as functions of $u$ at constant $\mu=0.1$. The leading $\ell=0$ term scales as $u^0\mu^0=const.$ while the $\ell=1$ dipole term scales as $(u/\mu)^2$. Note that the $\ell=0$ scaling ensues only in the strictly noncompact limit ({\it i.e.} when the integration goes all the way $\rho\to\infty$), while for a finite cutoff, resembling a crude approximation to a compact setting, there remains a weak dependence of the $\ell=0$ term on $\mu$, of the form $(u/\mu)^\delta$, where $\delta\to0$ for $\rho_{bulk}\to\infty$. For the example we have chosen $\epsilon\equiv N_w/N=0.1$.}\label{fig.2}
\end{center}
\end{figure}
We can now calculate
\bea
{\cal V}_{warped}^{(0)}&=&4\pi^2 \int_0^\pi d\theta_2\int_{\rho_{min}(\theta_2)}^{\rho_R}d\rho\sqrt{-g_4^{ind}}H_{\ell=0}(\rho)\\
\delta{\cal V}_{warped}^{\ell=1}(\delta\theta_2)&=&4\pi^2\int_0^\pi d\theta_2\int_{\rho_{min}(\theta_2)}^{\rho_R}d\rho\sqrt{-g_4^{ind}}\delta H_{\ell=1}(\rho,\theta_2+\delta\theta_2)
\eea
where $\rho_R\gg 1$ denotes a UV cutoff to compactify the resolved conifold geometry for the purpose of integration, and $\delta\theta_2$ denotes the angular misalignment of the D3-brane dipole configuration with respect to the four-cycle symmetry axis at $\theta_2=\pi/2$.

As $\sqrt{g_4^{ind}}$ is a symmetric function with respect to $\theta_2=\pi/2$ and $\delta H_{\ell=1}(\rho, \theta_2)$ is anti-symmetric with respect to $\theta_2=\pi_2$, we immediately find $\delta{\cal V}_{warped}(\delta\theta_2=0)=0$. So by fine-tuning a $\mathbb{Z}_2$-symmetric configuration we can forbid the $\ell=1$ term in the warped volume, whose corrections in this case start with the $\ell=2$ quadrupole terms.


We will now display the numerical results for the case $\delta\theta_2=-\pi/4$ in which the $\ell=1$ term will not vanish under the integral, and compare the scaling with $\mu$ between $\delta{\cal V}_{warped}(\delta\theta_2)$ and ${\cal V}_{warped}^{(0)}$. This is displayed in Fig.~\ref{fig.2}. We see clearly that the leading $\ell=0$ term scales as $u^0\mu^0=const.$ while the $\ell=1$ dipole term scales as $(u/\mu)^2$. Therefore, the $\ell=1$ dipole term has a parametric suppression $(u/\mu)^2$ relative to the leading $\ell=0$ term, and can therefore be made parametrically small (even in the non-$\mathbb{Z}_2$-symmetric general situation) in the limit where the four-cycle recedes far from the resolution $S^2$ ({\it i.e.} in the limit of large $\mu/u$).

Let us finally note that this relative suppression of the $\ell=1$ term with $(u/\mu)^2$ might have been guessed without any integration, as the integration boundary tells us that $\rho_{min}(\theta_2)\geq \rho_{min}(\pi/2)$, which corresponds to $r>2\mu^{3/2}$ or $\rho\gtrsim\mu$, and thus the relative scaling $u^2/\rho^2$ should be replaced by the scaling $u^2/\mu^2$.


\section{The Kaluza-Klein Spectrum}\label{a:KK}

In this appendix we obtain the (5+1)-dimensional effective action for a D5-brane wrapped on a two-cycle with $\int B \neq 0$. We show how a Kaluza-Klein reduction to four dimensions leads to masses that are suppressed with respect to the fluxless case. We then comment on the consequences of these light KK modes for axion monodromy inflation.

\subsection{The effective theory}

The DBI action for a D5-brane is
\be
S=T_5\int d^4x\,dy\,dz \sqrt{-{\rm{det}}\left(G^{ind}_{ab}+\FF_{ab}\right)}\,,
\ee
where $y,z$ are two coordinates in the internal space, which we take to be toroidal for the purpose of this derivation.  The indices are defined as follows: worldvolume indices are $a,b=0,\dots,5$; spacetime indices are $\mu,\nu=0,\dots,3$ as usual; ten-dimensional indices are $M,N=0,\dots,9$; six-dimensional compact indices are $m,n=4,\dots,9$; and indices transverse to the D5-brane are $i,j=6,\dots,9$.
We first expand the square root using
\be\label{sqrtexp}
\sqrt{{{\rm{det}}}(M_0+\delta M)}&=&\sqrt{{\rm{det}} M_0}\left\lbrace 1+\frac12 {\rm Tr}(M_0^{-1}\delta M)+\frac18 [{\rm Tr}(M_0^{-1}\delta M)]^2\right.\\
&& \qquad \left.-\frac14 {\rm Tr}(M_0^{-1}\delta M M_0^{-1}\delta M)+\dots \right\rbrace\,.
\ee
We will consider a background with two-form flux on the two-cycle
\be
\int \FF= \int B= \int dy\wedge dz B_{yz}(x,y,z)=b(x)=b\,,
\ee
{\it i.e.} the four-dimensional axion field $b(x)$ has a homogeneous expectation value that is approximately constant, up to terms suppressed by the slow-roll parameters. So the background is given by
\be
B_{MN}&=&b \delta_{My}\delta_{Nz}-b \delta_{Mz}\delta_{Ny}\,,\quad F_{ab}=0\,,\\ ds_{10}^{2}&=&g_{\mu\nu}dx^{\mu}{\nu}+g_{yy}dy^2+g_{zz}dz^2+2 g_{yz}dydz+g_{ij}dy^{i}dy^{j}\,.
\ee
Hence
\be
(M_0)_{ab}=\left( \begin{array}{ccc}
                   g_{\mu\nu}& &\\
		   &g_{yy}&g_{yz}+b\\
		   &g_{zy}-b&g_{zz}
                  \end{array}
 \right) \,.
\ee
The perturbations are
\be
(\delta M)_{ab}=\partial_a X^i\partial_b X^j (g_{ij}+B_{ij})+F_{ab}+\delta B_{ab}\,.
\ee
The calculation is simplified by the block-diagonal form of the background $M_0$. The $2\times2$ block is the sum of a symmetric and an antisymmetric piece that we call $S$ and $A$ respectively. We have that
\be
 {\rm{det}}(A+S)&=&{\rm{det}}(A)+{\rm{det}}(S)\,,\\
 (S+A)^{-1}&=&S^{-1} \frac{{\rm{det}}(S)}{{\rm{det}}(A)+{\rm{det}}(S)}+A^{-1} \frac{{\rm{det}}(A)}{{\rm{det}}(A)+{\rm{det}}(S)}\,,
\ee
which substantially simplifies the calculation. Using \eqref{sqrtexp} we get at leading order
\be
S&=&T_5\int d^4x\,dy\,dz \sqrt{-g_4}\sqrt{g_2+b^2}\Big[1+\frac12 \partial_\mu X^i \partial^\mu X_i \\
&& \quad+\frac12\frac{g_2}{g_2+b^2}\left(\partial_y X^i \partial^y X_i+\partial_z X^i \partial^z X_i\right) \\
&& \quad+\frac12\frac{2b}{g_2+b^2}\left(\partial_y X^i \partial_z X^j \delta B_{ij}+F_{yz}+\delta B_{yz}\right) \Big]+\dots\,,
\ee
where $g_2\equiv g_{yy}g_{zz}-g_{yz}^2$. After a KK reduction one finds the four-dimensional kinetic and potential terms, in the first line, as well as the Kaluza-Klein mass terms, in the second line.
The Kaluza-Klein masses in the presence of fluxes are
\be\label{suppressed}
m_{bKK}^2=\frac{g_2}{g_2+b^2}m_{KK}^2\,,
\ee
where $m_{KK}$ are the Kaluza-Klein masses in the absence of fluxes. This leads to the central point of this appendix: for $b\gg 1$, the Kaluza-Klein masses are {\it{suppressed}} by a factor of $\sqrt{g_2}/b\simeq L^2/b\ll 1$.\footnote{We have assumed for simplicity that the internal space is isotropic, with typical size $L\sqrt{\al}$.}
This phenomenon is intuitively understood in the T-dual picture in which flux becomes the angle of the D-brane. A large flux means that the T-dual brane winds around the torus many times, and thus becomes quite long. The Kaluza-Klein reduction of the fields living on the worldvolume of the T-dual brane therefore produces b-suppressed Kaluza-Klein masses.


\subsection{Effects of the light Kaluza-Klein modes}

Throughout this paper we have been careful to work in parameter ranges for which the typical Kaluza-Klein mass scale $m_{KK}$ obeys $m_{KK}\gg H$, as required for a consistent four-dimensional analysis of inflation. However, from (\ref{suppressed}) we learn that a subclass of Kaluza-Klein modes, namely those associated with transverse excitations of the fivebrane, have considerably smaller masses, $m_{bKK} \ll m_{KK}$.  For the numerical examples we have considered, we find that, very roughly, $m_{bKK} \sim (f_c/f)H$, where $f_c$ is a fiducial value of the decay constant, $f_c \sim 10^{-2}\Mpl$.  Therefore, for constructions with small values of $f$, the transverse excitations of the fivebrane can be lighter than $H$.

We leave a comprehensive study of this constraint for future work, as a proper implementation plausibly requires a more explicit compact model that we have been able to present in this work.  In particular, one should carefully compute the Kaluza-Klein mass, incorporating anisotropy in the geometry, warping, and, as we have explained above, the effect of worldvolume two-forms. To accomplish this, one needs a reasonably explicit construction of the warped throat region, of the two-cycle within the throat, and of the gluing of the throat into the compact space, which are beyond the scope of this work.

In this appendix, we will restrict ourselves to some qualitative statements that explain how our inflationary analysis can be consistent even in parameter regimes for which $m_{bKK}$ is slightly smaller than $H$.   Broadly speaking, one might worry about corrections to the inflationary Lagrangian, and about new contributions to the cosmological perturbations.  Concerning the first point, we remark that the excitations of the fivebrane depend on the inflaton expectation value only through their masses.  Therefore, the primary correction to the background evolution from these light modes would come if large numbers of Kaluza-Klein particles were produced by the time-dependent background.  In practice, the particle production is negligible, as can be seen by computing the adiabatic parameter $\dot m_{bKK} /m_{bKK}^2$ and substituting the constraints on the volume, and hence on the Kaluza-Klein mass, from \S\ref{s:const}.

More generally, let us stress that only a small subclass of the Kaluza-Klein modes  (a small portion of the tower of excitations of the fivebrane) have masses smaller than $H$.  From the viewpoint of the inflationary analysis, these fields constitute a small number of harmless spectators.  These light fields will fluctuate, absorbing energy, but this yields
a very small correction unless the number of fields approaches $(\Mpl/H)^2$.  Moreover, any entropy perturbations produced by these fields can turn into visible isocurvature perturbations only if their decays are distinct from that of the inflaton.  Although we have not specified a concrete reheating mechanism, one can argue that the most straightforward scenario involves visible sector degrees of freedom that are well-separated in the compact space from the inflationary fivebranes. Thus, we expect that excitations of the fivebranes will not give visible isocurvature perturbations, because they must first decay \cite{Barnaby:2004gg, Kofman:2005yz} to degrees of freedom localized in the inflationary throat, just as the inflaton does, and will plausibly do so  with rather similar couplings, as the modes correspond to small excitations of the NS5-brane that drives inflation.

\section{Numerical Examples}\label{a:num}

In this appendix, we specify two different sets of intersection numbers and show the relevant formulas for the volumes. For these two toy models, we explicitly performed the moduli stabilization outlined in \S\ref{ss:toy}, finding numerical values leading to the dot in figure \ref{point}.


\subsection{Intersection numbers: set I}

We consider as a toy-model Calabi-Yau manifold one with $H^{1,1}_+={\rm span}(\omega^L,\omega^+)$ for the orientifold-even homology two-cycles and $H^{1,1}_-={\rm span}(\omega^-)$ for the orientifold-odd homology two-cycles. We assume the following simple set of intersection numbers
\be
c_{LLL}=c_{LL+}=c_{+--}=1\,,
\ee
with all the others vanishing. We believe that, although very simplistic, the above toy model captures the relevant features of more realistic constructions. Notice that the intersection numbers in a basis for the homology of the covering space of the orientifold, {\it i.e.} without a definite parity with respect to the orientifold projection, are just linear combinations of those given above.

Using the standard relations
\be
\V_E&=&\frac16 c_{\alpha \beta \gamma} v^\alpha v^\beta v^\gamma\,,\quad \tau_\alpha=\partial_{v^\alpha}\V_E=\frac 12
c_{\alpha \beta \gamma} v^\beta v^\gamma\,,
\ee
one finds 
%
%
\be
v_L=\sqrt{2\tau_+},\,v_+=\frac{\tau_L-\tau_+}{\sqrt{2\tau_+}}\,,
\ee
and
\be
\V_E=\frac{\sqrt{2\tau_+}}{2}\tau_L-\frac{\sqrt{2}}{6}\tau_+^{3/2}\,.
\ee


\subsection{Intersection numbers: set II}

Again assuming $H^{1,1}_+={\rm span}(\omega^L,\omega^+)$ and $H^{1,1}_-={\rm span}(\omega^-)$, we consider the intersection numbers
\be
c_{LLL}=c_{L++}=c_{+--}=1\,,
\ee
with all the others vanishing. We find
\be
v_L=\frac{1}{\sqrt{2}}\left( (\tau_L+\tau_+)^{1/2}+(\tau_L-\tau_+)^{1/2}\right) ,\,v_+=\frac{1}{\sqrt{2}}\left( (\tau_L+\tau_+)^{1/2}-(\tau_L-\tau_+)^{1/2}\right)\,,
\ee
and
\be
\V_E=\frac{1}{3\sqrt{2}}\left( (\tau_L+\tau_+)^{3/2}+(\tau_L-\tau_+)^{3/2}\right)\,.
\ee


\end{document}